\documentclass[longauth]{aa} % for a printer version
\usepackage{graphicx}
\usepackage{txfonts}
\usepackage{natbib}
\def\hho{H$_2$O}
\def\hhoe{H$_2^{18}$O}
\def\hhos{H$_2^{17}$O}
\def\hh{H$_2$}
\def\hhhp{H$_3^+$}

\def\ammo{NH$_3$}
\def\thco{$^{13}$CO}
\newcommand{\amin}{$^{\prime}$}                   
\newcommand{\asec}{$^{\prime \prime}$}
\newcommand{\adeg}{$^{\circ}$}
\def\mic{$\mu$m}
\def\kms{km\,s$^{-1}$}
\def\pow#1#2{#1$\times$10$^{#2}$}
\def\eup{$E_{\rm up}$}
\def\scm{cm$^{-2}$}
\def\ccm{cm$^{-3}$}
\def\msol{M$_{\odot}$}
\def\lsol{L$_{\odot}$}
\def\dv{$\Delta${\it V}}
\def\vlsr{$V_{\rm LSR}$}
\def\gtsim{{_>\atop{^\sim}}}
\def\ltsim{{_<\atop{^\sim}}}

\def\tkin{$T_{\rm kin}$}
\def\tmb{$T_{\rm mb}$}
\def\menv{$M_{\rm env}$}
\def\lbol{$L_{\rm bol}$}
\def\newer#1{{#1}}
\def\new#1{{#1}}

\begin{document}
\title{Water in star-forming regions with Herschel (WISH).
\thanks{\textit{Herschel} is an ESA space observatory with science instruments provided
by European-led Principal Investigator consortia and with important participation from NASA}}
\subtitle{IV. A survey of low-$J$ \hho\ line profiles toward high-mass protostars}
\titlerunning{Low-$J$ \hho\ line profiles toward high-mass star-forming regions}

\author{F.F.S. van der Tak \inst{\ref{sron},\ref{rug}} \and
        L. Chavarr\'{\i}a  \inst{\ref{bordeaux},\ref{cnrs},\ref{madrid}} \and
        F. Herpin \inst{\ref{bordeaux},\ref{cnrs}} \and
        F. Wyrowski \inst{\ref{mpifr}} \and 
        C.M. Walmsley \inst{\ref{arcetri},\ref{dublin}} \and
        E.F. van Dishoeck \inst{\ref{leiden},\ref{mpe}} \and
        A.O. Benz \inst{\ref{eth}} \and
        E.A. Bergin \inst{\ref{mich}} \and
        P. Caselli \inst{\ref{leeds}} \and
        M.R. Hogerheijde \inst{\ref{leiden}} \and
        D. Johnstone \inst{\ref{jac}} \and
        L.E. Kristensen \inst{\ref{leiden},\ref{cfa}} \and
        R. Liseau \inst{\ref{onsala}} \and
        B. Nisini \inst{\ref{rome}} \and
        M. Tafalla \inst{\ref{oan}}
        }

\institute{SRON Netherlands Institute for Space Research, Landleven 12, 9747 AD Groningen, The Netherlands; \email{vdtak@sron.nl} \label{sron} \and
          Kapteyn Astronomical Institute, University of Groningen, The Netherlands \label{rug} \and
          Univ. Bordeaux, LAB, Floirac, France \label{bordeaux} \and
          CNRS, LAB, Floirac, France \label{cnrs} \and
          Centro de Astrobiolog\'{\i}a (CSIC-INTA), Madrid, Spain \label{madrid} \and
          Max-Planck-Institut f\"ur Radioastronomie, Bonn, Germany \label{mpifr} \and
          INAF -- Osservatorio Astrofisico di Arcetri, Firenze, Italy \label{arcetri} \and
          Dublin Institute for Advanced Studies, Dublin, Ireland \label{dublin} \and
          Leiden Observatory, Leiden University, The Netherlands \label{leiden} \and
          Max-Planck-Institut f\"ur Extraterrestrische Physik, Garching, Germany \label{mpe} \and
          Institute of Astronomy, ETH Z\"urich, Switzerland \label{eth} \and
          Department of Astronomy, University of Michigan, USA \label{mich} \and
          School of Physics and Astronomy, University of Leeds, UK \label{leeds} \and
          Joint Astronomy Center, Hilo, Hawaii, USA \label{jac} \and
          Harvard-Smithsonian Center for Astrophysics, Cambridge MA, USA \label{cfa} \and
          Onsala Space Observatory, Chalmers University of Technology, Sweden \label{onsala} \and
          INAF -- Osservatorio Astrofisico di Roma, Italy \label{rome} \and
          Observatorio Astron\'omico Nacional, Madrid, Spain \label{oan}
}

\date{Received 21 December 2012 / Accepted 8 April 2013}

\abstract
% context 
{Water is a key constituent of star-forming matter, but the origin of its line emission and absorption during high-mass star formation is not well understood.}
% aims 
{We study the velocity profiles of low-excitation \hho\ lines toward 19 high-mass star-forming regions and search for trends with luminosity, mass, and evolutionary stage.}
% methods 
{We decompose high-resolution \textit{Herschel}-HIFI line spectra near 990, 1110\,and 1670\,GHz into three distinct physical components. Dense cores (protostellar envelopes) are usually seen as narrow absorptions in the \hho\ 1113 and 1669\,GHz ground-state lines, the \hho\ 987\,GHz excited-state line, and the \hhoe\ 1102\,GHz ground-state line. In a few sources, the envelopes appear in emission in some or all studied lines, indicating higher temperatures or densities.
Broader features due to outflows are usually seen in absorption in the \hho\ 1113\,and 1669\,GHz lines, in 987\,GHz emission, and not seen in \hhoe, indicating a lower column density and a higher excitation temperature than the envelope component. A few outflows are detected in \hhoe, indicating higher column densities of shocked gas.
In addition, the \hho\ 1113 and 1669 GHz spectra show narrow absorptions by foreground clouds along the line of sight. The lack of corresponding features in the 987\,GHz and \hhoe\ lines indicates a low column density and a low excitation temperature for these clouds, although their derived \hho\ ortho/para ratios are close to 3.}
% results 
{The intensity of the ground state lines of \hho\  at 1113 and 1669 GHz does not show significant trends with source luminosity, envelope mass, or evolutionary state. 
In contrast, the flux in the excited-state 987 GHz line appears correlated with luminosity and the \hhoe\ line flux appears correlated with the envelope mass. 
Furthermore, appearance of the envelope in absorption in the 987 GHz and \hhoe\ lines seems to be a sign of an early evolutionary stage, as probed by the mid-infrared brightness and the \lbol/\menv\ ratio of the source.}
% conclusions 
{The ground state transitions of \hho\ trace the outer parts of the envelopes, so that the effects of star formation are mostly noticeable in the outflow wings.
These lines are heavily affected by absorption, so that line ratios \newer{of H$_2$O involving the ground states must be treated with caution, especially if multiple clouds are superposed as in the extragalactic case}.
The isotopic \hhoe\ line appears to trace the mass of the protostellar envelope, indicating that the average \hho\ abundance in high-mass protostellar envelopes does not change much with time.
The excited state line at 987 GHz increases in flux with luminosity and appears to be a good tracer of the mean weighted dust temperature of the source, \new{which may explain why it is readily seen in distant galaxies}.
%\old{Further evolutionary effects are probably masked by source confusion within the Herschel beam.}
}


\maketitle

\section{Introduction}
\label{s:intro}

%%%%%%%%%%%%%%%%%%%%%%%%%%%%%%%%%%%%%%%%%%%%%%

\begin{table*}[t]
\begin{flushleft}
\caption{Source sample.}
\label{t:sample}
%\resizebox{\hsize}{!}{   
\begin{tabular}{llllccc}
  \hline \hline
\noalign{\smallskip}
%   &   \multicolumn{2}{c}{Source Coordinates}    &   & \multicolumn{2}{c}{Properties} & \multicolumn{2}{c}{Settings} & Comments$^c$  \\
% \cline{2-3} \cline{5-6} \cline{7-8}       \\
Source     & RA  (J2000.0) & Dec  & $V_{\rm LSR}$  & $L_{\rm bol}$ & $d$ & Ref. \\
%HIFI$^a$   & PACS$^b$ &  \\
%\noalign{\smallskip}
&    h m s         & \adeg\ \amin\ \asec\ & km\,s$^{-1}$ & $L_\odot$ & kpc  & \\
%\noalign{\smallskip}
\noalign{\smallskip}
\hline
%\noalign{\smallskip}
%\noalign{\smallskip}
%{\bf Pre-stellar cores}&    &    &        &    &      &     &        &   \\
%\noalign{\smallskip}
%G11.11$-$0.12-NH$_{\rm 3}$-P1   & 18 10 33.9    &  $-$19 21 48   &  +30.4      & $-$     & 3.6 & 1+0  &0  &  \\
%G11.11$-$0.12-SCUBA-P1         & 18 10 28.4    &  $-$19 22 29   &  +29.2      & $-$     & 3.6 &  1+0 & 0 & \\
%G28.34+0.06-NH$_{\rm 3}$-P3     & 18 42 46.4    &  $-$04 04 12   &  +80.2      & $-$     & 4.8 &  1+0 & 0 &  \\
%G28.34+0.06-SCUBA-P2           & 18 42 52.4    &  $-$03 59 54   &  +78.5      & $-$     & 4.8 &  1+0 & 0 &  \\

\noalign{\smallskip}
{\bf mid-IR-quiet HMPOs}&    &    &        &    &      & \\
\noalign{\smallskip}
IRAS 05358+3543 & 05 39 13.1    &   +35 45 50   & $-$17.6   & $6.3\,\times\,10^3$     & 1.8   & 1, 1 \\
IRAS 16272--4837 & 16 30 58.7    & $-$48 43 55   & $-$46.2   & $2.4\,\times\,10^4$     & 3.4  & 2, 2 \\
NGC 6334I(N)        & 17 20 55.2    & $-$35 45 04      & $-$4.5    & $1.9\,\times\,10^3$     & 1.7  & 3, 4 \\
W43 MM1                & 18 47 47.0    & $-$01 54 28   & +98.8     & $2.3\,\times\,10^4$     & 5.5  & 5, 6 \\
DR21(OH)               & 20 39 00.8    &   +42 22 48   & $-$4.5    & $1.3\,\times\,10^4$     & 1.5  & 7, 8 \\

\noalign{\smallskip}
{\bf mid-IR-bright HMPOs}&    &    &        &    &     &  \\
\noalign{\smallskip}
W3 IRS5                   & 02 25 40.6    &   +62 05 51   & $-$38.4   & $1.7\,\times\,10^5$     & 2.0   & 9, 10 \\
IRAS 18089--1732 & 18 11 51.5    & $-$17 31 29   & +33.8     & $1.3\,\times\,10^4$     & 2.3  & 11, 12 \\
W33A                        & 18 14 39.1    & $-$17 52 07   & +37.5     & \new{$4.4\,\times\,10^4$}     & \new{2.4}  & 11, \new{30} \\
IRAS 18151--1208 & 18 17 58.0    & $-$12 07 27   & +32.0     & $2.0\,\times\,10^4$     & 2.9 & 13, 13 \\
AFGL 2591              & 20 29 24.7    &   +40 11 19   & $-$5.5    & $2.2\,\times\,10^5$     & 3.3  & 14, 8 \\

\noalign{\smallskip}
{\bf Hot Molecular Cores}&    &    &        &    &    &  \\
\noalign{\smallskip}
G327$-$0.6         & 15 53 08.8    & $-$54 37 01   & $-$45.0   & $5.0\,\times\,10^4$   & 3.3  & 15, 16 \\
NGC 6334I          & 17 20 53.3    & $-$35 47 00   & $-$7.7    & $2.6\,\times\,10^5$     & 1.7 & 3, 4 \\
G29.96$-$0.02   & 18 46 03.8    & $-$02 39 22   & +98.7     & $3.5\,\times\,10^5$     & 6.0  & 17, 18\\
G31.41+0.31       & 18 47 34.3    & $-$01 12 46   & +98.8     & $2.3\,\times\,10^5$     & 7.9  & 19, 20 \\
%(IRAS20126+4104)   &  20 14 25.1  & $+$41 13 32      & $-$3.8   & $1.0\,\times\,10^4$     &  1.7 &  13+2  & 18 &  \\

\noalign{\smallskip}
{\bf Ultracompact H{\sc II} Regions}&    &    &        &    &    &   \\
\noalign{\smallskip}
G5.89$-$0.39 (W28A) & 18 00 30.4    & $-$24 04 02   & +10.0     & $5.1\,\times\,10^4$     & 1.3  &  21, 22 \\
G10.47+0.03                 & 18 08 38.2    & $-$19 51 50   & +67.0     & $3.7\,\times\,10^5$     & 5.8  & 23, 20 \\
G34.26+0.15                 & 18 53 18.6    &   +01 14 58   & +57.2     & $3.2\,\times\,10^5$      & 3.3   & 24, 25 \\
W51N-e1                       & 19 23 43.8    &   +14 30 26   & +59.5     & $1.0\,\times\,10^5$     & 5.1 & 26, 27 \\
NGC 7538-IRS1           & 23 13 45.3    &   +61 28 10   & $-$57.4   & $1.3\,\times\,10^5$     & 2.7  & 28, 29 \\
\noalign{\smallskip}
\hline
\end{tabular}
%  }
\tablefoot{The first reference is for the luminosity, the second for the distance. If the distance measurement is more recent than the luminosity estimate, the luminosity has been scaled to the new distance. The text uses "short" source names, which is the part preceding the $+$ or $-$ sign. 
\\
References: (1) \citet{beuther2002}; (2) \citet{garay2002}; (3) \citet{sandell2000}; (4) \citet{neckel1978}; (5) \citet{nguyen2011}; (6) \citet{lester1985}; (7) \citet{jakob2007}; (8) \citet{rygl2012}; (9) \citet{ladd1993}; (10) \citet{hachisuka2006}; (11) \citet{faundez2004}; (12) \citet{xu2011}; (13) \citet{sridharan2002}; (14) \citet{vdtak1999}; (15) \citet{urquhart2012}; (16) \citet{minier2009}; (17) \citet{cesaroni1998}; (18) \citet{pratap1999}; (19) \citet{mueller2002}; (20) \citet{churchwell1990}; (21) \citet{vdtak2000} (22) \citet{motogi2011}; (23) \citet{hunter2000}; (24) \citet{hatchell2003} (25) \citet{kuchar1994}; (26) \citet{vdishoeck2011}; (27) \citet{xu2009}; (28) \citet{sandell2004}; (29) \citet{moscadelli2009}; \new{(30) \citet{immer2013}}.}
\end{flushleft}  
\end{table*}

% role of water in star formation 
The water molecule is a key constituent of star-forming matter with a great influence on the formation of stars and planets\footnote{This paper uses "water" to denote the chemical species, and \hho, \hhoe\ and HDO to denote specific isotopologues. Unless otherwise indicated, the gas phase is meant.}. In the gas phase, water acts as a coolant of collapsing dense interstellar clouds; in the solid state, it enhances the coagulation of dust grains in protoplanetary disks to make planetesimals; and as a liquid, it acts as a solvent bringing organic molecules together on planetary surfaces, which is a key step towards biogenic activity.
The first role is especially important for high-mass star formation which depends on the balance between the collapse of a massive gas cloud and its fragmentation \citep{zinnecker2007}. This balance depends strongly on the temperature (through the Jeans mass), and the sensitivity of the \hho\ abundance to the temperature, \new{much larger than for CO}, should make it a useful probe of the high-mass star formation process, which is mostly unexplored due to observational difficulties.

% pre-Herschel work
Interstellar \hho\ is well known from ground-based observations of maser emission at centimeter (22\,GHz; \citealt{cheung1969}), millimeter (183\,GHz; \citealt{cernicharo1990}), and submillimeter (325\,GHz; \citealt{menten1990}) wavelengths.
The high intrinsic brightness of maser emission makes it useful as a signpost of dense gas and for kinematic studies of protostellar environments (e.g., \citealt{trinidad2003, sanna:2591}).
With VLBI techniques, the proper motions and parallaxes of \hho\ masers can be measured to micro-arcsecond accuracy, leading to accurate distance estimates for star-forming regions as distant as $\sim$10\,kpc \citep{sanna:parallax} and a revised picture of Galactic structure \citep{reid2009}.

Thermal \hho\ lines are useful as probes of physical conditions and the chemical evolution of star-forming regions, but generally cannot be observed from the ground.
Before Herschel, space-based submm and far-IR observations of \hho\ lines were made with ISO \citep{vdishoeck1996}, SWAS \citep{melnick2005} and Odin \citep{bjerkeli2009}, but these data do not have sufficient angular resolution to determine the spatial distribution of \hho. 
In contrast, space-based mid-IR and ground-based mm-wave observations of thermal \hho\ and \hhoe\ lines have high angular resolution but only probe the small fraction of the gas at high temperatures \citep{vdtak2006,watson2007,joergensen2010,wang2012}.  

% work with Herschel
The first results from the Herschel mission have demonstrated the potential of \hho\ observations of high-mass star-forming regions at high spatial and spectral resolution. 
Mapping of the DR21 region shows orders of magnitude variations in \hho\ abundance between various physical components (envelope, outflow, foreground clouds) due to freeze-out, evaporation, warm gas-phase chemistry, and photodissociation \citep{vdtak2010}. 
Multi-line observations of the high-mass protostar W3 IRS5 show broad emission as well as blueshifted absorption by \hho\ in the outflow, in particular in lines from excited states of \hho, underlining the importance of shock chemistry for \hho\ \citep{chavarria2010}.
Observations of the p-\hho\ and p-\hhoe\ ground state lines and the p-\hho\ first excited state line at 987\,GHz toward four high-mass star-forming regions indicate \hho\ abundances ranging from \pow{5}{-10} to \pow{4}{-8} without a clear trend with the physical properties of the sources \citep{marseille:hifi}.
Finally, observations of multiple \hho, \hhoe\ and \hhos\ lines toward the massive star-forming region NGC 6334I show high-excitation line emission from hot ($\sim$200\,K) gas, and a \hho\ abundance ranging from $\approx$10$^{-8}$ in cold quiescent gas to \pow{4}{-5} in warm outflow material \citep{emprechtinger2010}.
However, these papers study either single sources or small sets of sources, which makes a trend analysis inconclusive, if not impossible. 

% goal of this work
This paper presents observations of low-excitation lines of \hho\ and \hhoe\ toward 19 regions of high-mass star formation (Table~\ref{t:sample}). The velocity-resolved spectra show a mixture of emission and absorption due to various physical components, which we disentangle by fitting Gaussian profiles. The resulting line fluxes for the protostellar envelopes and the molecular outflows are compared with basic source parameters such as luminosity and mass, as well as to the evolutionary stage of the source as estimated in various ways.

To probe the bulk of the material in the protostellar environment, this paper focuses on the lowest excitation lines of \hho.
The advantage of p-\hho\ is that its two lowest-excitation lines are close in frequency (Table~\ref{t:lines}) so that the telescope beam is similar in size.
This coincidence allows us to make estimates of the column density and the excitation temperature of \hho\ which are almost independent of assumptions on the size or the shape of the object.
In addition, we use observations of the o-\hho\ ground-state line at 1669\,GHz to probe the kinematics of the sources, which have a high continuum brightness at this high frequency, so that absorption lines are easily detected. This paper does not discuss observations of the other o-\hho\ ground-state line at 557\,GHz, because of the large difference in frequency and beam size. Maps of the 557\,GHz line toward our sources will be presented elsewhere.

% source sample
Table~\ref{t:sample} presents the sources, which have been selected from several surveys (\citealt{molinari1996}, \citealt{sridharan2002}, \citealt{wood1989}, \citealt{vdtak2000}), where preference is given to nearby ($\ltsim$2\,kpc) objects and 'clean' objects where one source dominates the emission on $\approx$30$''$ scales, which is the relevant scale for our observations. Furthermore, the sources were selected to cover a range of evolutionary stages. In high-mass protostellar objects (HMPOs), the central star is surrounded by a massive envelope with a centrally peaked temperature and density distribution. These objects show signs of active star formation such as outflows and masers. This paper distinguishes mid-IR-bright and -quiet HMPOs, with a boundary of 10 Jy at 12 $\mu$m, which may be an evolutionary difference \citep{vdtak2000,motte2007,lopez-sepulcre2010}.
The presence of a 'hot core' (submillimeter emission lines from complex organic molecules) or an 'ultracompact H{\sc II} region' (free-free continuum emission of more than a few mJy at a distance of a few kpc) are thought to be signposts of advanced stages of protostellar evolution. Hot cores occur when a star heats its surroundings such that ice mantles evaporate off dust grains which alters the chemical composition of the circumstellar gas. When the stellar atmosphere becomes hot enough for the production of significant ultraviolet radiation, the circumstellar gas is ionized and an ultracompact H{\sc II} region is created.

This paper is organized as follows: Section 2 describes the observations and Section 3 their results. Section 4 presents a search for trends in the results with luminosity, mass or age, and Section 5 provides the discussion and our conclusions.
In addition to this overview paper, our group is preparing detailed multi-line studies of individual sources.

%%%%%%%%%%%%%%%%%%%%%%%%%%%%%%%%%%%%%%%%%%%%%%
\section{Observations and data reduction}
\label{s:obs}

\begin{table}
\caption{Observed lines.}
\label{t:lines}
\begin{tabular}{cccccc}
\hline \hline
\noalign{\smallskip}
Molecule & Transition  & Frequency  & $E_{\rm up} $ & $\theta_{\rm mb} $ & $\sigma_{\rm mb}$ \\
                 &                      & GHz             & K                       & arcsec                       & mK \\ 
\noalign{\smallskip}
\hline
\noalign{\smallskip}
\hho\      & $1_{11}$--$0_{00}$ & 1113.343  & \phantom{1}53.4   & 19.2 &  40 \\
\hhoe\    & $1_{11}$--$0_{00}$ & 1101.698  & \phantom{1}53.4   & 19.2 &  40 \\
\hho\      &  $2_{02}$--$1_{11}$ & \phantom{1}987.927   & 100.8 &  21.5 &  50 \\
\hho\      &  $2_{12}$--$1_{01}$ & 1669.905 & 114.4                       &  12.7 & 75 \\
\noalign{\smallskip}
\hline
\noalign{\smallskip}
\end{tabular}
\tablefoot{The rms noise level in the last column is for HRS velocity resolution (0.5~MHz).}
\end{table}

The sources were observed with the Heterodyne Instrument for the Far-Infrared (HIFI; \citealt{degraauw2010}) onboard ESA's \textit{Herschel} Space Observatory \citep{pilbratt2010} in the course of 2010 and 2011.
Spectra were taken in double sideband mode using receiver bands 4b, 4a and 6b, respectively, for the $1_{11}$--$0_{00}$, $2_{02}$--$1_{11}$ and $2_{12}$--$1_{01}$ lines (Table~\ref{t:lines}). The $1_{11}$--$0_{00}$ lines of \hho\ and \hhoe\ were observed simultaneously. 
All data were taken using the double beam switch observing mode with a throw of $2\farcm5$ to the SW, as part of the Guaranteed Time key program \textit{Water in Star-forming regions with Herschel} (WISH; \citealt{vdishoeck2011}). See Appendix~\ref{s:obslog} for a detailed observing log.

Data were simultaneously taken with the acousto-optical Wide-Band Spectrometer (WBS) 
and the correlator-based High-Resolution Spectrometer (HRS), in both horizontal and vertical polarization.
This paper focuses on the HRS data, which cover 230\,MHz bandwidth at 0.48\,MHz ($\sim$0.13\,\kms) resolution,
although for a few sources where the 1669 GHz line is too broad to fit in the HRS backend, the WBS data are used, which cover 1140\,MHz bandwidth at 1.1\,MHz ($\sim$0.30\,\kms) resolution.
Spectroscopic parameters of the lines are taken from the JPL \citep{pickett1998} and CDMS \citep{mueller2001} catalogs. 
The adopted FWHM beam sizes are taken from \citet{roelfsema2012} and scaled to our observing frequencies.
System temperatures are 340--360\,K (DSB) at 1113\,GHz, 400--450\,K at 988\,GHz, and 1400-1500\,K at 1670\,GHz. 
Integration times are 6.4 minutes at 1113\,GHz (ON+OFF), 12 minutes at 988\,GHz and 34 minutes at 1670\,GHz.

The calibration of the data was performed in the \textit{Herschel} Interactive Processing
Environment (HIPE) version 6 or higher; further analysis was done within the CLASS\footnote{\tt http://www.iram.fr/IRAMFR/GILDAS} package, version of January 2011. 
The intensity scale was converted to $T_{mb}$ scale using main beam efficiencies from \citet{roelfsema2012}, and linear baselines were subtracted.
After inspection, data from the two polarizations were averaged together to obtain rms noise levels reported in the last column of Table~\ref{t:lines}. The calibration uncertainty is 15\% at $\sim$1000~GHz and 20\% at $\sim$1700~GHz (V. Ossenkopf and R. Shipman, priv. comm.)\footnote{See {\tt http://herschel.esac.esa.int/twiki/bin/view/} {\tt Public/HifiCalibrationWeb?template=viewprint}}.

\section{Results}
\label{s:res}

\begin{table}
\caption{SSB continuum flux densities (Jy/beam).}
\label{t:cont}
\begin{tabular}{lccc}
\hline \hline
\noalign{\smallskip}
Source & 987 GHz & 1108 GHz & 1669 GHz \\
              & (304 \mic) & (271 \mic) & (180 \mic) \\
\noalign{\smallskip}
\hline
\noalign{\smallskip}
IRAS 05358      & 146 & 235 & 447 \\
IRAS 16272      & 382 & 579 & 787 \\  
NGC 6334 I(N) & 892 & 1100 & 887 \\ 
W43 MM1          & 772 & 995 & 1073 \\  
DR21(OH)         & 896 & 1248 & 1678  \\ 
W3 IRS5              & 546 & 760 & 1896 \\
IRAS 18089        & 393 & 507 & 837 \\
W33A                   & 415 & 593 & 1216 \\ 
IRAS 18151        & 106 & 217 & 308 \\
AFGL 2591         & 310 & 467 & 830 \\ % 987 = actually at 995 GHz; scaling old point would be 293 Jy
G327-0.6             & 1206 & 1426 & 1621 \\ 
NGC 6334I          & 2186 & 2601 & 4624 \\ % 1669 = my previous point scaled up by 12%
G29.96                 & 510 & 688 & 1372 \\ 
% 987 = actually at 995 GHz, scaling old point would be 495 Jy; 1108 = actually at 1097 GHz; 1669 = my scaled point 
G31.41                 & 492 & 684 & 948 \\  
G5.89                   & 1020 & 1359 & 2908 \\ % 1108 = actually at 1097 GHz
G10.47                 & 914 & 1340 & 2368 \\ % 1108 = actually at 1097 GHz
G34.26                 & 1566 & 2272 & 3549 \\ 
W51N-e1             & 1894 & 2388 & 2468 \\ 
NGC 7538 IRS1 & 474 & 633 & 1037 \\ 
\noalign{\smallskip}
\hline
\noalign{\smallskip}
\end{tabular}
\tablefoot{The values for 1108\,GHz are averages of measurements near the \hho\ 1113 GHz and the \hhoe\ lines. }
\end{table}

\begin{figure}[t]
\centering
\includegraphics[width=6cm,angle=-90]{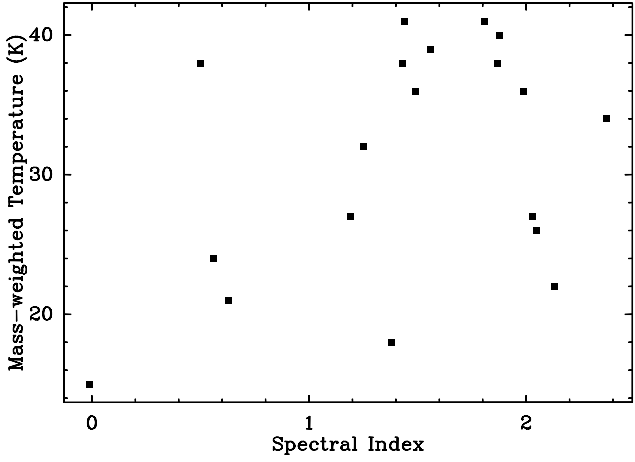}
\caption{\new{Spectral indices of our sources, calculated from the HIFI continuum levels at 987 and 1669 GHz, versus the mass-weighted temperatures calculated from the models in \S~\ref{ss:luis}.}}
\label{f:spindex}
\end{figure} 

\subsection{Continuum emission}
\label{ss:cont}

We have measured the continuum levels in the HIFI data by fitting zeroth- or first-order polynomials to the velocity ranges in the spectra without any appreciable line signal. 
In the cases of G327-0.6, NGC 6334I, and NGC6334I(N), the HRS backend is not broad enough to contain the 1669\,GHz line, so that we have used the WBS data instead. 
In general, the differences between continuum levels from the two backends \new{(and within polarization channels from a given backend)} are within 5\% near 1000\,GHz and within 10\% at 1670\,GHz, which is within the calibration uncertainty (\S\ref{s:obs}).
The flux densities $S_\nu$ in Jy/beam were derived from the measured DSB continuum levels by 
$$ S_\nu = \frac{ 8k T'_A}{2 \pi \eta_a D^2}$$ 
(\citealt{rohlfs-wilson}, Eq. 7.18) where the effective telescope diameter $D=3.28$\,m and the aperture efficiency $\eta_a \approx 0.65$ \citep{roelfsema2012} with a small frequency dependence due to Ruze losses. 
The factor 2 converts DSB to SSB signal, and $T'_A = \eta_f T_A^*$ with the forward efficiency $\eta_f = 0.96$ \citep{roelfsema2012}. 

As seen from Table~\ref{t:cont}, the continuum flux densities of our sources generally increase with frequency, as expected for thermal emission from warm dust. 
The free-free emission of our sources is $\ltsim$1~Jy which is negligible \citep{vdtak2005}.
For optically thick dust emission at a temperature $T_d$, $S_\nu = B_\nu$($T_d$) so that \new{in the Rayleigh-Jeans limit}, the spectral index $\alpha$ (defined as $S_\nu \propto \nu^\alpha$) equals 2, while for optically thin emission, the spectrum steepens to $\alpha = 2 + \beta$ = 3--4, depending on the far-infrared dust opacity index $\beta$, which is 1--2 in most dust models \citep{ossenkopf1994}. 
\new{However, at the high frequencies of our observations and the low average temperatures of our sources (15--40~K: Table~\ref{t:models}), deviations from the Rayleigh-Jeans approximation are significant, which reduce these expected spectral indices by $\approx$1 at $T_d = 40$~K and by $\approx$2 at $T_d = 15$~K.}

The spectral indices of our sources, calculated between 987 and 1670 GHz, are mostly between 1.2 and 2.4 \new{(Fig.~\ref{f:spindex})}, which indicates that the emission is close to optically thick.
\new{The models in \S\ref{ss:luis} also indicate continuum optical depths around unity at $\sim$1000 GHz.}
Four sources (NGC 6334I(N), W43 MM1, G327-0.6, and W51e) have lower spectral indices (between 0.0 and 0.5), which for compact sources would suggest low dust temperatures.
For extended sources (size $\gtsim$15$''$), the emission would be resolved at the highest frequencies, which would artificially flatten the spectrum. 
The model calculations in \S\ref{ss:luis} indicate median source sizes of 17.6, 14.2, and 8.5$''$ FWHM at 987, 1113, and 1669 GHz, which is 83\%, 75\%, and 67\% (i.e., most) of the corresponding beam size, suggesting that size effects play a minor role. 
\new{The importance of dust temperature is also suggested by Fig.~\ref{f:spindex} which shows that the sources with low spectral indices have low mass-averaged temperatures as calculated from the models in  \S\ref{ss:luis}.}
We conclude that our observed variations in the spectral index of the HIFI continuum emission are not due to variations in the source size, but rather to a low dust temperature,
\new{both directly and through deviation from the Rayleigh-Jeans law}.

\subsection{\hho\ lines}
\label{ss:profs}

\begin{figure}[t]
\centering
\includegraphics[width=7cm,angle=0]{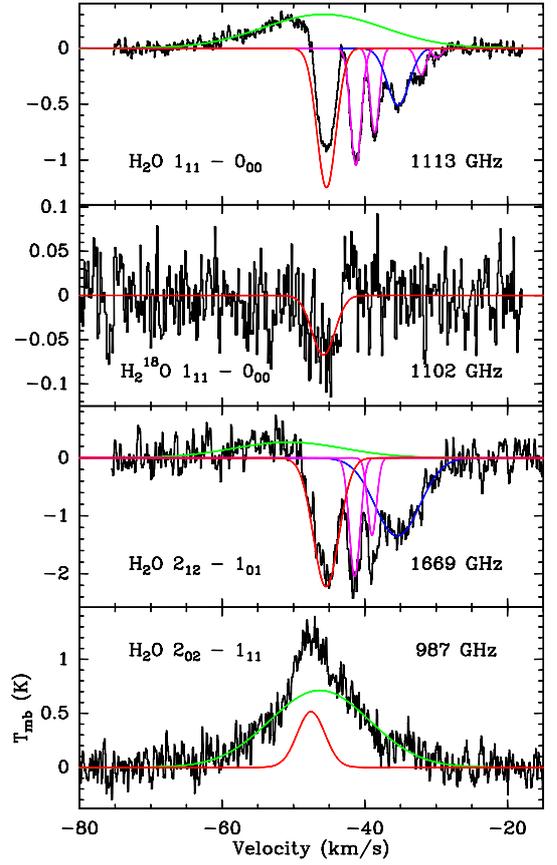}
\caption{Gaussian decomposition of the observed \hho\ line profiles (after continuum subtraction) toward IRAS 16272. The envelope component is drawn in red, the broad outflow in green, the narrow outflow in blue, and foreground clouds in purple.}
\label{f:decomp}
\end{figure} 

\begin{figure}[t]
\centering
\includegraphics[width=8cm,angle=0]{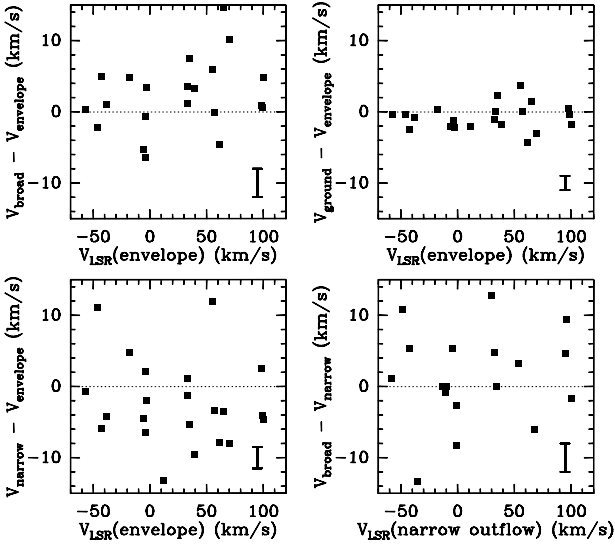}
\caption{Central velocities of the envelope, narrow outflow, and broad outflow components of the observed \hho\ lines compared with each other and with the systemic velocity from ground-based data. \new{In the bottom right-hand corner, the typical error bar on the velocities is shown}.}
\label{f:velocity}
\end{figure} 

Figures~\ref{f:05358}--\ref{f:7538} in Appendix~\ref{app:profs} show the observed line profiles. 
All four lines are detected toward all nineteen sources (except \hhoe\ toward IRAS 18151), and the line profiles are a mixture of emission and absorption features. 
Analogous to the case of DR21 \citep{vdtak2010}, three physical components appear from these observations (Fig.~\ref{f:decomp}).

First are the protostellar envelopes (also sometimes called dense cores) which usually appear in all four lines in absorption at an LSR velocity known from ground-based millimeter-wave emission line data (Fig.~\ref{f:velocity}, top right); the widths are also similar to the ground-based values. 
Several sources show two envelope components, due to binarity (e.g., W3 IRS5) or infall/outflow profiles (e.g., W33A, G34.26). 
The general appearance in absorption indicates a high \hho\ column density and a low excitation temperature for this component, which is further discussed in \S\ref{ss:env}.

Second are the molecular outflows which usually appear in 1113\,and 1669\,GHz absorption and 987\,GHz emission at velocity offsets of 5--10\,\kms\ from the envelope signal and with widths of 10--20\,\kms\ FWHM or sometimes more, up to 40\,\kms. 
Most sources show two outflow components: a broad (\dv\ $\sim$20\,\kms) and a narrower (\dv\ $\sim$10\,\kms) feature.
In some cases, the width of the narrow outflow feature is similar to that of the corresponding envelope, but the comparison to the ground-based velocity and the presence or absence of \hhoe\ emission or absorption allows for an unambiguous assignment.
The lack of \hhoe\ signal and the appearance in 987\,GHz emission suggest that the outflows have lower \hho\ column densities and higher excitation temperatures than the envelopes. 
Section~\ref{ss:outflow} discusses the outflow component in detail.

Third are foreground clouds, which appear as narrow (\dv$<$5\,\kms) 1113\,and 1669\,GHz absorption features at velocities offset by 5-50\,\kms\ from the envelope signal. 
The lack of signal in the 987\,GHz and \hhoe\ lines indicates a low \hho\ column density and a low excitation temperature for these components, without exception.
The number of foreground clouds \new{within the HRS bandwidth} ranges from zero toward IRAS 05358 and NGC 7538 to seven toward W51e.
Many of these foreground clouds are known from ground-based mm-wave observations of ground-state molecular lines such as CS, HCO$^+$ and HCN \citep{greaves1994,godard2010} and from [H{\sc I}] 21~cm observations \citep{fish2003,pandian2008}. 
Other foreground clouds have first been seen with HIFI, such as 
\new{the $V$=+24\,\kms\ cloud toward W51e which is seen in HF \citep{sonnentrucker2010} but not in our \hho\ data,}
and the $V$=+13\,\kms\ cloud toward AFGL 2591 \new{which is seen in \hho\ and HF but not in CO}
\citep{emprechtinger2012,vdwiel2012,choi2012}. 
\new{Some foreground clouds are intervening objects on the line of sight, while others represent the `outer envelopes' of large-scale molecular cloud complexes, such as the $V$=+65\,\kms\ cloud toward W51e \citep{koo1997} and two clouds toward NGC 6334 \citep{vdwiel2010}}.
See \S\ref{ss:fg-clouds} for further discussion of the foreground clouds.

Our decomposition of the line profiles is similar to the description of \hho\ line profiles toward low-mass protostars in terms of 'narrow', 'medium' and 'broad' components \citep{kristensen2010}, 
where narrow = envelope, medium = narrow outflow, and broad = broad outflow.
Because of the large range in luminosity and mass within our sample, we base the assignment of features to envelope or outflow not only on the line width, but also on the appearance in emission or absorption in the various lines.

\subsection{Profile decomposition and flux extraction}
\label{ss:decomp}

\begin{figure}[t]
\centering
\includegraphics[width=8cm,angle=0]{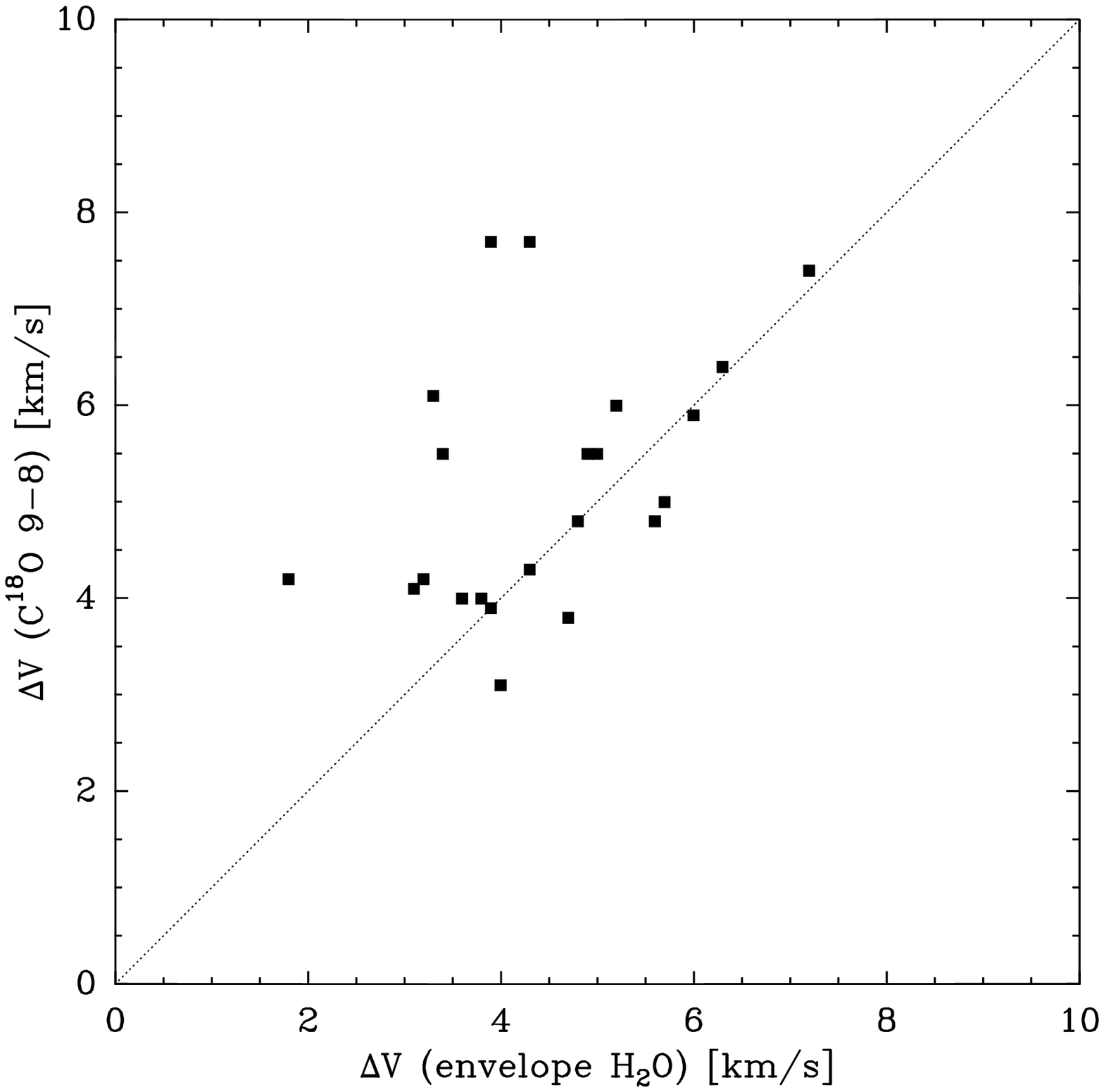}
\caption{Line widths (FWHM) of the envelope components of the observed \hho\ lines compared to the widths of the C$^{18}$O 9--8 line measured with HIFI \citep{irene2013}. Error bars are smaller than the plotting symbols.}
\label{f:irene}
\end{figure} 

\begin{figure}[t]
\centering
\includegraphics[width=7cm,angle=0]{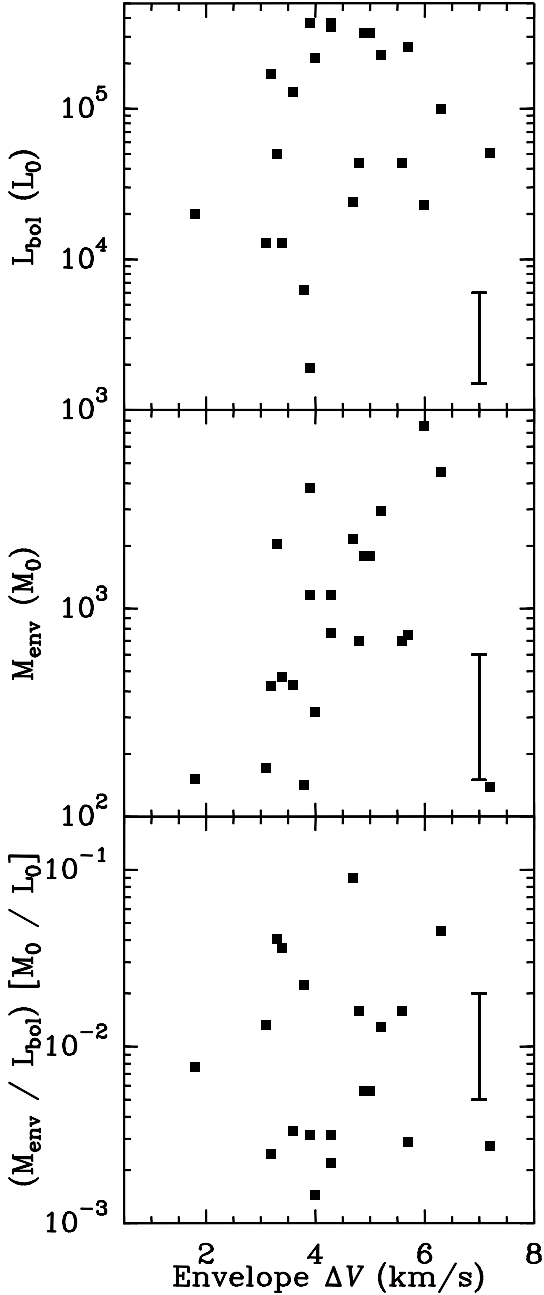}
\caption{Envelope line width measured from our spectra versus bolometric luminosity (top), envelope mass (middle), and mass/luminosity ratio (bottom). Sources with two envelope features are shown with two data points. In the bottom right-hand corner, the typical uncertainty in the masses, luminosities, and mass/luminosity ratios is shown.}
\label{f:lwidth}
\end{figure} 

We have measured the fluxes ($\int T_{\rm mb} dV$; in emission) and absorbances ($\int \tau dV$; in absorption) of the lines by fitting Gaussians to the observed profiles \new{for each line independently}. 
Figure~\ref{f:decomp} shows IRAS 16272 as an example of the procedure, and the results are presented in Table~\ref{t:env} for the envelopes, Table~\ref{t:flow} for the outflows, and Table~\ref{t:fg} for the foreground clouds. 
\new{Our fitting procedure is similar to that by \citet{flagey2013}, where the reader can find further details.}

Since some of the emission and absorption features in our data have non-Gaussian shapes, we have verified the results of the Gaussian fitting procedure by a "moment" analysis, which just adds up the flux in all spectrometer channels within a given velocity range. 
This alternative method extracts the flux of a spectral component independent of its shape, but it cannot handle blending of components, which occurs frequently in our data, unless further assumptions are made.
Another disadvantage of the moment method is that the definitions of the velocity ranges are often somewhat arbitrary.
After some experimenting, we conclude that fitting Gaussians gives more reliable results than taking moments, if we accept that in some cases, one physical component (usually the outflow) is represented by several (up to three) Gaussians. 
Another consequence of the non-Gaussian line shape is that the extracted central position and the width of a given component vary somewhat from line to line: \new{up to 2\,\kms\ for the envelopes and up to 5\,\kms\ for the broad outflow component}. 
The uncertainties on the line positions and widths in Tables~\ref{t:env}, \ref{t:flow} and \ref{t:fg} represent this line-to-line variation, whereas the uncertainties on the fluxes and absorbances are dominated by the overall calibration uncertainty (\S\ref{s:obs}), which is always larger than the formal error of the Gaussian fit caused by the deviation of the observed profile from a Gaussian shape. 

For absorption features, Tables \ref{t:env}--\ref{t:fg} give the absorbance (= the velocity-integrated optical depth):

$$ \int \tau dV = -1.06 \Delta V \ln (\frac{T_c - T_l}{T_c}) $$

where $T_l$ is the peak depth of the feature, $T_c$ is the SSB continuum temperature, and $\Delta V$ is the FWHM line width. 
For small optical depth, the error on $T_l$ propagates linearly into the absorbance, but if the absorption is almost saturated ($T_l \to T_c$), the error on the absorbance is much larger than that on the measured absorption depth. 
Cases where the error exceeds the absorbance due to saturation, which occurs mostly for envelope components, are marked with lower limits in Tables \ref{t:env}--\ref{t:fg}.

In the frequent case that absorption and emission features are partially blended, we have estimated the line contribution to the absorption background from the spectra, and added the result to the continuum brightness. This procedure assumes that the absorber is located in front of both the continuum and the line emitting source, which is plausible since in most cases, the absorption depth significantly exceeds the continuum brightness alone.

The model calculations in \S\ref{ss:luis} indicate median source sizes of 17.6, 14.2, and 8.5 arc seconds FWHM at 987, 1113, and 1669 GHz, which is 83\%, 75\%, and 67\% of the corresponding beam size. The filling factor would be the square of this fraction which is 68\%, 56\%, and 45\% respectively, so that our absorption line measurements will be affected by covering factor effects if the absorbers are much smaller in size than the beam. In this case, the beam-averaged optical depth would be below the source-averaged value.

\subsection{Envelopes}
\label{ss:env}

The \hho\ line spectra of all sources show envelope components, and the velocities and widths of these components in Table~\ref{t:env} agree well with previous measurements for these sources in other molecules from the ground (Fig.~\ref{f:velocity}). 
The widths also mostly agree with the widths of the C$^{18}$O 9--8 line observed with HIFI in a similar beam (Fig.~\ref{f:irene}); the few cases where \dv(C$^{18}$O) $>$ \dv(\hho) may be influenced by an outflow contribution to the C$^{18}$O 9--8 line.
The measured \hho\ line widths range over a factor of 4, and show a clear correlation with the envelope masses calculated in \S\ref{ss:luis} (Fig.~\ref{f:lwidth}).

The appearance of the envelopes in the four lines differs markedly from source to source, which can be grouped into four types. 
Seven envelopes appear in absorption in all four lines: G31.41, DR21OH, G327, NGC6334I(N), IRAS 05358, W43MM1, and one of the G34.26 envelopes. 
Six envelopes show emission in the 987 GHz line and absorption in the other lines: G10.47, IRAS 18089, W51e, NGC 6334I, IRAS 16272, and one of the W33A envelopes. 
A further six envelopes appear in \hho\ 987 GHz and in \hhoe\ emission, and in \hho\ 1113 and 1669 GHz absorption: G29.96, NGC 7538, W3 IRS5, G5.89, and the second envelopes of W33A and G34.26. 
Finally, one source, AFGL 2591, appears in emission in all four lines.
One curious case is IRAS 18151, where the non-detection of the \hhoe\ line leaves ambiguous whether it belongs in the second or third of the above source types; or perhaps it is in a transition from the second to the third type.

One possible origin for the diverse appearance of the envelope features is related to temperature, as appearance in emission  (absorption) requires an excitation temperature above (below) the continuum brightness. 
In this scenario, pure absorption sources would have the most massive and coldest envelopes, sources where the 987 GHz line appears in emission would be somewhat warmer, sources where both \hho\ 987 GHz and \hhoe\ appear in emission would be even warmer, and the pure emission source would be the warmest.
Alternatively, the appearance of the envelope features in emission or absorption may be a matter of continuum brightness, as detectability in absorption requires a background source whereas emission does not. 
In particular, the presence of outflow cavities in the HIFI beam would decrease the continuum compared with a beam filled with envelope emission. 
However, the continuum flux densities in Table~\ref{t:cont} do not support this option: the four envelope types above have very similar median continuum brightness levels. 
%This does not rule out line-of-sight effects: the water may be well mixed with the dust or be located in front of  it or behind, which will change the appearance of the lines but not the continuum level.
%However, such line-of-sight effects cannot be the full story because a \hho\ cloud in front of a dust cloud may appear in emission or absorption, depending on its excitation temperature.
\new{Outflow cavities may also affect the \hho\ line signals, but the net effect may either be emission or absorption, depending on the difference between the \hho\ excitation temperature and the dust temperature, and on the relative location of dust and water along the line of sight.}
We conclude that heating effects are more likely than geometric effects to be responsible for the variation in envelope appearance between our sources.

\begin{figure}[t]
\centering
\includegraphics[width=8cm,angle=0]{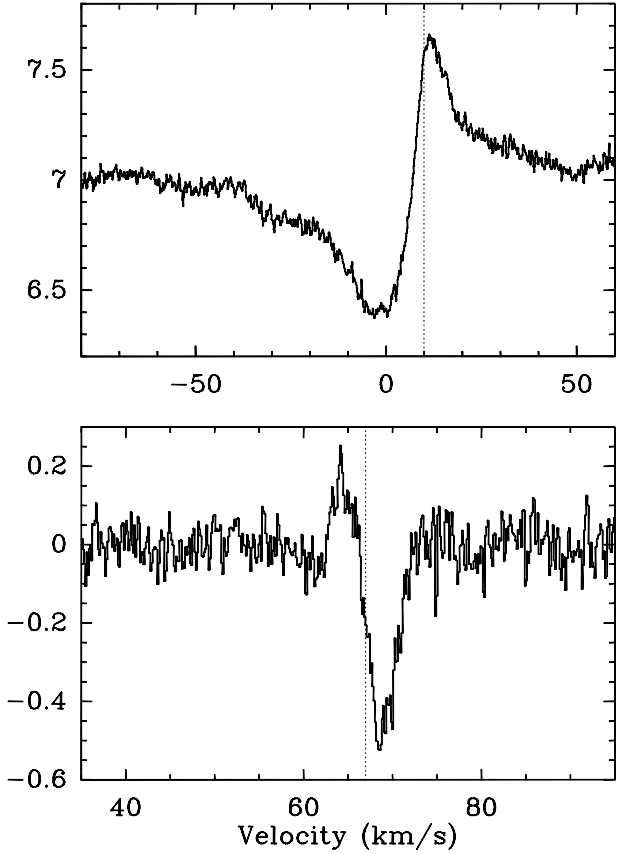}
\caption{Top: \hhoe\ line profile toward G5.89, showing a P~Cygni profile. Bottom: \hhoe\ line profile toward G10.47, showing an inverse P~Cygni profile. The bottom spectrum is continuum subtracted. Vertical scale is \tmb\ in K, and the dotted line denotes the systemic velocity from ground-based data.}
\label{f:pcp-ipc}
\end{figure} 

\begin{figure}[t]
\centering
\includegraphics[width=5cm,angle=0]{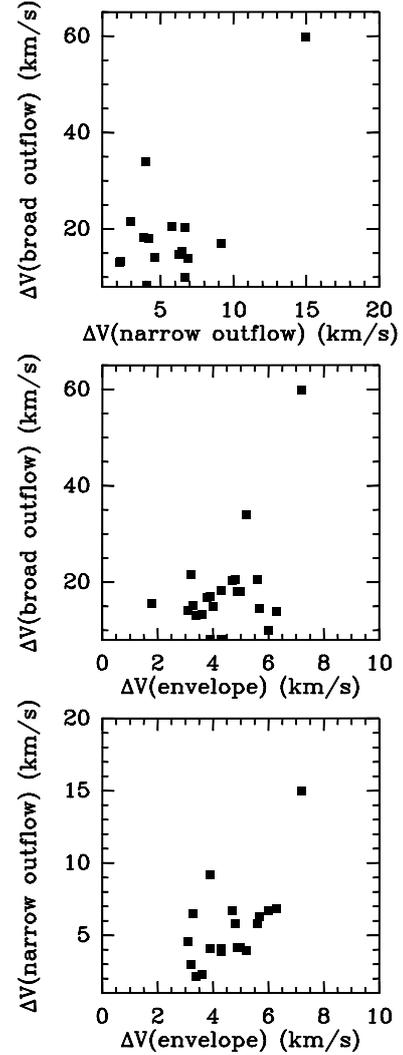}
\caption{Line widths (FWHM) of the envelope, narrow outflow, and broad outflow components of the observed \hho\ lines compared with each other. Error bars are smaller than the plotting symbols.}
\label{f:linewidth}
\end{figure} 

Some sources show two envelope features in some or all \hho\ line profiles, which in certain cases may be due to the presence of common-envelope binaries at $\sim$1000~AU separation, as suggested for W3~IRS5 \citep{chavarria2010}.
In other cases, the lines have P~Cygni or inverse P~Cygni profiles, which are due to expansion and infall, respectively (Fig.~\ref{f:pcp-ipc}). 
Such profiles have also been observed in the \hho\ line spectra of low-mass protostars \citep{kristensen2012} and for the low-mass pre-stellar core L1544 \citep{caselli2012}. 
Interestingly, the (inverse) P~Cygni signature is usually not seen in all lines toward a given source. 
The sources G10.47 and G34.26 show inverse P~Cygni profiles in the \hho\ 987\,GHz and \hhoe\ lines, suggesting that infall motion only takes place in their inner envelopes. 
For G34.26, this result is confirmed by SOFIA observations of redshifted \ammo\ line absorption at 1810\,GHz \citep{wyrowski2012}, and for G10.47 by SMA observations of vibrationally excited HCN \citep{rolffs2011}.
In contrast, G5.89 and DR21OH show P~Cygni profiles in the \hho\ 1113 and 1669\,GHz lines and in \hhoe, but not in the \hho\ 987\,GHz line, suggesting that expanding motions only take place in their outer envelopes. 
In W33A, where only the \hhoe\ line may show a P~Cygni profile, the expanding motion may take place further inside the envelope. 
For W43~MM1, the low-$J$ lines do not show infall or outflow motions, consistent with the observation by \citet{herpin2012} that only high-$J$ lines show infall motions in this source, i.e., that infall is confined to the inner envelope. 
Similar qualitative kinematic variations with radius have been observed in HCN lines toward the Sgr~B2 envelope, where the emission peak shifts from blue to red with increasing $J$, suggesting infall in the outer and expansion in the inner envelope \citep{rolffs2010}. 

\subsection{Outflows}
\label{ss:outflow}

\begin{figure}[t]
\centering
\includegraphics[width=8cm,angle=0]{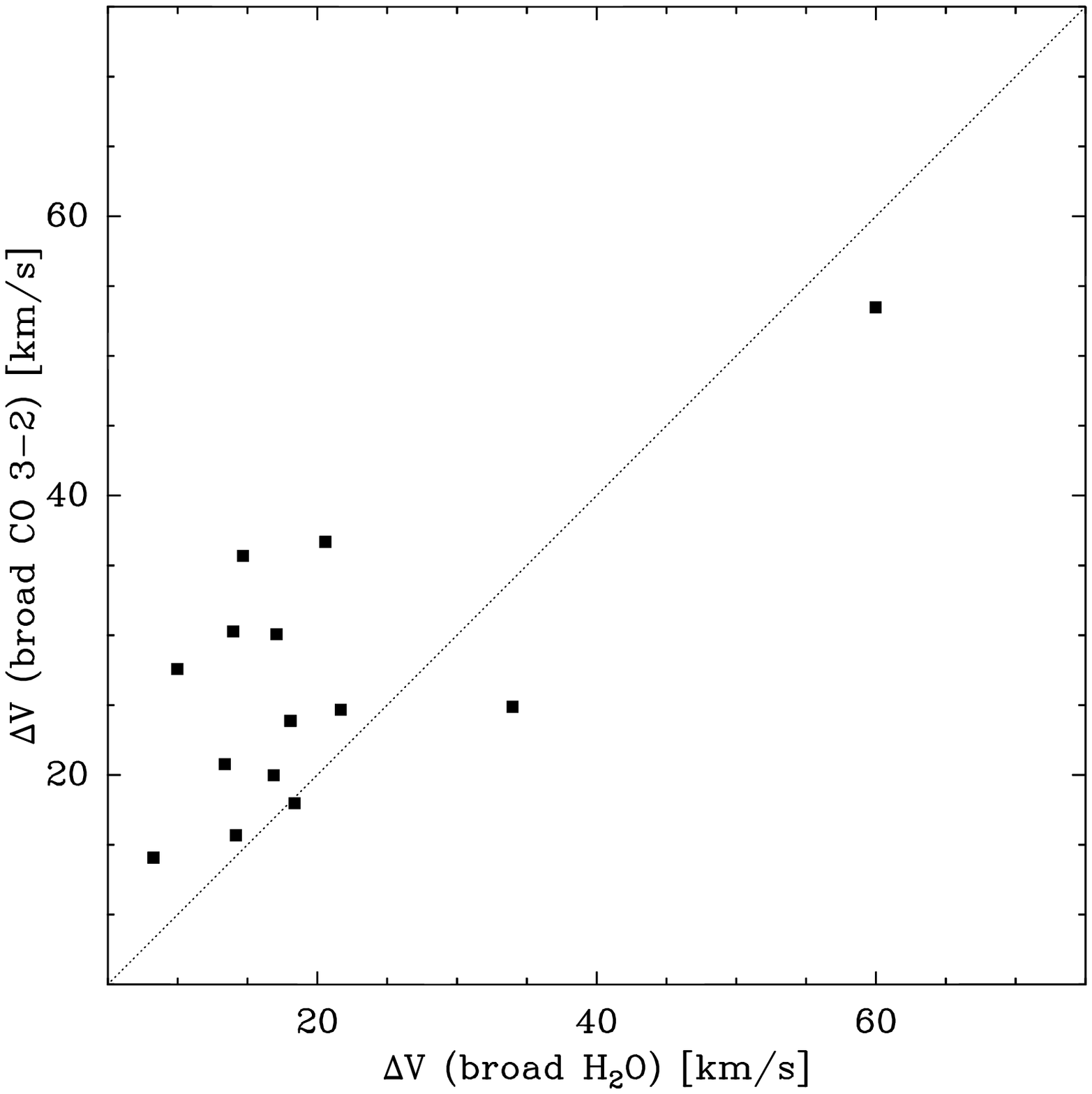}
\caption{Line widths (FWHM) of the broad outflow components of the observed \hho\ lines compared to the widths of the CO 3--2 lines measured with the JCMT \citep{irene2013}. Error bars are smaller than the plotting symbols.}
\label{f:irene2}
\end{figure} 

The outflows usually show up in the \hho\ line profiles as two components: a broad (FWHM $\sim$10--20\,\kms) emission component visible in all lines, and a narrower (FWHM $\sim$5--10\,\kms) absorption component seen in all but the 987 GHz line. Exceptions are AFGL 2591, IRAS 18151 and IRAS 05358, which only show a broad component, and G5.89, which has an exceptionally broad and complex line profile; this source is known to have a very massive and powerful outflow \citep{acord1997}. For the sources DR21OH, W3~IRS5 and NGC 7538, the widths of the narrower components are so small (2--3\,\kms) that these may be `second envelopes', as also seen toward W33A and G34.26, and as also advocated by \citet{chavarria2010} for W3~IRS5. The mm-wave continuum emission from these sources indeed shows multiple spatial components \citep{woody1989,vdtak:w3irs5}, sometimes with mid-infrared counterparts \citep{megeath2005,dewit2009}.

For most of our sources, the broad outflow component appears in emission in all four lines, but there are some exceptions. For AFGL 2591, NGC 6334I, \new{W51e} and G34.26, the broad outflow component appears in absorption in the 1113 and 1669 GHz lines, and in emission in the 987 GHz line; for G10.47, this component appears in emission in the 1113 and 987 GHz lines, and in absorption in the 1669 GHz line. \new{Toward W51e, a `very broad' component appears in emission in all four lines between +30 and +90 \kms}. It is unclear in which category G31.41 belongs, since the broad outflow component is only detected in the 987 GHz line, \new{perhaps because in the ground state lines, emission and absorption cancel each other out}. The shape of the broad outflow component on the 987 GHz line profile is often non-Gaussian, but the outflow contribution can be modeled with two Gaussians of about the same height and width. In this case the velocity in Table~\ref{t:flow} is the mean of the values for the two Gaussians, while the width and the flux are the sums.

The narrow outflow component usually appears in absorption in the \hho\ 1113 and 1669 GHz lines and the \hhoe\ line, and is not seen in the \hho\ 987 GHz line, but again, exceptions occur. For W3~IRS5, G327 and G31.41, the narrow outflow component appears in 1113 and 1669 GHz absorption and in 987 GHz emission; for IRAS 18089 and G34.26, this component appears in 1113 and 1669 GHz absorption and in \hhoe\ emission, while for G10.47, it appears in emission in all four lines. Unlike the envelopes, the outflow absorptions are usually not saturated, with G327 as only exception.

We have used the measured velocities of the outflow components to test if our sources have a preferred orientation in the sky. In particular, for a face-on geometry, the outflow would be directed toward us, and thus appear blueshifted relative to the envelope component if seen in absorption. 
The \new{narrow} outflow component is the one that is usually seen in absorption, and Figure~\ref{f:velocity} shows that this component is indeed usually blueshifted relative to the envelope. 
The exceptions are IRAS 16272, DR21(OH), G29.96, and \new{W51e}, and Fig.~\ref{f:decomp} illustrates the case of IRAS 16272, especially for the 1669~GHz line. 
The widths of these redshifted absorption features are too large to be due to infall motions, and we suggest that these sources have binaries or other deviations from centrosymmetric geometry that allow their receding outflow lobes to show in absorption.
In addition, Figure~\ref{f:linewidth} shows that there are no clear trends between the widths of the envelope and outflow components. 

Figure~\ref{f:irene2} compares the widths of the broad outflow components to the widths observed in CO 3--2 with the JCMT \citep{irene2013}. 
While sources with broader \hho\ outflows also seem to have broader CO outflows, the width for any given source is larger in CO than in \hho\ by about 10\,\kms. 
This result is contrary to the case of low-mass protostars, where the \hho\ line widths exceed those of CO \citep{kristensen2012}.
For our sources, the \hho\ may be confined to denser gas where velocities are lower. 
Alternatively, the difference is excitation: our \hho\ values are averages over ground-state and excited-state lines, while the outflow line widths increase with $J$ or \eup. 
Indeed, the widths of the excited-state \hho\ lines are very close to those measured in CO, as also found by \citet{chavarria2010} and \citet{herpin2012}.

\begin{table}[t]
\caption{Parameters of continuum models.}            
\label{t:models}      
\centering                          
\begin{tabular}{lcccc}     
\hline \hline                 
\noalign{\smallskip}
& \multicolumn{2}{c}{Input parameters}  & \multicolumn{2}{c}{Output parameters}  \\
\noalign{\smallskip}
\hline
\noalign{\smallskip}
Source  &    $R_{\rm out}$ &   $p$    &  \menv\        & $\bar{T}$\tablefootmark{a} \\      
             &    ( 10$^4$ AU)   &         &  (\msol)   & (K)    \\
\noalign{\smallskip}
\hline
\noalign{\smallskip}
IRAS 05358      & \phantom{1}5.4    &  1.5\phantom{5} &    \phantom{5}142 & 22 \\
IRAS 16272      &  17.0                       &  1.5\phantom{5} &    2170                    & 18 \\
NGC 6334I(N)  &  \phantom{1}7.2    &  1.3\phantom{5} &   3826                    & 15 \\
W43 MM1          &  14.7                       &  1.5\phantom{5} &   7550                    & 21 \\
DR21(OH)         &  \phantom{1}5.0    &  1.35                    &   \phantom{5}472 & 27 \\
W3 IRS5            &  11.3                        & 1.4\phantom{5}  &  \phantom{5}424 &  34 \\
IRAS 18089      &   \phantom{1}3.9    &  1.9\phantom{5} &  \phantom{5}172 & 41 \\
W33A                 &  \phantom{1}7.1    & 1.2\phantom{5}   &  \phantom{5}700 & 26 \\
IRAS 18151      &  \phantom{1}5.0     & 1.4\phantom{5}  & \phantom{5}153  & 27 \\
AFGL 2591       &   \phantom{1}7.1    &  1.0\phantom{5} &  \phantom{5}320 &  38 \\
G327	         &   \phantom{1}8.0   &  1.2\phantom{5}  &  2044                     & 24 \\   
NGC 6334I       &   \phantom{1}6.8   &  1.5\phantom{5}  &  \phantom{5}750  & 38 \\  
G29.96              &   \phantom{1}9.6   &  1.4\phantom{5}  &  \phantom{5}768  & 40 \\  
G31.41              & 11.9                         &  1.4\phantom{5}  &   2968                    & 32 \\  
G5.89                &    \phantom{1}3.6   &  1.3\phantom{5}  &  \phantom{5}140 & 36 \\    
G10.47              &    \phantom{1}6.0   &  1.4\phantom{5}  & 1168                     & 41 \\  
G34.26	        &    \phantom{1}8.1    & 1.4\phantom{5}  &   1792                    & 39 \\  
W51N-e1         &    14.0                        &  1.4\phantom{5}  &   4530                   & 38 \\  
NGC 7538       &    \phantom{1}6.4    &  1.0\phantom{5}  &  \phantom{5}433 & 36 \\  
\noalign{\smallskip}
\hline
\noalign{\smallskip}
\end{tabular}
\tablefoottext{a}{\new{Mass-weighted temperature of the envelope model.}} 
\end{table}

\subsection{Foreground clouds}
\label{ss:fg-clouds}

\begin{figure}[t]
\centering
\includegraphics[width=8cm]{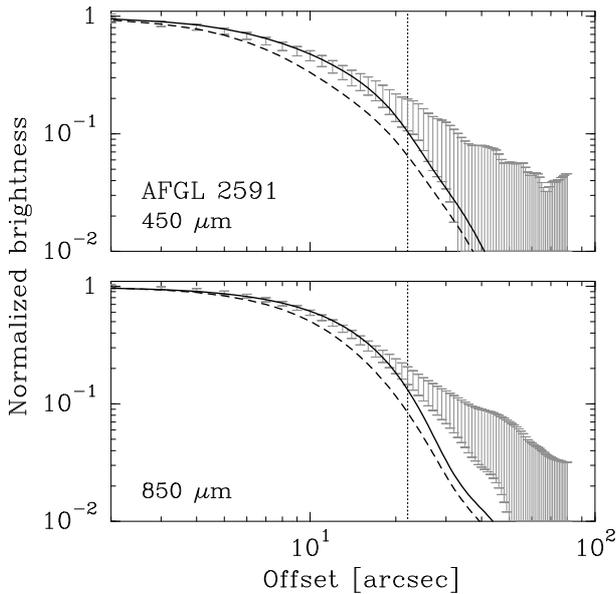}
\caption{Submillimeter brightness profiles of AFGL 2591 measured with JCMT/SCUBA (error bars), and synthetic brightness profiles derived from WR models and convolved by the SCUBA beams for $p$=1.0 (solid line) and $p$=1.5 (dashed line). The vertical dotted line indicates the adopted source radius. For this source, we choose a density profile of $p$=1.0.}
\label{f:model-radius}
\end{figure}

The foreground clouds appear as narrow (FWHM = 1--4\,\kms) absorption features in the 1113 and 1669 GHz line spectra of our sources. 
While one cloud also shows \hhoe\ absorption, the general lack of signal in the \hho\ 987 GHz and \hhoe\ lines indicates a low column density and a low volume density for these clouds.
The last two columns of Table~\ref{t:fg} list estimates of the \hho\ column densities and ortho-para ratios of the foreground clouds. 
These values were obtained from the velocity-integrated optical depths in the table by assuming that in these clouds, all \hho\ molecules are in their ortho and para ground states, i.e., that excitation is negligible.
As discussed by \citet{emprechtinger2013}, this assumption generally holds for column density estimates from the 1113 and 1669 GHz lines, but fails for the \hho\ 557 GHz line.

The \hho\ column densities of the foreground clouds are found to lie mostly in the range \pow{1}{12} -- \pow{1}{13} \scm, with a few outliers at the upper and lower end. These values are comparable to earlier estimates \citep{vdtak2010,lis2010} and indicate that the \hho\ abundance in these clouds is limited by photodissociation.
The derived ortho/para ratios for \hho\ lie mostly in the range between 3 and 6, again with a few outliers to higher and lower values. Our derived o/p ratios are very similar to the values by \citet{lis2010} and \citet{herpin2012}. 
\new{Comparison to the detailed multi-line study by \citet{flagey2013} for the 5 overlapping sources suggests that our estimated column densities and o/p ratios for individual clouds have uncertainties of a factor of $\sim$2, which is an upper limit since not exactly the same positions were observed.}

\new{The ortho-para ratio of \hho\ is expected to rise from $\approx$1.0 at low temperatures ($\approx$15\,K) to $\approx$3 at high temperatures ($\gtsim$50\,K) as shown by \citet{mumma1987}}. 
Even though the values derived here are probably uncertain to a factor of $\sim$2, our results are inconsistent with the low-temperature limit. 
If the \hho\ molecules were formed on cold (10 K) grain surfaces, their spin temperature does not reflect these conditions, and may instead be set by the grain temperature shortly before the \hho\ ice mantle evaporated (at $T \approx$100\,K). 
Alternatively, \new{the \hho\ molecules may have formed in the warm (\tkin$\approx$50\,K) gas phase by ion-molecule reactions, or}
the o/p ratio of the evaporated \hho\ may have equilibrated with that of \hh\ in the warm gas phase in reactions with \hhhp. 
%For further discussion of \hho\ in foreground clouds, we refer the reader to \citet{flagey2013}.

\section{Analysis and discussion}
\label{s:lysis}

\begin{figure}[t]
\centering
\includegraphics[width=8cm]{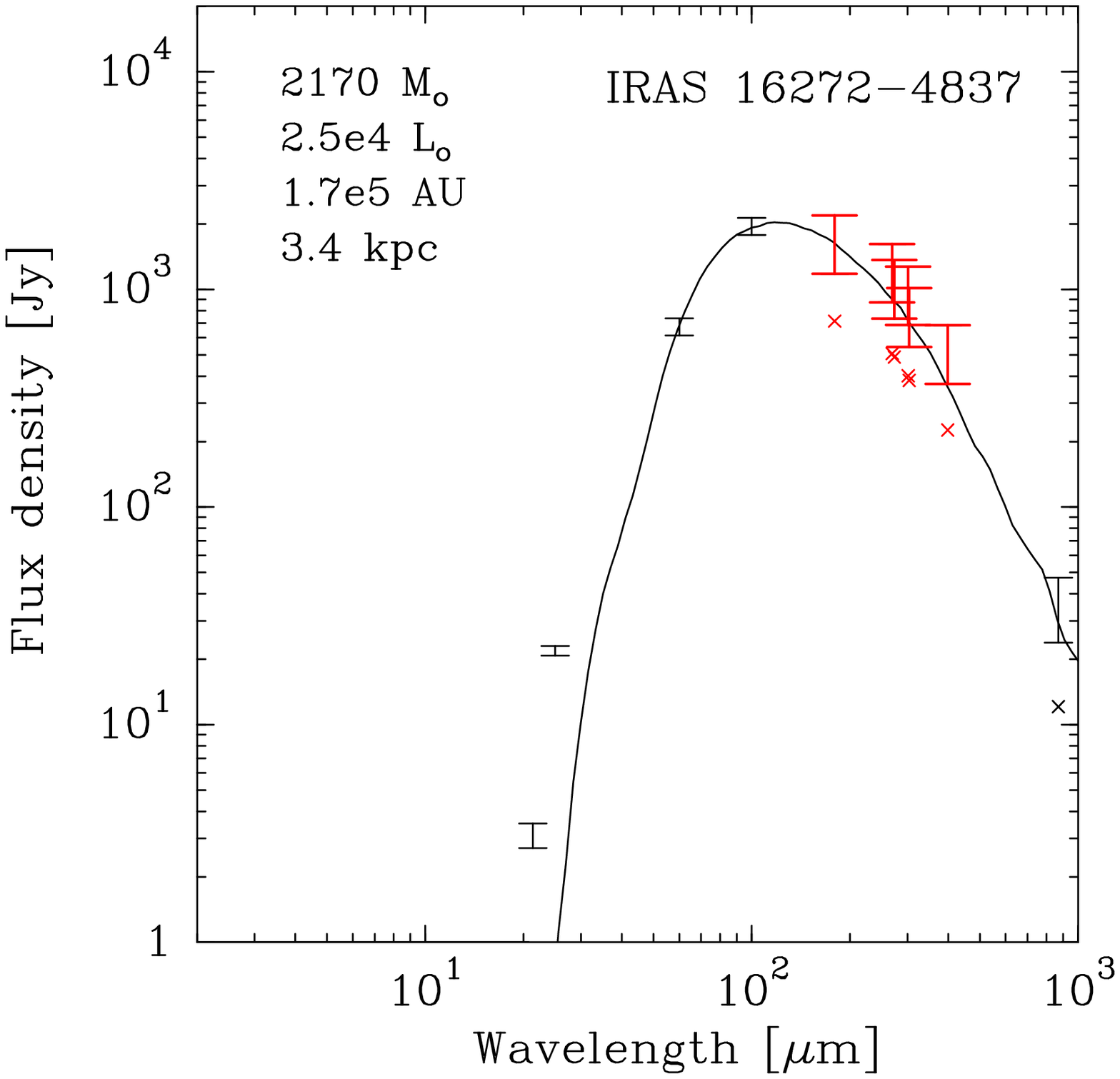}
\bigskip
\includegraphics[width=8cm]{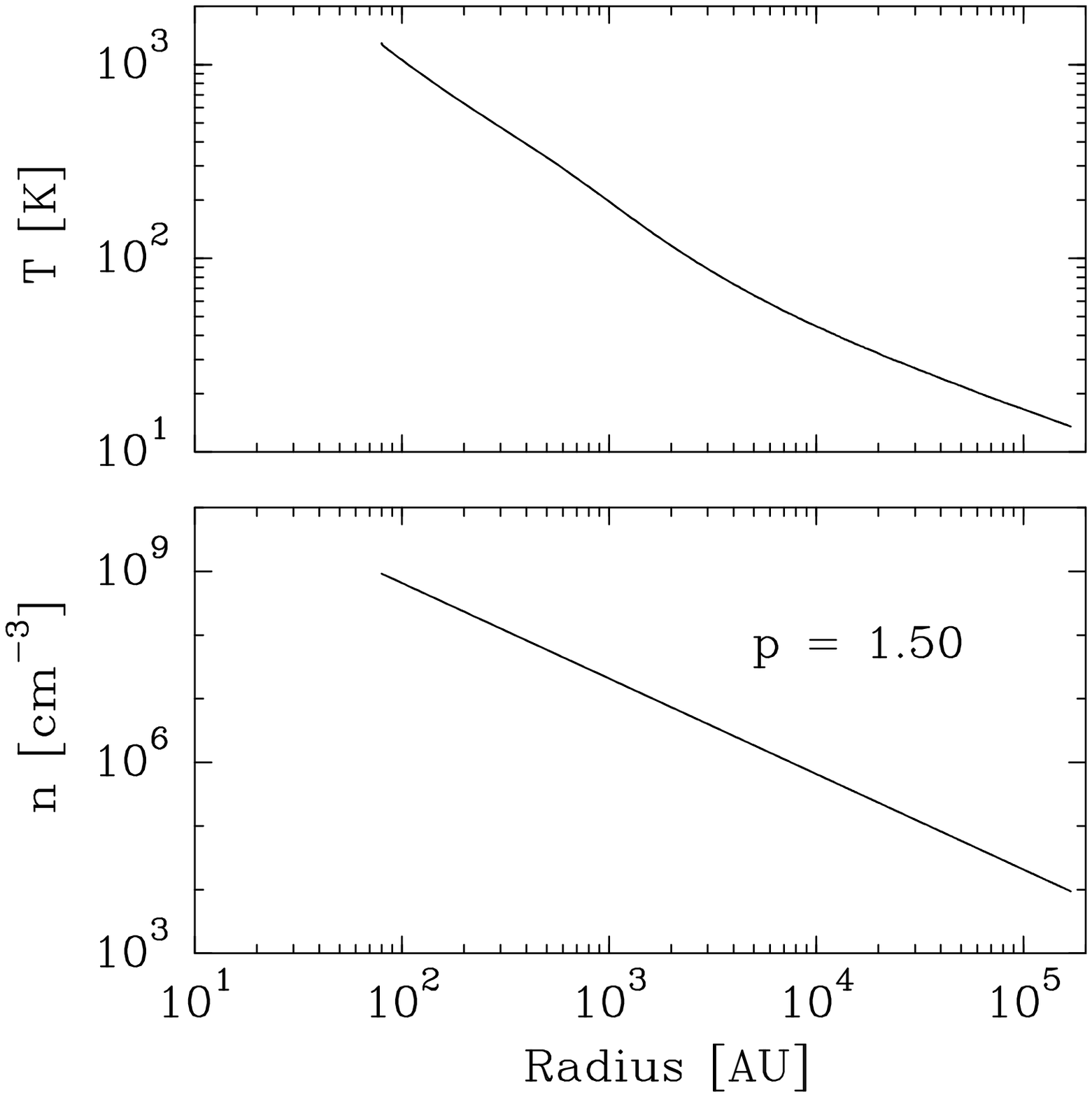}
\caption{Top: Model (continuous line) and observed SED fluxes (error bars) for IRAS 16272. 
The SED points are normalized by the source size for $\lambda > 100$~microns, as shown by the "X" symbols. 
Continuum flux densities from our HIFI data are labeled in red. 
Bottom: temperature and density profiles for IRAS 16272.}
\label{f:sed_example}
\end{figure}

\subsection{Mass estimation}
\label{ss:luis}

\begin{figure}[t]
\centering
\includegraphics[width=8cm]{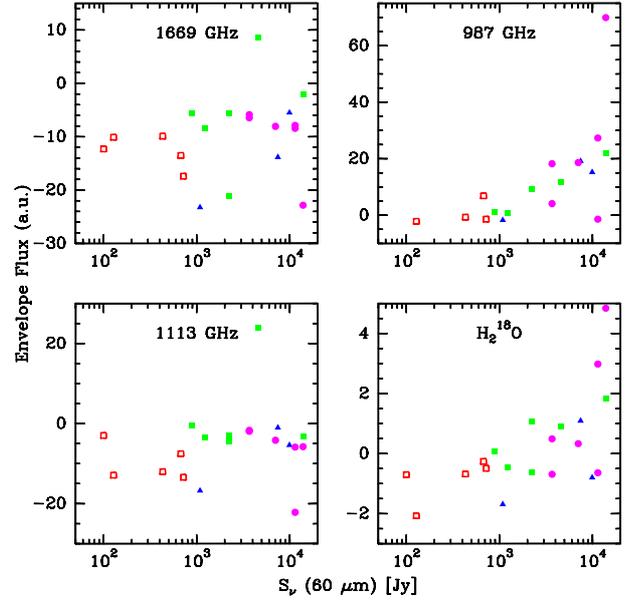}
\caption{Envelope line flux versus 60\,\mic\ flux density. Uncertainties are $\approx$10\% for HIFI data and $\approx$20\% for IRAS data. \new{Open red and filled green squares are mid-IR-quiet and -bright protostellar objects, filled blue triangles are hot molecular cores, and filled purple circles are ultracompact H{\sc II} regions.}}
\label{f:987vs60}
\end{figure}

Estimates of the envelope masses exist in the literature for many of our sources. However, these mass estimations were performed by different authors, using different techniques and based on data from different telescopes. 
In order to have a homogeneous mass determination over our sample, we derive the envelope mass by comparing models of the continuum spectral energy distribution (SED) with the available observations for each source.

We use a modified version of the 3D continuum radiative transfer code HOCHUNK3D \citep{whitney2013,robitaille2011} (hereafter WR) to find the envelope temperature profile and derive a model SED for each source. Our modified version of the code uses dust opacities from \citet{ossenkopf1994}, model 5 with thin ice mantles and a gas density of $10^6$~\ccm, suitable for high-mass protostars \citep{vdtak1999}, and allows to vary the radial density power-law exponent $p$. The WR code allows for a detailed description of the protostellar environment in terms of a star, a disk, a cavity and an envelope. To decrease the number of free parameters, we use a simple spherically symmetric model with no cavity and no disk. This simplification has its main effect in the near to mid-IR region of the spectrum, while the sub-millimeter and millimeter-wave regions, where the dust is mainly emitting, will stay almost unchanged. 

The envelope size and density exponent $p$ are derived from archival sub-millimeter images from the JCMT/SCUBA\footnote{The James Clerk Maxwell Telescope is operated by the Joint Astronomy Centre on behalf of the Science and Technology Facilities Council of the United Kingdom, the Netherlands Organisation for Scientific Research, and the National Research Council of Canada.} and APEX/LABOCA \new{\citep{siringo2009}} instruments. For the envelope size, we choose the radius of the contour at the 3$\sigma$ noise level, while for the density exponent, we use the WR code to create synthetic sub-millimeter maps for various $p$--values. We determine the best-fit density exponent by minimizing the difference between the observed and synthetic brightness profiles (see Fig.~\ref{f:model-radius}). The synthetic maps are created using the corresponding instrumental pass-band and convolved with the instrumental beam profile including the error-beam.

With the outer radius and the density profile fixed, the only remaining parameters are the envelope mass and the bolometric luminosity.
The observed fluxes from the literature correspond to the peak emission at the source position. 
In most cases, the beam size of the observed fluxes is smaller than the source size. 
To correct for the missing flux, we normalize the observed fluxes by the ratio between the flux from the whole model and the flux from a region as big as the beam size. 
As an example, Figure~\ref{f:sed_example} shows the observed and derived SED for the source IRAS 16272 as well as the temperature profile. 
Table~\ref{t:models} lists the adopted and the derived physical parameters for each source, \new{and Appendix~\ref{app:models} presents the results in graphical form}. 

\subsection{Correlations with luminosity and mass}
\label{ss:env-corr}

\begin{figure}[t]
\centering
\includegraphics[width=6cm,angle=-90]{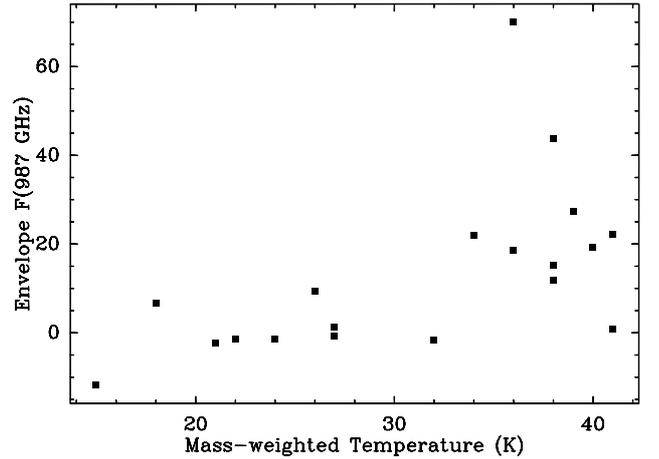}
\caption{\new{Envelope \hho\ 987\,GHz line flux versus mass-weighted temperature in our envelope models.}}
\label{f:987vstbar}
\end{figure}

To understand the origin of the observed \hho\ lines, we now investigate if the measured line fluxes show trends with the luminosity of the source \lbol\ (Table~\ref{t:sample}) and the mass of the envelope \menv\ (Table~\ref{t:models}). 
\new{In addition, we have searched for trends with distance but did not find any.}
For the luminosity, we use the results of the model fit in \S\ref{ss:luis} so that the source flux at all wavelengths refers to the same area on the sky. 
Figures \ref{f:env-flux-corr}--\ref{f:broad-flux-corr} in Appendix~\ref{app:corr} show scatter plots of the line fluxes of the envelope and the narrow and broad outflow components versus \lbol\ and \menv. 
In these figures, absorption is plotted as negative emission, and 'double envelope' features are plotted as two data points. 
We identify possible trends by visual inspection of these plots, and test these (if applicable) by computing the correlation coefficient $r$ and the probability of false correlation $P$, which depends on $r$ and the number of data points $N$ (see Appendix of \citealt{marseille:meth}).

For the envelope component, the visual inspection reveals two trends.
First, the 987\,GHz line flux increases with \lbol, but only if the line appears in emission. This trend is not a correlation, but more a 'threshold effect': strong 987\,GHz line emission is only seen at high luminosities (\lbol\ $\gtsim$10$^5$\,\lsol). Second, the \hhoe\ line flux decreases with increasing \menv, but again not as a correlation but in the sense that the line is seen in emission at low \menv\ and absorption for high \menv. Such behaviour is expected if the line optical depth, and thus the \hho\ column density, increases with envelope mass. An optical depth of $\sim$unity for the \hhoe\ line is confirmed by detections of \hhos\ line emission in several of our sources \citep{herpin2012,choi2012}. Apparently the average \hho\ abundance is rather constant across all envelopes in our source sample. The studies mentioned in \S\ref{s:intro} indicate outer envelope abundances of $\sim$10$^{-9}$--10$^{-8}$.

The increase of the 987 GHz line flux with \lbol\ is probably due to a higher temperature of the inner envelopes of sources with a higher luminosity. 
This idea is confirmed by the trend between 987\,GHz line flux with far-infrared continuum flux density (Fig.~\ref{f:987vs60}). 
Such a trend is seen both at 60 and at 100\,\mic, and like the trend with \lbol, is more a threshold effect than a correlation.
The large beam size of IRAS may influence these trends quantitatively, but probably has no strong qualitative effect because our sources were selected to be relatively isolated (\S\ref{s:intro}).
Far-infrared pumping could influence the results for individual \hho\ lines, but is unlikely to dominate here, because the effect is seen in several lines.

\new{Further evidence that the \hho\ 987 GHz line flux is due to envelope heating is presented in Fig.~\ref{f:987vstbar}, which is a plot of the flux in this line versus the mass-weighted temperature from the envelope models in \S\ref{ss:luis}. While a 1:1 correlation cannot be claimed, the points show a clear trend toward higher line fluxes at higher average temperatures.}

Figures~\ref{f:medium-flux-corr} and~\ref{f:broad-flux-corr} show a similar analysis for the narrow and broad outflow components in the \hho\ line profiles of our sources. The most significant trend is that the \hhoe\ line flux of the broad outflow component decreases with increasing \lbol\ ($r$=--0.683, $P$=3\%), but only in the 11 cases where this component is detected in \hhoe. Second, the 987 GHz line flux of the broad outflow component appears to increase with increasing \menv, but with $r$=0.174 and $P$=14\%, this trend is less significant than the first one. The line fluxes of the narrow outflow component do not show any clear trends with \lbol\ or \menv.

The lack of evolutionary trends in the \hho\ line fluxes of our sources is in contrast with the study of \citet{lopez-sepulcre2011}, who observed the SiO $J$=2--1 and 3--2 lines toward 57 high-mass star-forming regions, and found that the SiO line luminosity drops with evolutionary stage as measured by increasing values of \lbol/\menv. This drop suggests a decline of the jet activity with the evolutionary state of the source, as also observed for low-mass sources \citep{bontemps1996}, although excitation effects also may play a role.

One reason for this difference may be that even the \hho\ line signals from the outflows may be somewhat optically thick, so that saturation masks any evolutionary trends. Therefore, we consider next trends of the line width with \lbol\ and \menv. The line width is less sensitive to the optical depth than the line intensity, and has the additional advantage that the appearance of the line in emission or absorption does not influence its measurement. Section~\ref{ss:env} already shows the trend of the envelope \dv\ with \menv, which with $r$=0.407 and $P$=4\% for $N$=22 is likely real.

Figures~\ref{f:medium-width-corr} and~\ref{f:broad-width-corr} show scatter plots of the widths of the narrow and broad outflow components versus \lbol\ and \menv. The only significant trend that appears is that \dv\ of the narrow outflow component increases with \menv. A relation of this component with the envelope has been found before for low-mass protostars \citep{kristensen2012}, and suggests that the narrow outflow component is due to outflow-envelope interactions.

\subsection{Evolutionary trends}
\label{ss:evol}

\begin{figure}[t]
\centering
\includegraphics[width=8cm,angle=0]{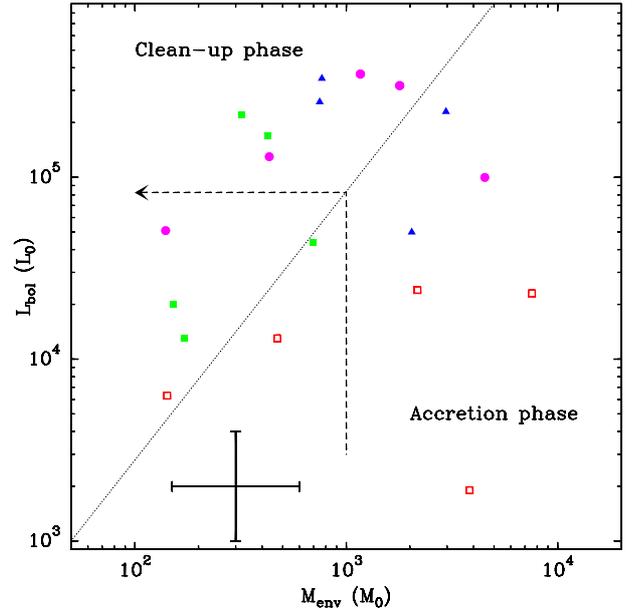}
\caption{Bolometric luminosity versus envelope mass for our sources, where the dotted line divides the two phases of high-mass protostellar evolution suggested by \citet{molinari2008} and the dashed arrow is a typical evolutionary track. The typical uncertainty of a factor of 2 in both luminosity and mass is indicated at the bottom. \new{Symbols are as in Fig.~\ref{f:987vs60}}.}
\label{f:molinari}
\end{figure} 

We now discuss several proposed evolutionary indicators for high-mass star formation and test them on our source sample. We will use the ratio of \lbol\ over \menv\ as primary evolutionary indicator, because of previous success \citep{vdtak2000,motte2007,lopez-sepulcre2010}, because it is a continuous variable, and because it has a physical basis (declining \new{accretion} activity and envelope dispersal). 
In particular, \citet{molinari2008} have proposed an evolutionary sequence for high-mass protostars in terms of two parameters: the envelope mass and the bolometric luminosity. They suggest a two-step evolutionary sequence: protostars first accrete mass from their envelopes, and later disperse their envelopes by winds and/or radiation. In the diagram in Fig.~\ref{f:molinari}, their calculated evolutionary tracks start at the bottom, rise upward until they reach the dotted line, and then proceed horizontally to the left. The positions of our sources in this diagram suggests that the subsample of mid-IR-quiet protostellar objects is the youngest, and that the other subsamples contain more evolved sources. The same effect is seen in Figs.~\ref{f:env-flux-corr}--\ref{f:broad-flux-corr}, where the mid-IR-quiet protostellar objects occupy the left (low-\lbol) sides of the diagrams. 
On the other hand, the presence of a hot molecular core and/or an ultracompact H{\sc II} region is unrelated to mid-IR brightness and \lbol/\menv\ ratio, at least for our sample. The occurrence of these phenomena may depend more on absolute luminosity than on evolutionary stage.

To test the use of \hho\ line profiles as evolutionary indicators, we use the appearance of the envelope component in the \hho\ ground-state line profiles, as introduced in \S\ref{ss:env}. In Type~1 sources, all lines appear in absorption; Type~2 sources show emission in the 987~GHz line and absorption in the others; in Type~3 sources, both the 987~GHz and the \hhoe\ lines appear in emission; and Type~4 sources show all lines in emission. While the Type~1 sources all seem to be in early evolutionary stages as probed by a low \lbol/\menv\ ratio, the other source types are all mixed. 
Thus, an \hho\ line profile with pure absorption appears to be a sign of an early evolutionary stage, whereas emission is a sign of a more evolved stage.

The appearance of the broad and narrow outflow components in the \hho\ ground-state line profiles differs between our sources, and for each component, four types may be defined as for the envelopes (\S\ref{ss:outflow}). While these emission/absorption types may be useful to trace protostellar evolution, our data do not allow to test this hypothesis, since most of our sources belong to just one of these types, with the broad outflow appearing in absorption and the narrow outflow in emission. This issue may be revisited after HIFI data become available for a larger sample of sources \new{outside the WISH program}.

Besides the shape of the line profile, the fluxes and widths of the \hho\ line may also be evolutionary indicators. The bottom panels of Figs.~\ref{f:env-flux-corr}--\ref{f:broad-flux-corr} show scatter plots of our observed line fluxes in the envelope and outflow components versus the ratio \menv/\lbol\ which as discussed above is sometimes used as evolutionary indicator for high-mass protostars. No trends appear, and the same is true for the plots of the line width versus \menv/\lbol\ (Figs.~\ref{f:lwidth}, \ref{f:medium-width-corr} and~\ref{f:broad-width-corr}).

\section{Summary and conclusions}
\label{s:concl}

This paper presents spectroscopic observations of low-$J$ \hho\ transitions observed toward 19 regions of high-mass star formation. The line profiles show contributions from protostellar envelopes, bipolar outflows, and foreground clouds. These contributions, both in emission and in absorption, are disentangled by fitting multiple Gaussians to the line profiles.

Our observations show that the ground state lines of \hho\ are powerful tracers of the various physical and kinematic components of the high-mass protostellar environment, as well as of foreground clouds. 
The \new{envelope components of the} ground state lines of \hho\  at 1113 and 1669 GHz do not show significant trends with bolometric luminosity, envelope mass, or far-infrared flux density or colour, indicating that these lines trace the outer parts of the envelopes where the effects of star formation are hardly noticeable.
\new{On the other hand, broad wings due to outflows are clearly seen in these lines, which are a clear effect of star formation.
The competing effects of line emission and absorption in the ground-state lines lead to low integrated line fluxes, which may explain the difficulty in detecting these lines in distant galaxies. }

Unlike the ground-state lines, the flux in the excited-state 987 GHz line \new{increases systematically} with luminosity and with the far-infrared flux density, both at 60 and at 100 \mic. 
This line therefore appears to be a good tracer of the average temperature of the source.
\new{Indeed, this line and other excited-state lines have proven to be more easily detected at high redshift \citep{omont2011,vanderwerf2011}.}

Furthermore, the \hhoe\ ground-state line \new{turns from emission into absorption} with increasing envelope mass. This trend indicates that the \hho\ abundances in protostellar envelopes do not change much with time.

The appearance of the \hho\ line profiles in our sample changes with evolutionary state, in the sense that the youngest sources (as probed by mid-infrared brightness and \lbol/\menv\ ratio) show pure absorption profiles for all lines, while for more evolved sources, one or more of our studied lines appears in emission. This effect is visible for the envelope component but not for the outflows. In addition, the presence of infall or expansion, as probed by (inverse) P~Cygni profiles, does not appear related to evolutionary stage.

In the future, our team will make detailed estimates of the \hho\ abundances in these sources as a function of radius, based on HIFI observations of $\approx$20 lines of \hho, \hhoe\ and \hhos. 
\new{The Herschel/PACS instrument will be used to study higher-excitation \hho\ lines and to improve estimates of the bolometric temperatures of our sources, which may act as evolutionary indicator.}
Maps of selected lines with HIFI and PACS will clarify the spatial distribution of \hho, both on the scale of protostellar envelopes and outflows as on larger (protostellar cluster) scales. 
On smaller scales, ALMA will provide images of high-excitation \hho\ and \hhoe\ lines and thus zoom into the close vicinity of the central protostars.

\begin{acknowledgements}
This paper is dedicated to the memory of Annemieke Gloudemans-Boonman, a pioneer of \hho\ observations with ISO, who passed away on 11 October 2010 at the age of 35.
We thank Umut Yildiz for maintaining the WISH data archive (the 'live show'), \new{and the referee for carefully reading our manuscript and writing a constructive report}. 
This work has used the spectroscopic databases of JPL \citep{pickett1998} and CDMS \citep{mueller2001}, \new{and the facilities of the Canadian Astronomy Data Centre operated by the the National Research Council of Canada with the support of the Canadian Space Agency.}

HIFI has been designed and built by a consortium of institutes and university departments from across Europe, Canada and the US under the leadership of SRON Netherlands Institute for Space Research, Groningen, The Netherlands with major contributions from Germany, France and the US. Consortium members are: Canada: CSA, U.Waterloo; France: CESR, LAB, LERMA, IRAM; Germany: KOSMA, MPIfR, MPS; Ireland, NUI Maynooth; Italy: ASI, IFSI-INAF, Arcetri-INAF; Netherlands: SRON, TUD; Poland: CAMK, CBK; Spain: Observatorio Astron\'omico Nacional (IGN), Centro de Astrobiolog\'{\i}a (CSIC-INTA); Sweden: Chalmers University of Technology - MC2, RSS \& GARD, Onsala Space Observatory, Swedish National Space Board, Stockholm University - Stockholm Observatory; Switzerland: ETH Z\"urich, FHNW; USA: Caltech, JPL, NHSC.
\end{acknowledgements}

\bibliographystyle{aa}
\bibliography{survey}

\clearpage
\newpage

\appendix

\section{Observation log and line flux tables}
\label{s:obslog}

\begin{table}
\caption{List of observations.}
\label{t:obsid}
\begin{tabular}{llll}
\hline \hline
\noalign{\smallskip}
Source & \hho\ \& \hhoe\ & \hho\   & \hho\  \\
              & $1_{11}-0_{00}$ & $2_{02}-1_{11}$ & $2_{12}-1_{01}$ \\
\noalign{\smallskip}
\hline
\noalign{\smallskip}
IRAS 05358     & 206126 & 204510 & 203954 \\
IRAS 16272     & 214419 & 203168 & 192584 \\
NGC 6334I(N) & 206385 & 204520 & 214454 \\
W43-MM1        & 191670 & 191616 & 192575 \\
DR21(OH)       & 194794 & 195026 & 192569 \\
W3 IRS5           & 191658 & 191612 & 192566 \\
IRAS 18089     & 229883 & 215911 & 216696 \\
W33A                & 191638 & 191636 & 192576 \\
IRAS 18151     & 229880 & 218211 & 218909 \\
AFGL 2591      & 197973 & 195019 & 192570 \\
G327                 &  214425 & 203170 & 192585 \\
NGC 6334I       & 206383 & 204519 & 214455 \\
G29.96              & 191668 & 191617 & 192574 \\
G31.41              & 191671 & 191615 & 192573 \\
G5.89                & 229888 & 218120 & 216694 \\
G10.47             & 229884 & 215914 & 216695 \\
G34.26             & 191674 & 194995 & 192572 \\
W51e                & 194801 & 195014 & 192571 \\
NGC 7538       & 191663 & 201599 & 200758 \\
               
\noalign{\smallskip}
\hline
\noalign{\smallskip}
\end{tabular}
\tablefoot{The preceding 1342 has been omitted from all obsIDs.
}
\end{table}

%\clearpage
%\newpage

\begin{table*}
\caption{Central velocities, FWHM widths, and integrated line fluxes measured for the protostellar envelopes.}
\label{t:env}
\begin{tabular}{lcccccc}
\hline \hline
\noalign{\smallskip}
Source & \vlsr\ & \dv\ & 1113\,GHz & \hhoe\ & 1669\,GHz & 987\,GHz \\
\noalign{\smallskip}
\hline
\noalign{\smallskip}
IRAS 05358      & $-$17.9(6)      & 3.8(14) & $-$13.4 & $-$0.49 & $>$17.4 & $-$1.39 \\
IRAS 16272      & $-$45.9(7)      & 4.7(20) & $-$7.51 & $-$0.26 & $>$13.5 & $+$6.83 \\
NGC 6334I(N)  & $-$2.4(5)        & 3.9(4)    & $-$2.96 & $-$0.70 & $>$12.3 & $-$11.8 \\
W43 MM1          & $+$100.5(7)  & 6.0(15) & $-$12.9 & $-$2.07 & $-$10.1  & $-$2.17 \\
DR21(OH)         & $-$3.3(4)        & 3.4(13) & $>$12.0& $-$0.68 & $>$9.9   & $-$0.69 \\
W3 IRS5                & $-$37.7(12) & 3.2(6)    & $-$3.06 & $+$1.84 & $-$2.00 & $+$21.9 \\
IRAS 18089          & $+$33.7(11) & 3.1(11) & $-$3.47 & $-$0.45 & $-$8.44 & $+$0.90 \\
W33A                     & $+$35.1(12) & 5.6(13) & $-$4.34 & $-$0.61 & $>$21.0  & $<$0.60 \\ %MM1?
\tablefootmark{a} & $+$39.3(6)   & 4.8(14) & $-$2.85 & $+$1.09 & $-$5.66  & $+$9.46 \\ %MM2?
IRAS 18151          & $+$33.1(2)   & 1.8(10) & $-$0.52 & $<$0.07 & $>$5.63 & $+$1.35 \\ % 2nd envelope at V=35.34 km/s?
AFGL 2591           & $-$3.6(14)    & 4.0(10) & $+$24.0 & $+$0.92 & $+$8.67 & $+$11.8 \\
G327-0.6           & $-$42.5(5)  & 3.3(9)    & $-$1.96   & $-$0.62  & $>$15.3 & $-$1.43 \\
NGC 6334I        & $-$5.7(16)  & 5.7(7)    & $-$5.31   & $-$0.79  & $-$5.46  & $+$15.3 \\
G29.96               & $+$98.2(6) & 4.3(13)  & $-$0.97   & $+$1.10 & $>$13.8 & $+$19.2 \\
G31.41               & $+$99.1(4)  & 5.2(14) & $>$16.7  & $-$1.68  & $>$23.2 & $-$1.67 \\
G5.89                    & $+$12.0(4)    & 7.2(20) & $-$5.71 & $+$4.85 & $>$22.8 & $+$70.0 \\
G10.47                  & $+$70.0(10) & 3.9(4)    & $-$1.88 & $-$0.68 & $-$5.82   & $+$4.17 \\
\tablefootmark{a} & $+$65.5(9)   & 4.3(13)  & $-$1.63 & $+$0.50 & $-$6.43  & $+$18.3 \\
G34.26                  & $+$61.5(8)   & 4.9(11) & $-$22.1 & $-$0.63 & $-$8.40   & $-$1.35 \\ %second
\tablefootmark{a} & $+$57.1(4)    &  5.0(6)  & $-$5.81 & $+$2.99 &  $-$7.85 & $+$27.4 \\   %first
W51e                     & $+$55.7(17) & 6.3(12) & $-$12.5 & $-$0.81 & $-$10.2  & $+$43.8 \\
NGC 7538            & $-$57.0(3)     & 3.6(10) & $-$4.10 & $+$0.34 & $-$8.02  & $+$18.7 \\
\noalign{\smallskip}
\hline
\noalign{\smallskip}
\end{tabular}
\tablefoot{Positive fluxes denote integrated intensities ($\int$\tmb d$V$ in K\,\kms) of emission features, where upper limits denote non-detections. Negative fluxes denote integrated optical depths ($\int$$\tau$d$V$ in \kms) of absorption features, where lower limits denote saturated absorptions. Central velocities and FWHM line widths are in units of \kms, and numbers in parentheses are 1$\sigma$ errors in units of the last decimal. Line fluxes have uncertainties of $\approx$10\%. \\
\tablefoottext{a}{Source with two envelope components.} 
}
\end{table*}

\clearpage
\newpage

\begin{table*}
\caption{As Table~\ref{t:env}, for the outflows.}
\label{t:flow}
\begin{tabular}{lcccccc}
\hline \hline
\noalign{\smallskip}
Source & \vlsr\ & \dv\ & 1113\,GHz & \hhoe\ & 1669\,GHz & 987\,GHz \\
\noalign{\smallskip}
\hline
\noalign{\smallskip}
IRAS 05358      & --13.13(54)   & 16.9(46)  & $+$14.2 & +0.42 & +19.0   & +23.3 \\
IRAS 16272      & --34.8(3)       &   6.72(3)   & $-$1.73  & n/d      & --6.75    & n/m \\ 
                            & --48.1(28)     & 20.4(49)  & $+$6.16 & n/d      & +3.79     & +11.4 \\ % take together with next?
NGC6334I(N)   &  --4.3(19)      & 9.2(66)    & $-$3.58 & --2.06  & $>$14.2 &  n/m \\ % narrow abs (but H2O-18 is broad)
                            &  +1.0(30)      & 17.1(39)  & +21.6    & +4.33   & n/m         & +41.0 \\ %broad em
W43 MM1          &  +95.85(35) & 6.65(81)  & --8.25    & n/d       & --11.3  & n/m \\ %abs
                             &  +105.4(35) & 10.0(48) & +5.62     & n/d       & +3.41   & +29.9 \\ %em
DR21(OH)         & --1.15(60)    & 2.17(7)    & $-$2.61  & --0.24 & --2.50   & n/m \\ %narrow absorption
                            & --3.9(46)       & 13.1(64) & $+$41.0 & +1.21 & +21.9   & +80.8 \\% broad emission
W3 IRS5             & --41.88(90) & 3.0(14)   & --0.17   & +1.74 & --4.57    & +2.85 \\ % narrow (envelope?)
                             & --36.6(11)  & 21.7(73)  & +26.0   &  n/m    & +65.5     & +66.3 \\ % broad
IRAS 18089       & +32.5(27)  & 4.6(9)       & --10.8  & +0.59  & $>$17.0 & n/m \\
                             & +37.3(27)  & 14.2(28)  & +5.41   &  n/m    & +5.67      & +10.5 \\  % broad
W33A                  & +29.8(15)  &   5.8(26)   & --16.80 & --0.28 & --6.34    & n/m \\ %narrow abs
                             & +42.6(45) & 20.6(91)   & +11.1    & +1.09 & +10.7    & +15.2 \\ % broad em
IRAS 18151       & +34.3(4)   & 15.6(51)   & +3.98    & n/d      & +4.21    & +4.20 \\
AFGL 2591        & --10.0(17) & 15.0(16)   & --20.6 & n/d      & --38.3 & +6.30 \\
G327-0.6            & --48.3(15)     & 6.5(26)    & $-$6.91 & $-$0.71 & $>$18.44 & +3.75 \\ %narrow
                             & --37.5(50)     & 15.3(47) & +2.29      & +2.51 & n/m & +32.6 \\ %broad
NGC 6334I        & --10.2(49)      &  6.3(23)  & $-$2.21  & $-$1.19 & n/m        & $-$1.22 \\ % narrow 
                             &--11.0(56)      & 14.7(54) & $-$1.57  & $-$0.29 & $-$21.8 & +55.2 \\ % broad
G29.96                & +100.7(56) & 3.9(15)   & --16.96 & n/d & --2.98 & n/m \\ %narrow = abs
                             & +99.13(39)   & 18.4(24) & +16.1 & n/d & +24.6 & +20.2 \\ % broad = em
G31.41                & +95.0(4)       & 4.0(19)    & --5.36 & n/d & --2.34 & +6.91 \\  % narrow
                             & +99.65(3)     & 34.0(3)    & n/m & n/d & n/m & +12.8 \\  %broad
G5.89                  & +44.1(30) & 14.8(4)   & +7.55    & n/m    & +18.46 & n/m \\
                             & +27.7(32) & 19.3(65) & +20.5 & +3.86 & +35.71 & +119.4 \\
                             & --9.4(28)   & 30(13)    & --3.78   & --1.64 & +34.14 & +65.7 \\
                             & --26.9(5)   & 13.9(45) & --0.70   & --0.62 & n/m       & n/m \\
                             & --49.5(85) & 16.2(72) & --1.14   & --0.30 & --1.53  & +4.83 \\
G10.47                & +62.0(5)   & 4.1(5)     & +2.70   & n/m      & n/m       & +4.45 \\  %narrow
                             & +80.2(51) & 8.3(6)     & +4.71   & n/m     & --1.36   & +5.47 \\  %broad
G34.26                & +53.7(77) &  4.2(13)  & --9.60  & +1.20  & --9.14   & n/m \\ % narrow
                             & +57.0(85) & 18.1(81) & --1.66  & --0.96  & --10.7 & +25.8 \\ % broad
                             % ambiguous assignment .. redo fits to 1113 & H2O-18 lines?
W51e                  & +67.7(13) & 6.9(28)    & --14.3  & --0.25 & --28.2    & --1.44 \\ % narrow
                             & +61.7(13) & 14(11)     & --17.6  & --0.91 & --4.70    & +62.5 \\ % broad
NGC 7538 IRS1& --57.7(51) & 2.25(43)  & --2.36 & --0.13 & --0.80    & n/m     \\ % narrow
                             & --56.8(58) & 13.4(24)  & +10.3  & +1.67 & +2.66    & +17.4 \\ % broad (1669 = sum em/abs)
\noalign{\smallskip}
\hline
\noalign{\smallskip}
\end{tabular}
\tablefoot{The symbol n/d means that the outflow is not \newer{at all} detected in this line; the symbol n/m means that \newer{only} this component has no match in this line.}
\end{table*}

\clearpage
\newpage

\begin{table*}[p]
\caption{As Table~\ref{t:env}, for the foreground clouds.}
\label{t:fg}
\begin{tabular}{lcccccc}
\hline \hline
\noalign{\smallskip}
Source & \vlsr\ & \dv\ & 1113\,GHz & 1669\,GHz & $N$(\hho) & o/p \\
\noalign{\smallskip}
\hline
\noalign{\smallskip}
%IRAS 05358      & (none)
IRAS 16272      & --41.38(8) & 1.67(3) & --2.16 & $>$6.33 & $>$35 & $>$5.9 \\
                            & --38.80(23) & 1.94(5) & --1.36 & --2.50  & 15 & 3.7 \\
NGC6334I(N)   & +6.86(13) & 1.49(12) & --1.52 & $>$5.12 & $>$27 & $>$6.6 \\
W43 MM1          & +62.1(6) & 1.04(8) & --0.094 & --0.11 & 0.73 & 2.4 \\
                            & +67.21(4) & 1.81(8) & --1.13 & --1.80 & 11 & 3.2 \\
                            & +70.96(8) & 1.6(2) & --0.73 & --1.15 & 7.1 & 3.2 \\
                            & +77.(7) & 4.1(6) & --1.63 & --1.11 & 9.0 & 1.4 \\
                            & +79.7(1) & 1.93(5) & --1.48 & --3.67 & 21 & 5.0 \\  
                            & +82.4(2) & 2.49(5) & --1.63 & --2.58 & 16 & 3.2 \\  
                            & +87.83(12) & 2.1(4) & --0.20 & --0.49 & 2.75 & 4.9 \\
DR21(OH)         & +14.9(1) & 1.5(2) & --0.35 & --0.90 & 5.0 & 5.1 \\
                            & +12.3(1) & 1.8(2) & --0.47 & --2.03  & 11 & 8.6 \\
                            & +10.0(1) & 1.7(7) & --0.51 & --2.53 & 13 & 9.9 \\
                            &  +8.4(4) &  1.9(2) & --3.78 & --0.80 & 13 & 0.4 \\
                            &  +6.9(1) & 1.22(1) & --0.51 & --0.38 & 3.0 & 1.5\\
W3 IRS5            & --20.8(1)    & 1.04(8) & --0.61     & $>$3.56 & $>$18 & $>$12 \\
IRAS 18089      & +17.6(1)    & 3.4(6)   & --2.81      & --7.24     & 40.3    & 5.2 \\           
                            & +19.7(3)    & 2.8(11) & --0.91     & --2.34     & 13.0     & 5.1 \\ 
                            & +21.1(1)    & 2.2(4)   & --0.95     & --1.42     & 8.8       & 3.0 \\ 
                            & +23.9(2)    & 1.9(1)   & --0.82     & --2.68     & 14.4     & 6.5 \\ 
                            & +25.7(2)    & 2.67(7) & --1.46     & --4.00     & 22.0    & 5.5 \\  % +one more at 1669 only (V0 = +10 km/s)                            
W33A                 & +24.01(7)  & 1.5(1)   & --0.66     & --1.26      & 7.4      & 3.8 \\
IRAS 18151      & +27.7(2)    & 0.7(2)   & --0.36      & --1.06     & 5.8       & 5.9 \\           
AFGL 2591        & --0.67(13) & 3.0(6)   & $>$9.54 & $>$7.63 & $>$58 & $\sim$1.6 \\
                            & +13\tablefootmark{a} \\
G327--0.6\tablefootmark{d}            & --38.02(4) & 1.18(33) & --0.06 & --0.19 & 1.1 & 6.0 \\ 
NGC 6334I         & +0.48(11) & 2.58(52) & --1.45 & --3.42 & 13.8 & 2.7 \\
                             & +6.63(17) & 1.61(16) & --2.11 & $>$5.63 & $>$31.2 & $>$5.3 \\
                             & +8.72(39) & 1.27(52) & --0.20 & --0.17 & 1.3 & 1.7 \\
G29.96               & +103.5(1) & 1.7(1) & --1.09 & --2.91 & 16 & 5.3 \\
                             & +91.6(2) & 3.4(5) & --0.67 & --0.59 & 4.3 & 1.8 \\
G31.41                & +83.0(3) & 3.4(6) & --0.48 & --1.27 & 7.0 & 5.3 \\ %sum of two clouds?
G5.89                  & +19.85(1) & 4.4(6)   & --1.90 & --25.17\tablefootmark{b} & 121.7\tablefootmark{b} & 26.5\tablefootmark{b} \\           
G10.47                & +42.3(1)   & 1.1(3)   & --0.16 & --0.24   & 1.5      &  3.0 \\  
                             & +47.06(4) & 1.35(7) & --0.09 & --0.19   & 1.1      & 4.5 \\  
                             & +84.9(1)   & 2.6(10) & --0.19 & --1.67\tablefootmark{b}   & 8.2\tablefootmark{b}      & 17.7\tablefootmark{b} \\  
                             & +91.21(6) & 1.4(4)   & --0.08 & --0.32   & 1.7      & 8.3  \\  % two more clouds seen at 1669 only
G34.26                & +11.4(1) & 2.6(1) & --6.43 & --14.0 & 80 & 4.4 \\
                             & +14.26(3) & 1.18(1) & --0.35 & --0.55 & 3.4 & 3.1 \\
                             & +27.09(1) & 1.6(3) & --1.52 & --5.48 & 29 & 7.2 \\
W51e                  & +48.66(1) & 2.27(1) & --0.42 & --1.14 & 6.3 & 5.4 \\
                             & +45.57(9) & 1.2(1) & --2.51 & $>$4.42 & $>$26 & $>$3.5 \\
                             & +12.7\tablefootmark{c} \\ 
                             & +6.48(5) & 2.4(2) & --2.63 & --2.99 & 20 & 2.3 \\
                             & +4.87(1) & 1.1(2) & --0.27 & --0.84 & 4.5 & 6.2 \\
%NGC 7538 IRS1& (none)
\noalign{\smallskip}
\hline
\noalign{\smallskip}
\end{tabular}
\tablefoot{The last two columns list the derived \hho\ column density (in $10^{12}$ cm$^{-2}$) and ortho/para ratio. \\
\tablefoottext{a}{Blended with \hho\ 1661 GHz line emission from other sideband; see \citet{choi2012}}
\\
\tablefoottext{b}{Highly uncertain because 1669 GHz line is saturated.}
\\
\tablefoottext{c}{\new{Only detected at 1113 GHz.}}
\\
\tablefoottext{d}{\new{WBS spectra of this source show spatially extended \hho\ foreground absorbers between --20 and --3 \kms\ (S. Leurini, priv. comm.).}}
}
\end{table*}

%%%%%%%%%%%%%%%%

\clearpage
\newpage

\section{Observed line profiles}
\label{app:profs}

\begin{figure}[]
\centering
\includegraphics[width=7cm,angle=0]{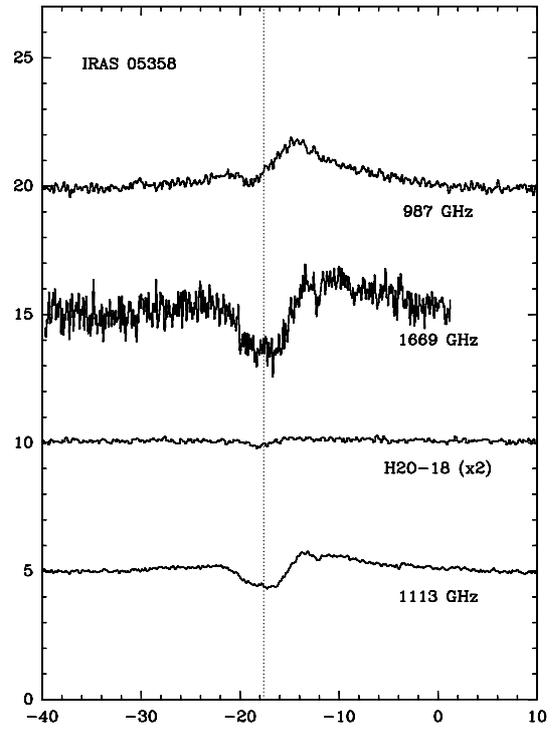}  
\caption{Spectra of the \hho\ $1_{11}-0_{00}$ (bottom), \hhoe\ $1_{11}-0_{00}$ (second from bottom), \hho\ $2_{12}$--$1_{01}$ (third from bottom) and \hho\ $2_{02}$--$1_{11}$ (top) lines toward IRAS 05358. The vertical scale is \tmb\ in K, and the dashed line denotes the systemic velocity determined from ground-based observations. \newer{The spectra have been shifted vertically for clarity.}}
\label{f:05358}
\end{figure} 

\begin{figure}[]
\centering
\includegraphics[width=7cm,angle=0]{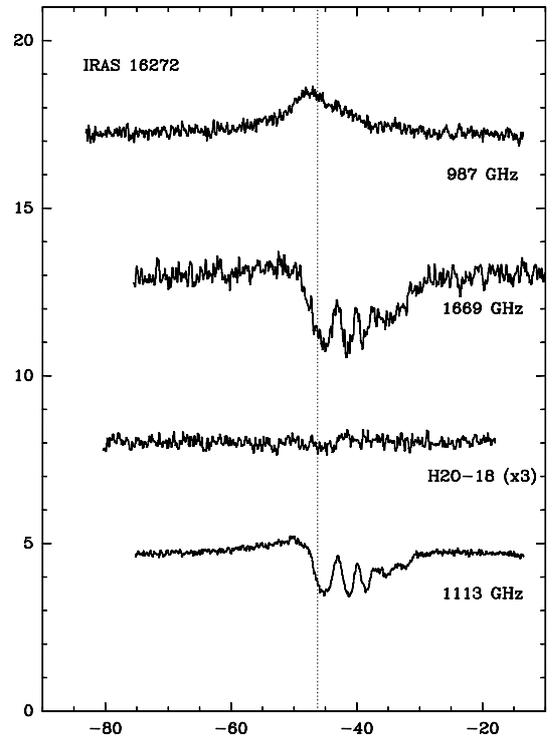}  
\caption{As Figure~\ref{f:05358}, toward IRAS 16272.}
\label{f:16272}
\end{figure} 

\begin{figure}[]
\centering
\includegraphics[width=7cm,angle=0]{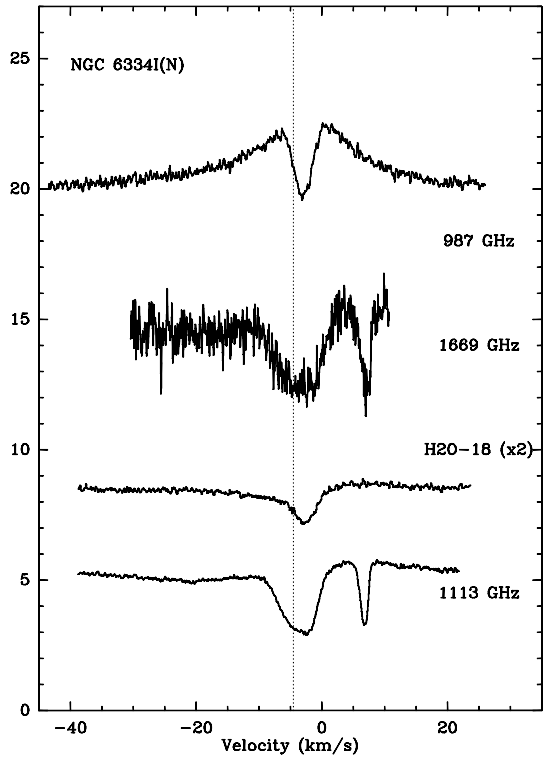}  
\caption{As Figure~\ref{f:05358}, toward NGC 6334I(N).}
\label{f:6334in}
\end{figure} 

\begin{figure}[]
\centering
\includegraphics[width=7cm,angle=0]{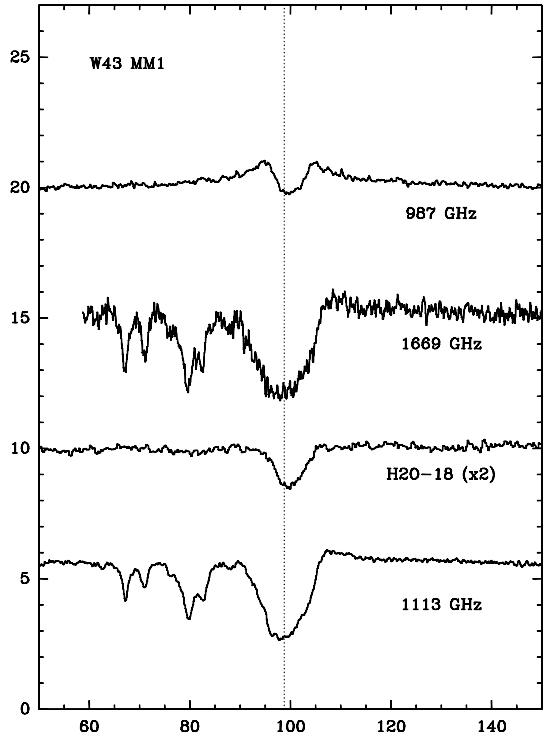}  
\caption{As Figure~\ref{f:05358}, toward W43 MM1.}
\label{f:w43mm1}
\end{figure} 

\clearpage
\newpage

\begin{figure}[]
\centering
\includegraphics[width=7cm,angle=0]{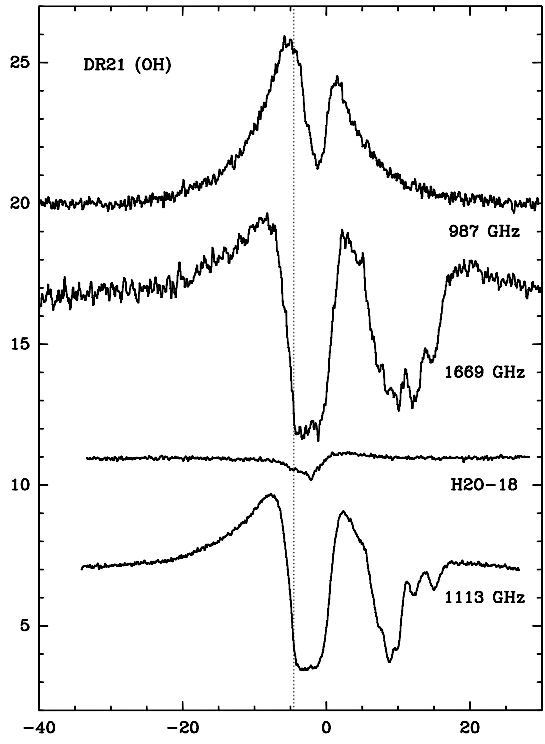}  
\caption{As Figure~\ref{f:05358}, toward DR21(OH).}
\label{f:dr21oh}
\end{figure} 

\begin{figure}[]
\centering
\includegraphics[width=7cm,angle=0]{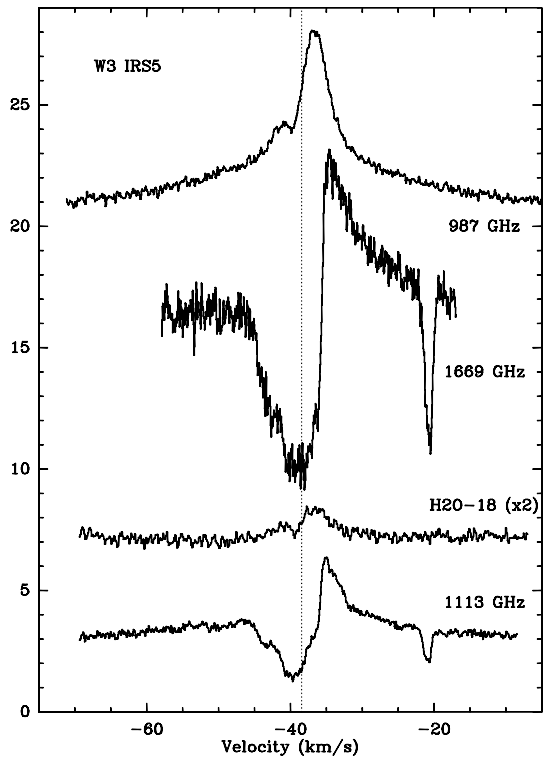}  
\caption{As Figure~\ref{f:05358}, toward W3 IRS5.}
\label{f:w3irs5}
\end{figure} 

\begin{figure}[]
\centering
\includegraphics[width=7cm,angle=0]{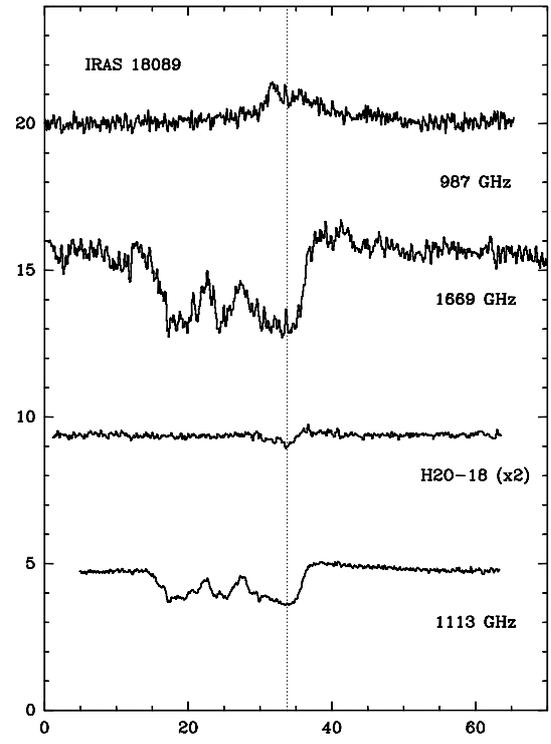}  
\caption{As Figure~\ref{f:05358}, toward IRAS 18089.}
\label{f:18089}
\end{figure} 

\begin{figure}[]
\centering
\includegraphics[width=7cm,angle=0]{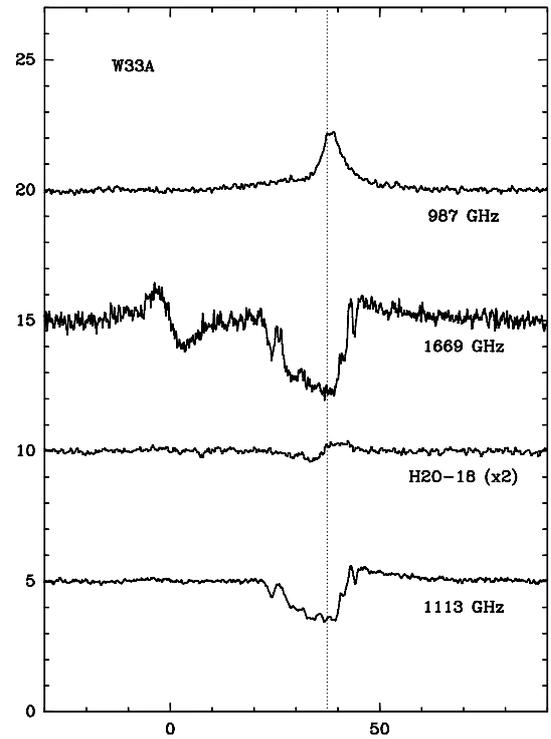}  
\caption{As Figure~\ref{f:05358}, toward W33A. \new{The emission/absorption feature near 0 \kms\ in the 1669 GHz spectrum is the \hho\ 1661 GHz line from the other sideband.}}
\label{f:w33a}
\end{figure} 

\clearpage
\newpage

\begin{figure}[]
\centering
\includegraphics[width=7cm,angle=0]{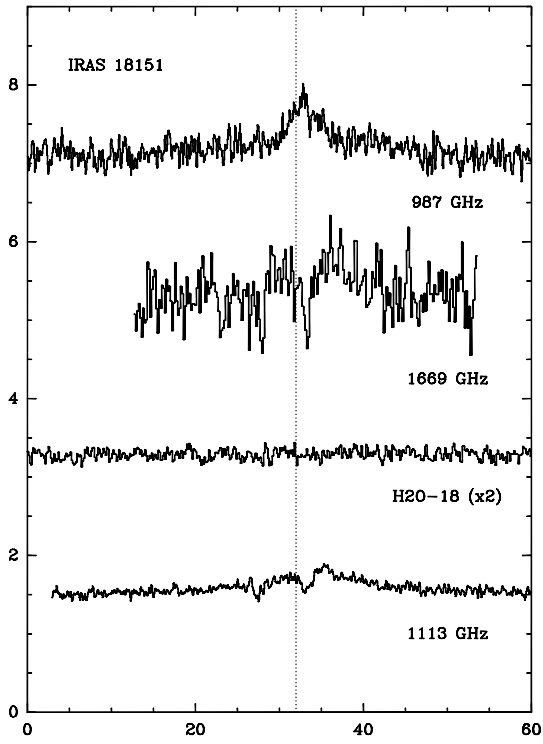}  
\caption{As Figure~\ref{f:05358}, toward IRAS 18151.}
\label{f:18151}
\end{figure} 

\begin{figure}[]
\centering
\includegraphics[width=7cm,angle=0]{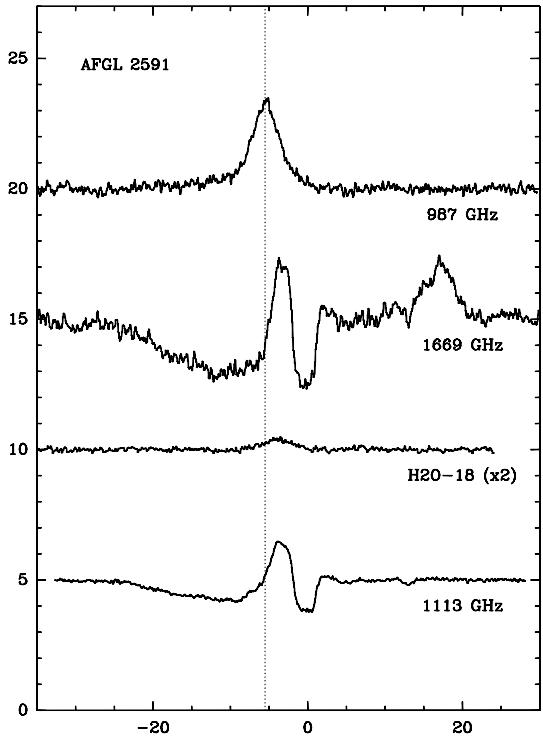}  
\caption{As Figure~\ref{f:05358}, toward AFGL 2591. \new{The emission feature near +18 \kms\ in the 1669 GHz spectrum is the \hho\ 1661 GHz line from the other sideband.}}
\label{f:2591}
\end{figure} 

\begin{figure}[]
\centering
\includegraphics[width=7cm,angle=0]{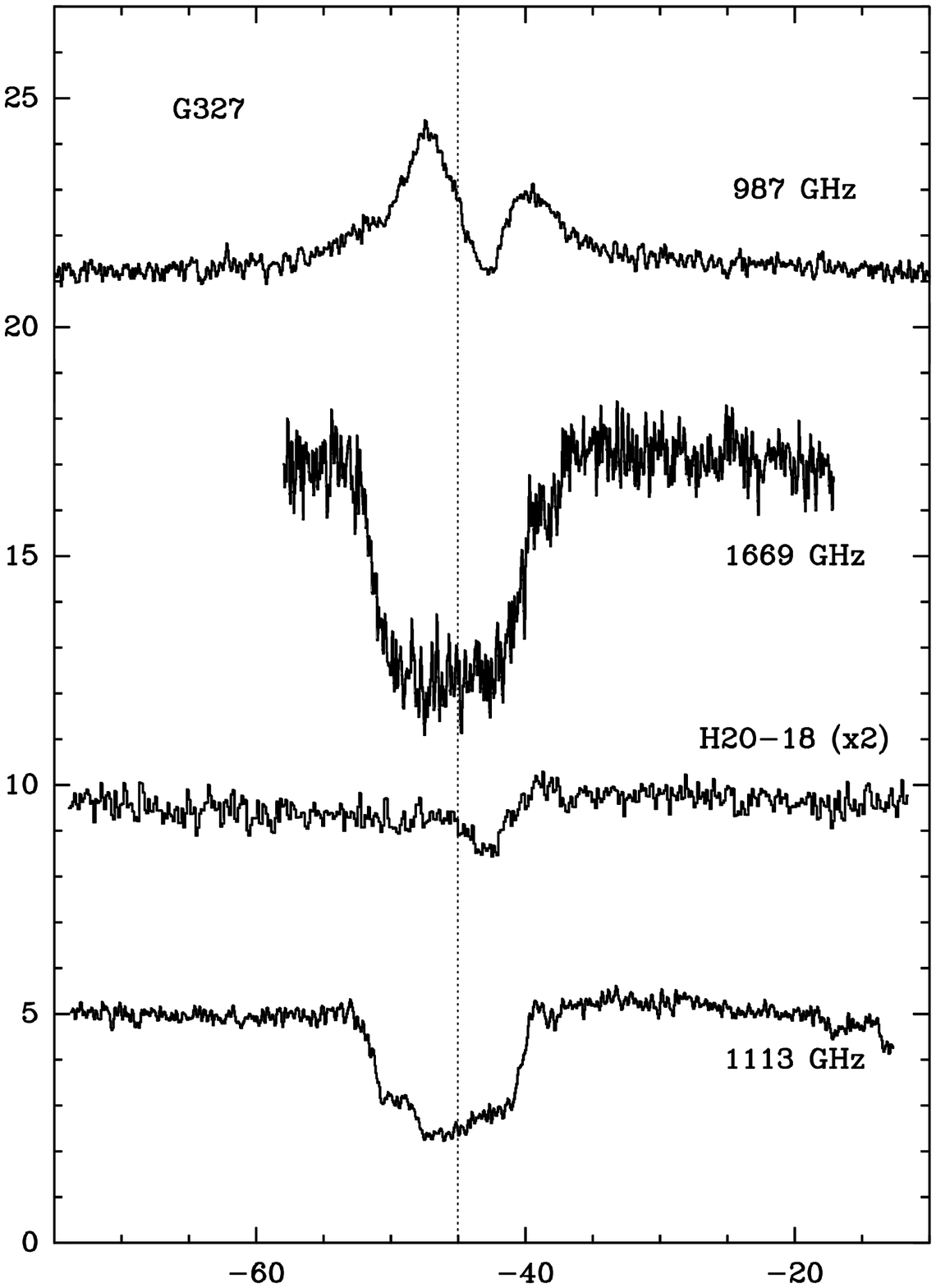}  
\caption{As Figure~\ref{f:05358}, toward G~327.}
\label{f:g327}
\end{figure} 

\begin{figure}[]
\centering
\includegraphics[width=7cm,angle=0]{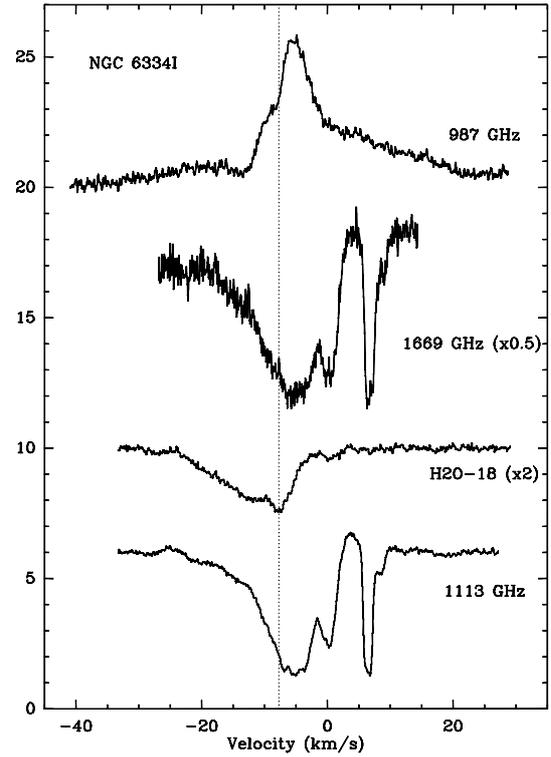}  
\caption{As Figure~\ref{f:05358}, toward NGC 6334I.}
\label{f:6334i}
\end{figure} 

\clearpage
\newpage

\begin{figure}[]
\centering
\includegraphics[width=7cm,angle=0]{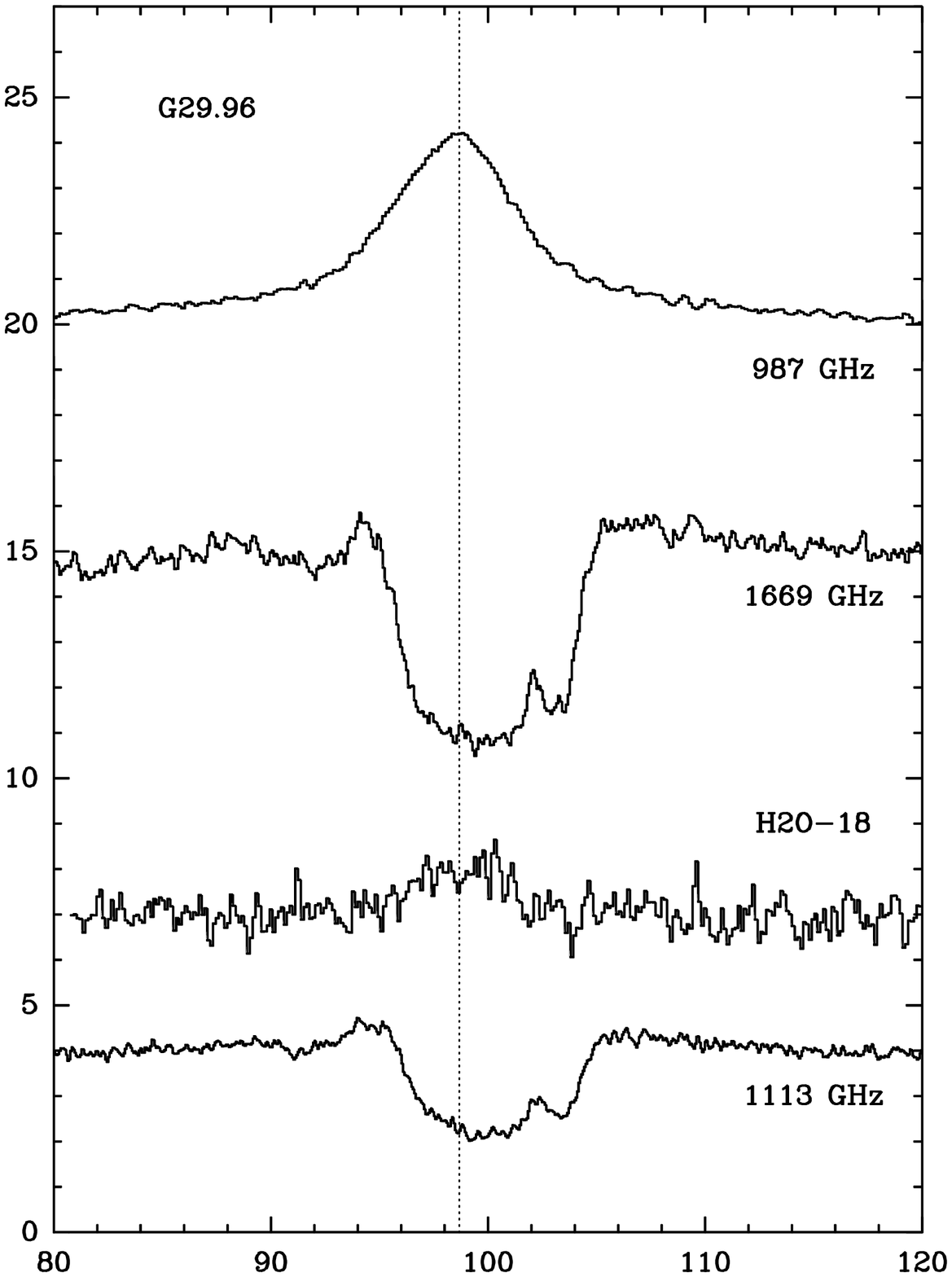}  
\caption{As Figure~\ref{f:05358}, toward G29.96.}
\label{f:g2996}
\end{figure} 

\begin{figure}[]
\centering
\includegraphics[width=7cm,angle=0]{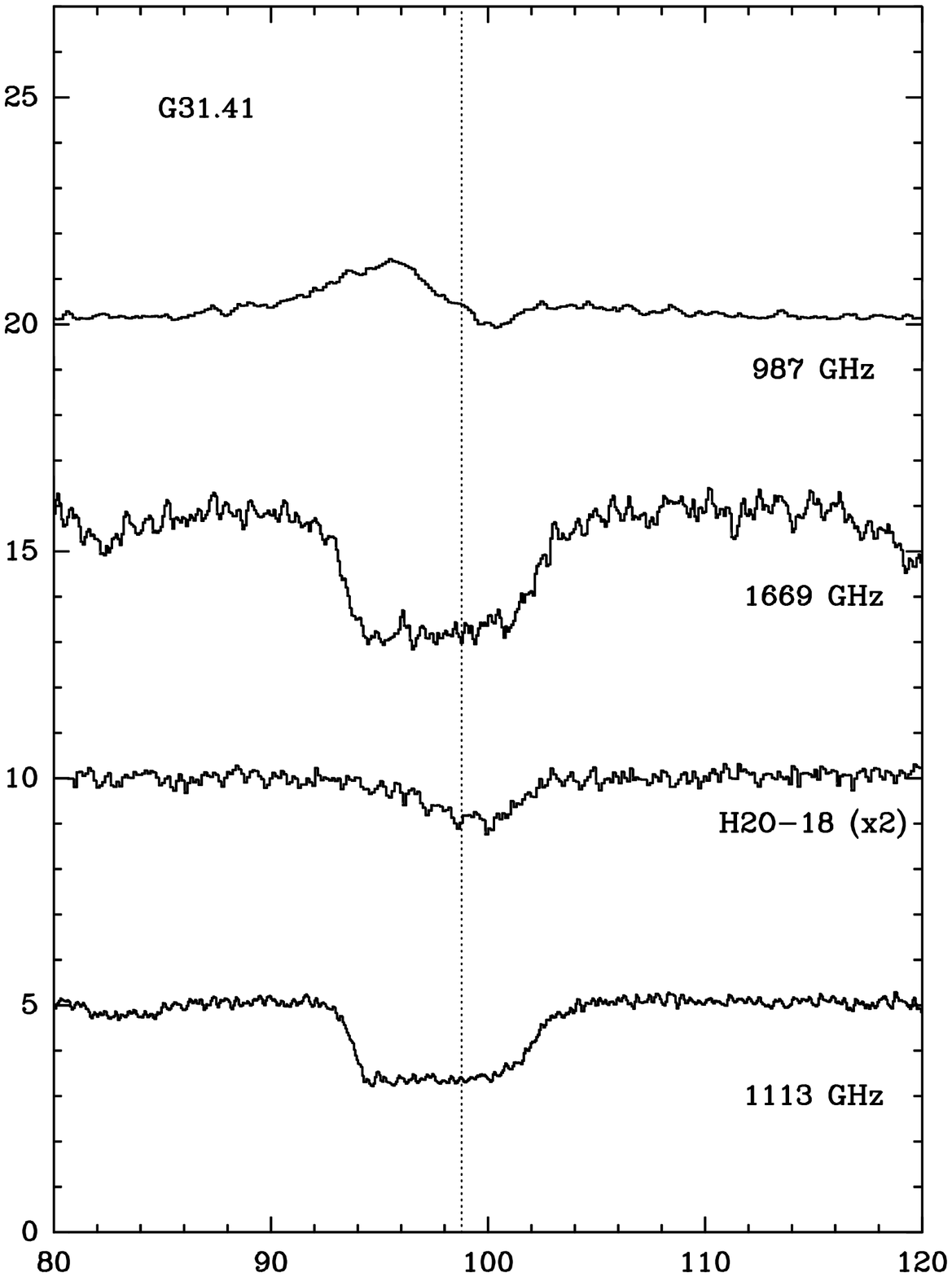}  
\caption{As Figure~\ref{f:05358}, toward G31.41.}
\label{f:g3141}
\end{figure} 

\begin{figure}[]
\centering
\includegraphics[width=7cm,angle=0]{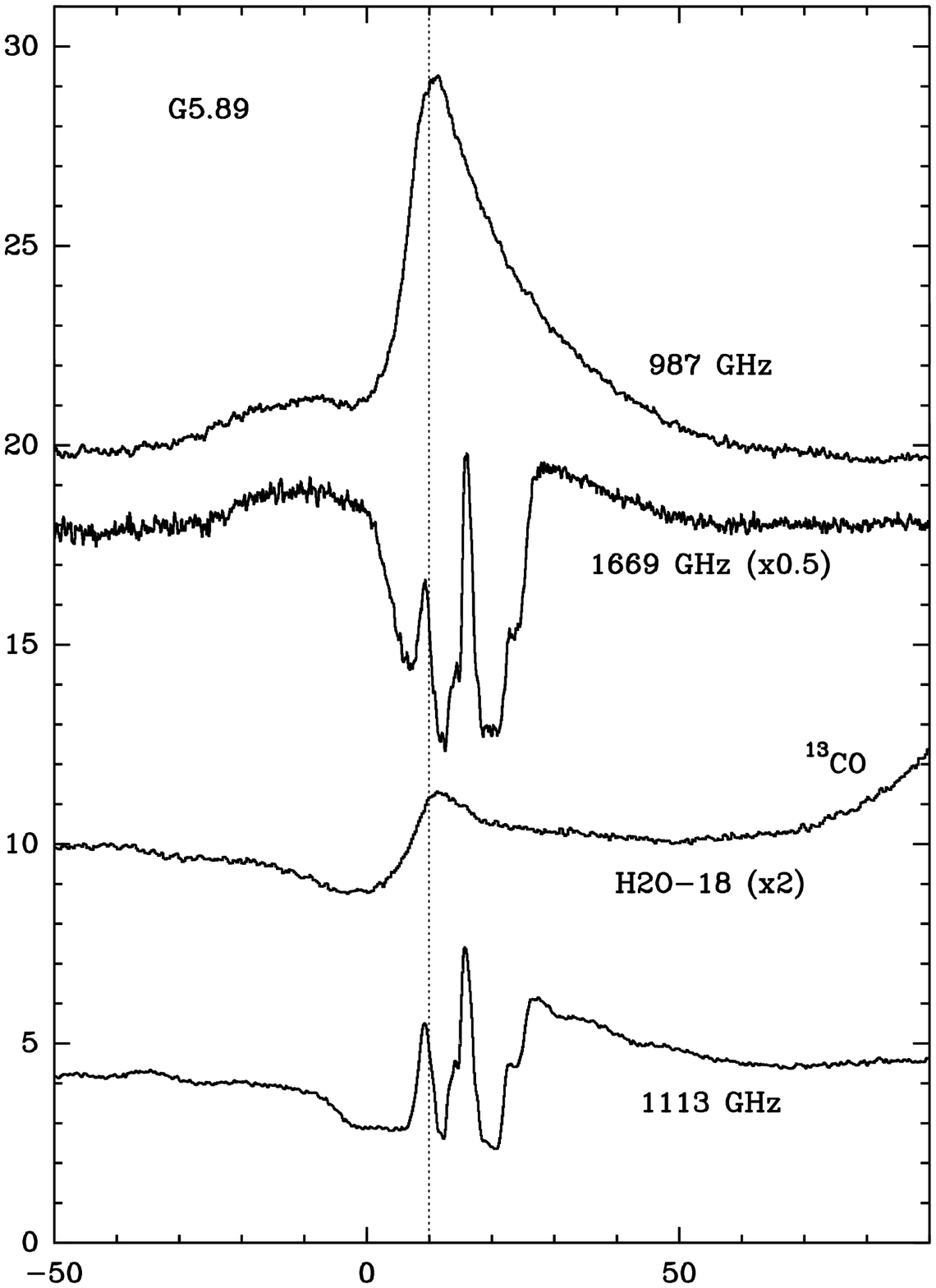}  
\caption{As Figure~\ref{f:05358}, toward G5.89. \new{The emission feature near +90 \kms\ in the \hhoe\ spectrum is the \thco\ 10--9 line.}}
\label{f:g589}
\end{figure} 

\begin{figure}[]
\centering
\includegraphics[width=7cm,angle=0]{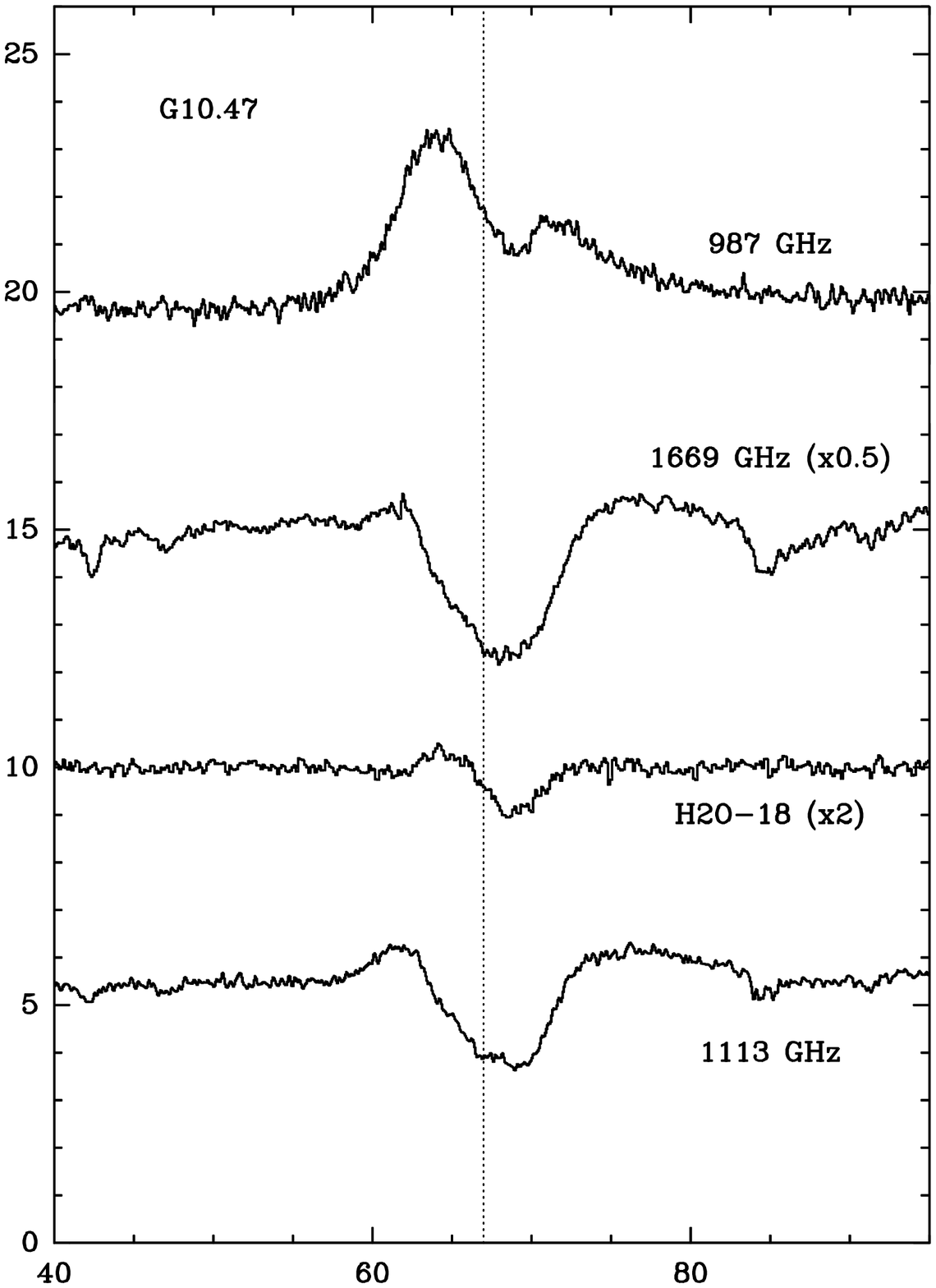}  
\caption{As Figure~\ref{f:05358}, toward G10.47.}
\label{f:g1047}
\end{figure} 

\clearpage
\newpage

\begin{figure}[]
\centering
\includegraphics[width=7cm,angle=0]{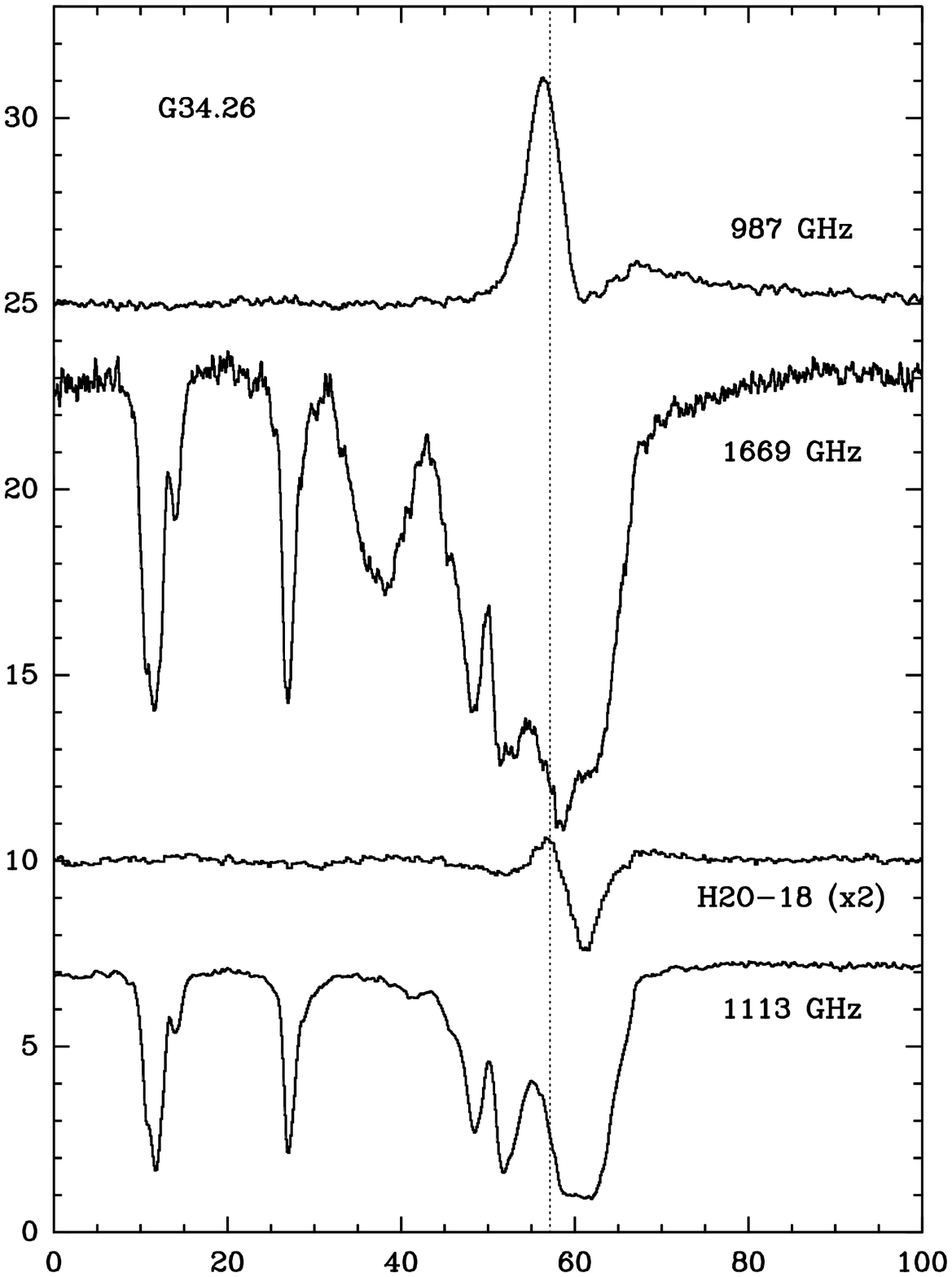}  
\caption{As Figure~\ref{f:05358}, toward G34.26.}
\label{f:g3426}
\end{figure}

\newpage

\begin{figure}[]
\centering
\includegraphics[width=7cm,angle=0]{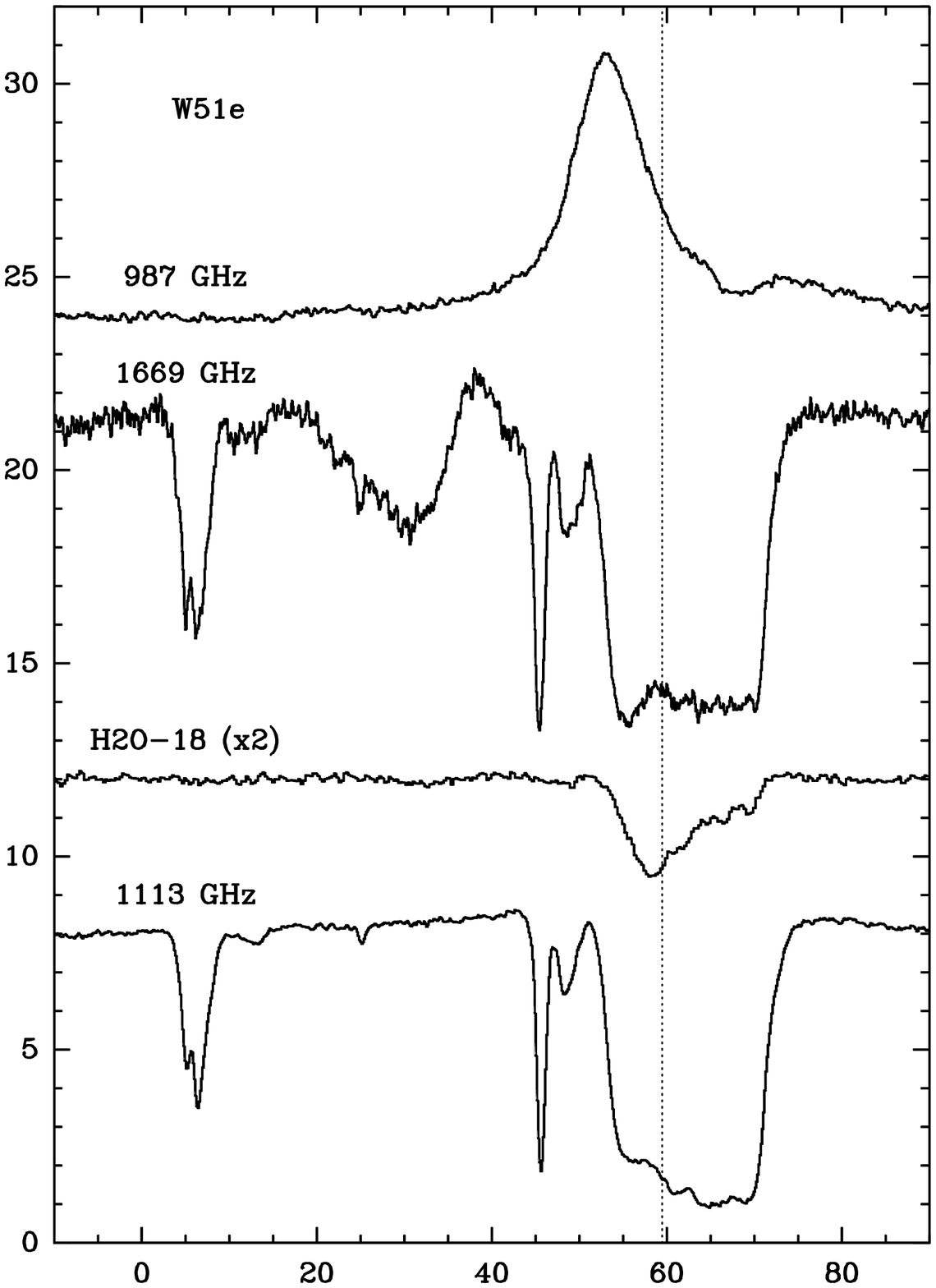}  
\caption{As Figure~\ref{f:05358}, toward W51e.}
\label{f:w51e}
\end{figure} 

\newpage

\begin{figure}[]
\centering
\includegraphics[width=7cm,angle=0]{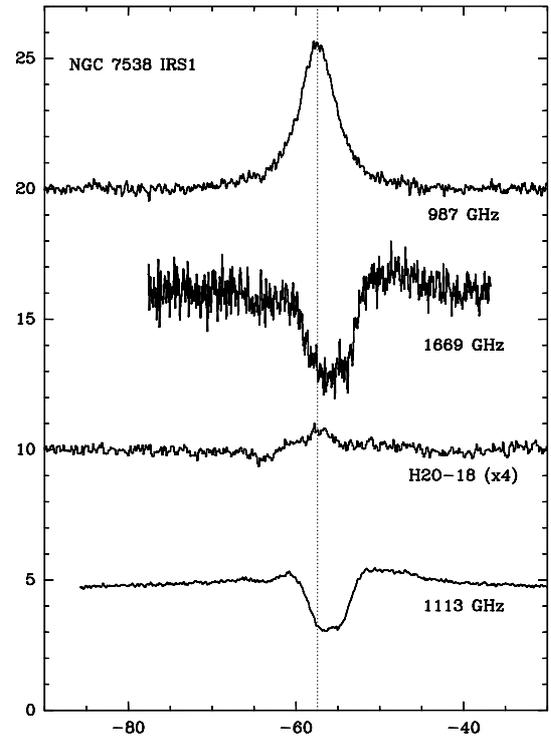}  
\caption{As Figure~\ref{f:05358}, toward NGC 7538 IRS1.}
\label{f:7538}
\end{figure} 

\clearpage
\newpage

\section{Continuum models}
\label{app:models}

\begin{figure*}[t]
\centering
\includegraphics[width=10cm,angle=0]{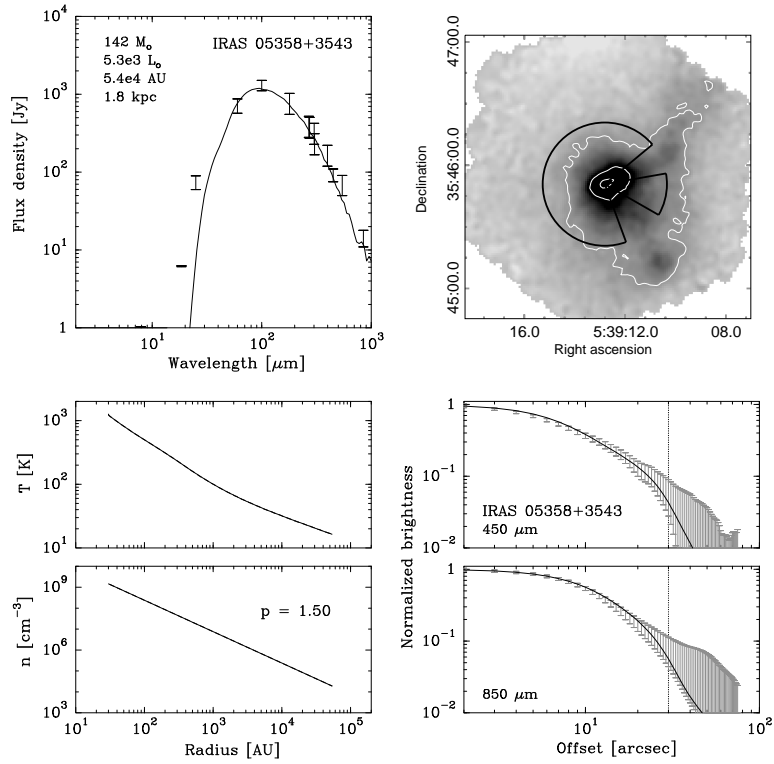}
\caption{Continuum model for IRAS 05358. Top left: Spectral energy distribution. Top right: Submillimeter image with 3$\sigma$ contour marked in white and useable map sectors marked in black. Bottom left: Temperature and density structure as a function of radius. Bottom right: Submillimeter emission profiles. \newer{The numbers in the top left panel are the modeled envelope mass, the modeled luminosity, the adopted envelope size and the adopted distance.}}
\label{f:05358-model}
\end{figure*} 

\begin{figure*}[t]
\centering
\includegraphics[width=10cm,angle=0]{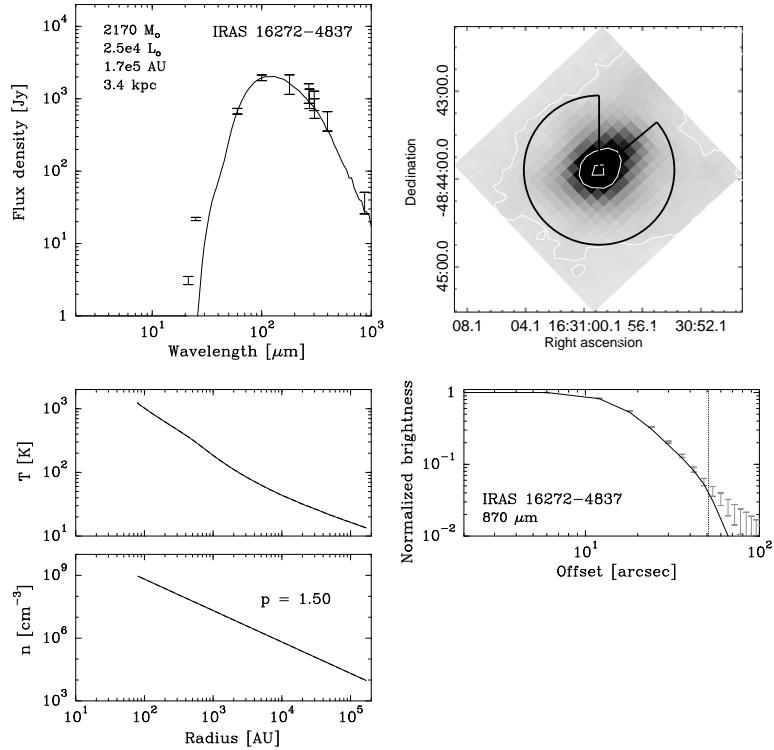}
\caption{As Figure~\ref{f:05358-model}, for IRAS 16272.}
\end{figure*} 

\clearpage
\newpage

\begin{figure*}[t]
\centering
\includegraphics[width=10cm,angle=0]{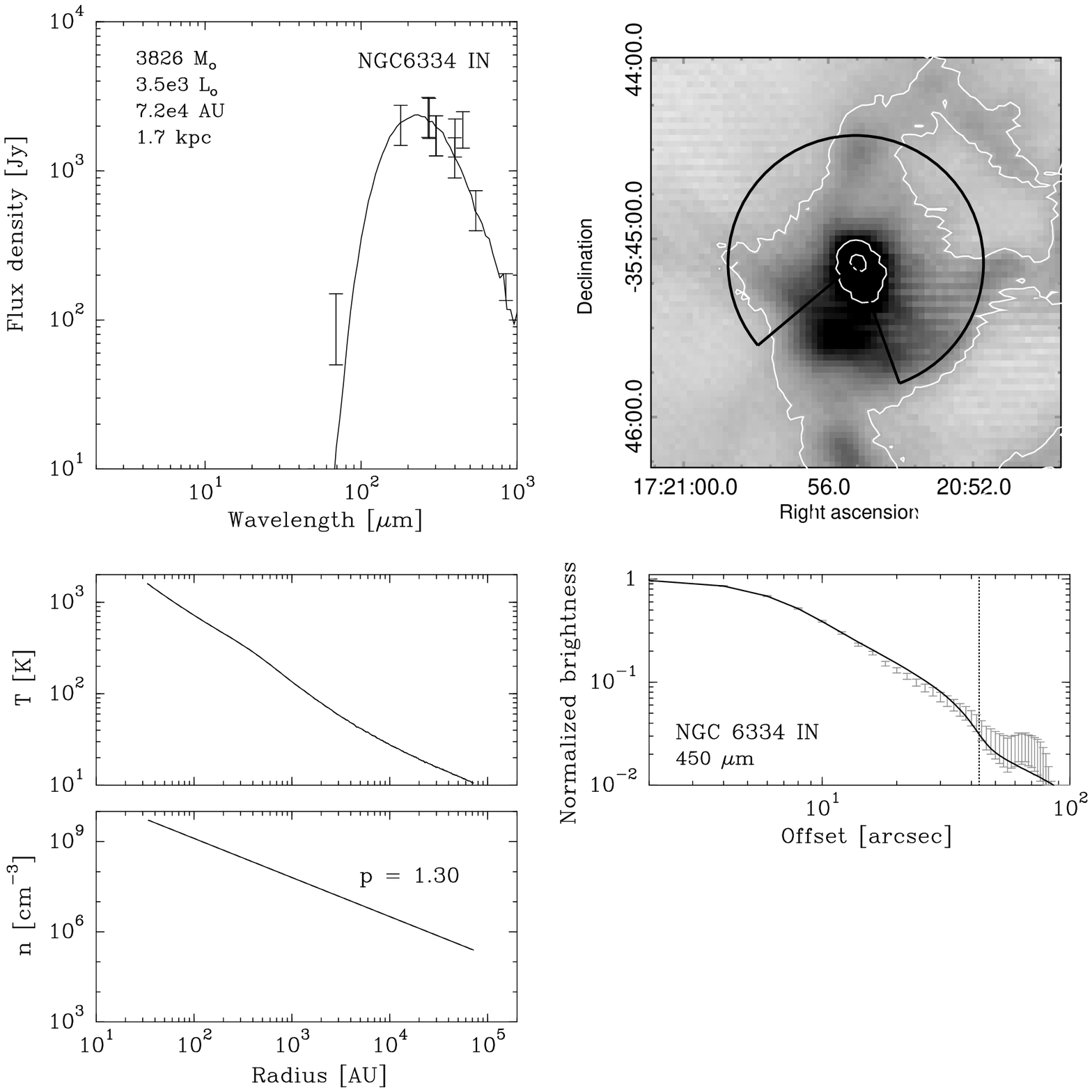}
\caption{As Figure~\ref{f:05358-model}, for NGC 6334I(N).}
\end{figure*} 

\begin{figure*}[t]
\centering
\includegraphics[width=10cm,angle=0]{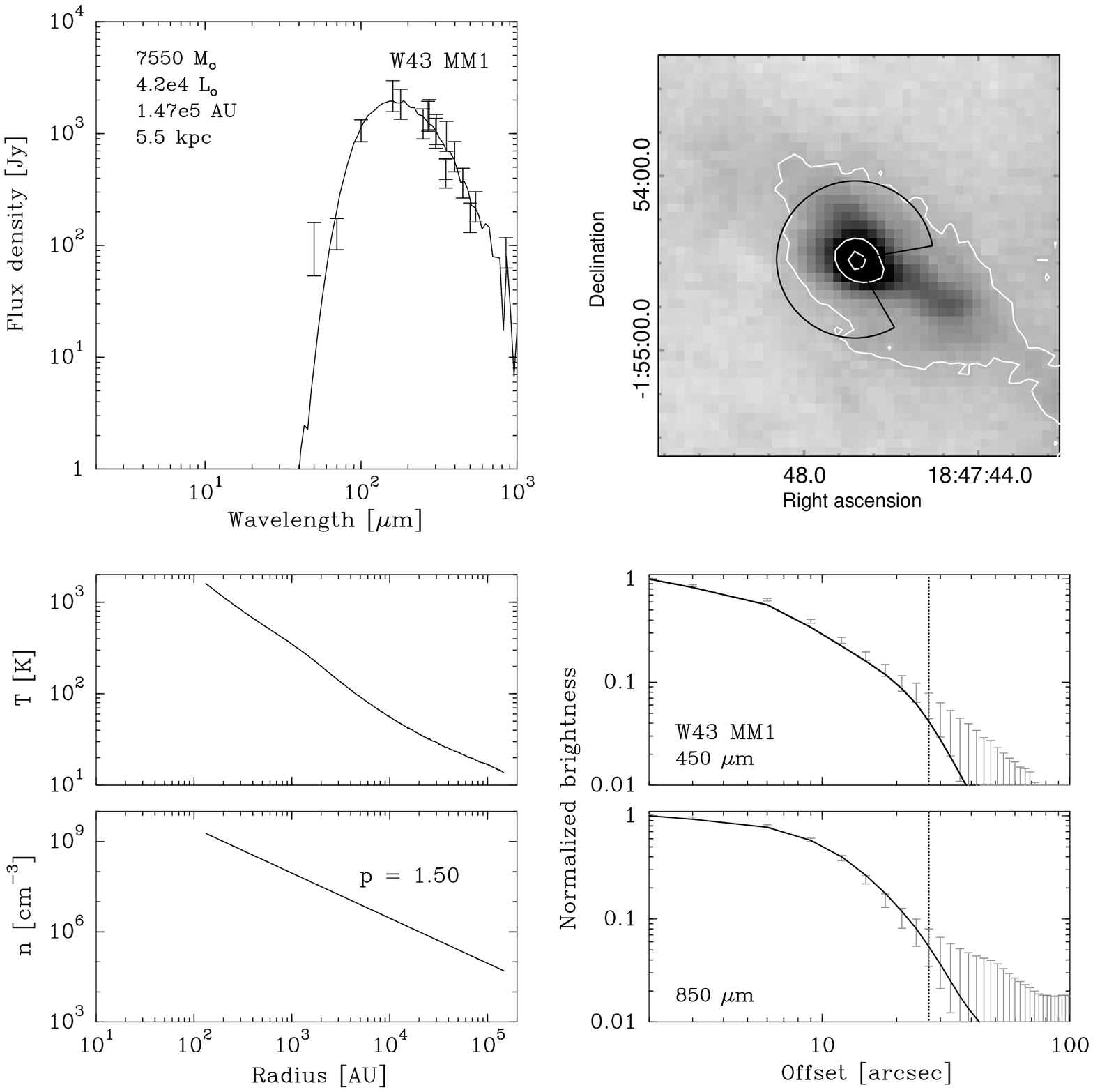}
\caption{As Figure~\ref{f:05358-model}, for W43-MM1.}
\end{figure*} 

\clearpage
\newpage

\begin{figure*}[t]
\centering
\includegraphics[width=10cm,angle=0]{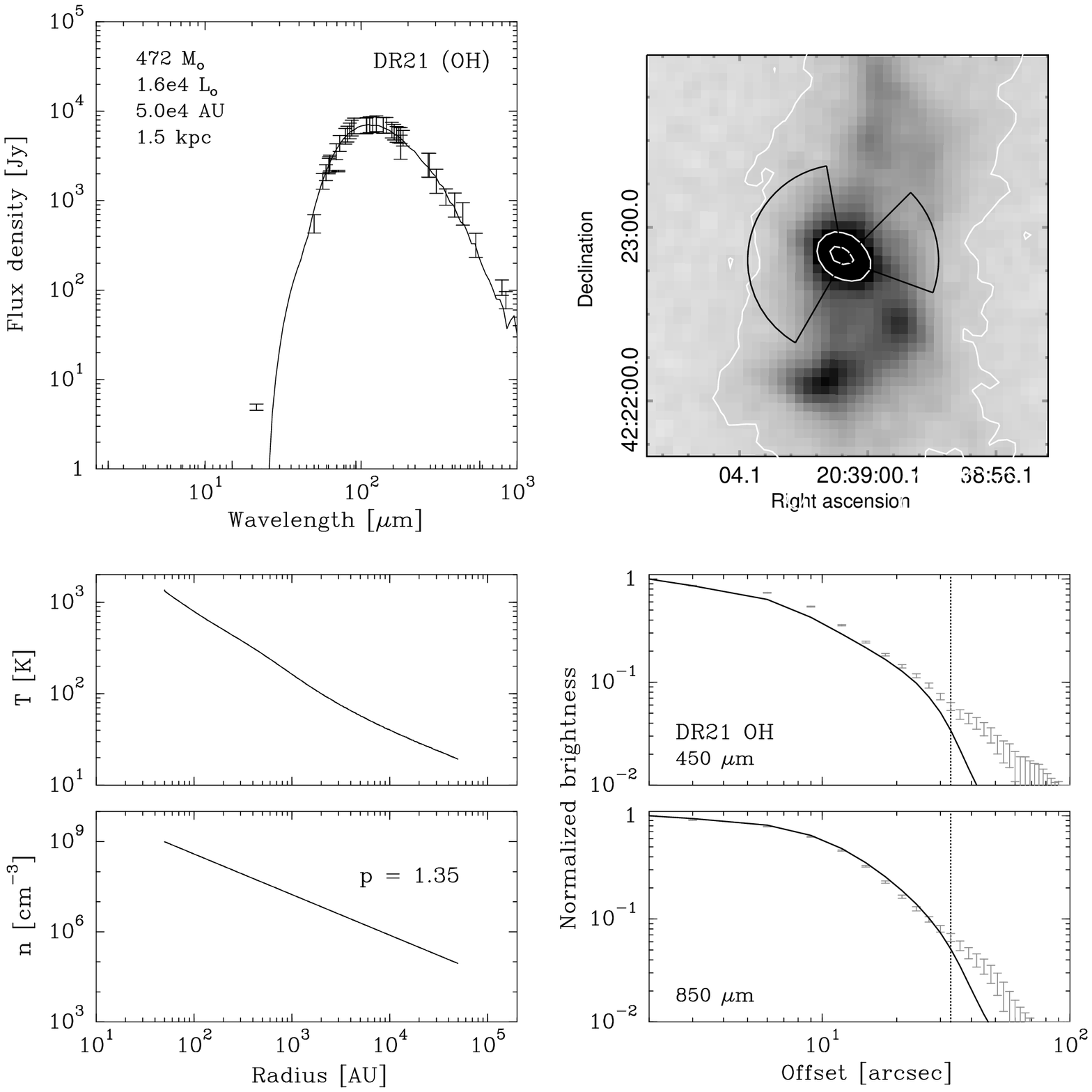}
\caption{As Figure~\ref{f:05358-model}, for DR21(OH).}
\end{figure*} 

\begin{figure*}[t]
\centering
\includegraphics[width=10cm,angle=0]{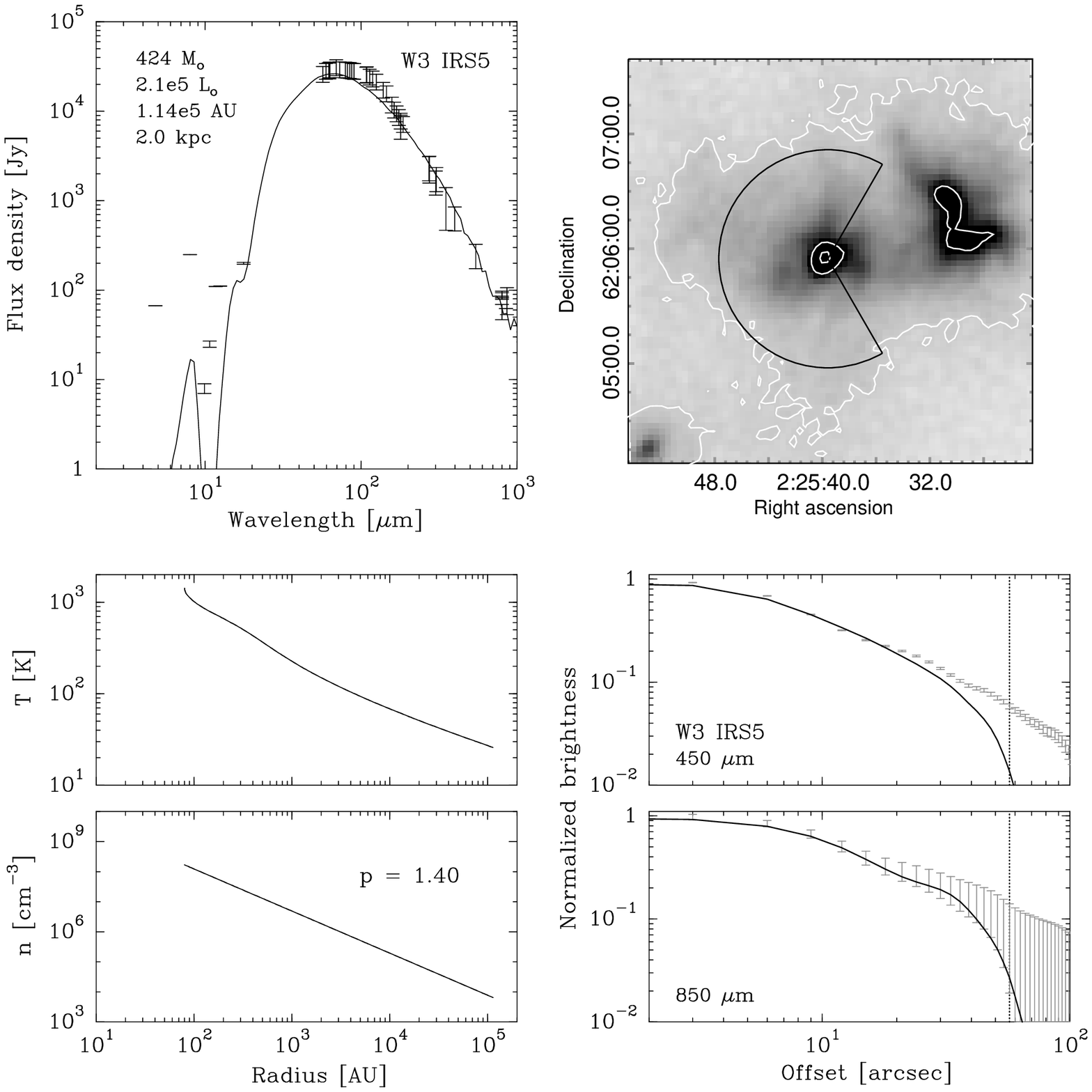}
\caption{As Figure~\ref{f:05358-model}, for W3 IRS5.}
\end{figure*} 

\clearpage
\newpage

\begin{figure*}[t]
\centering
\includegraphics[width=10cm,angle=0]{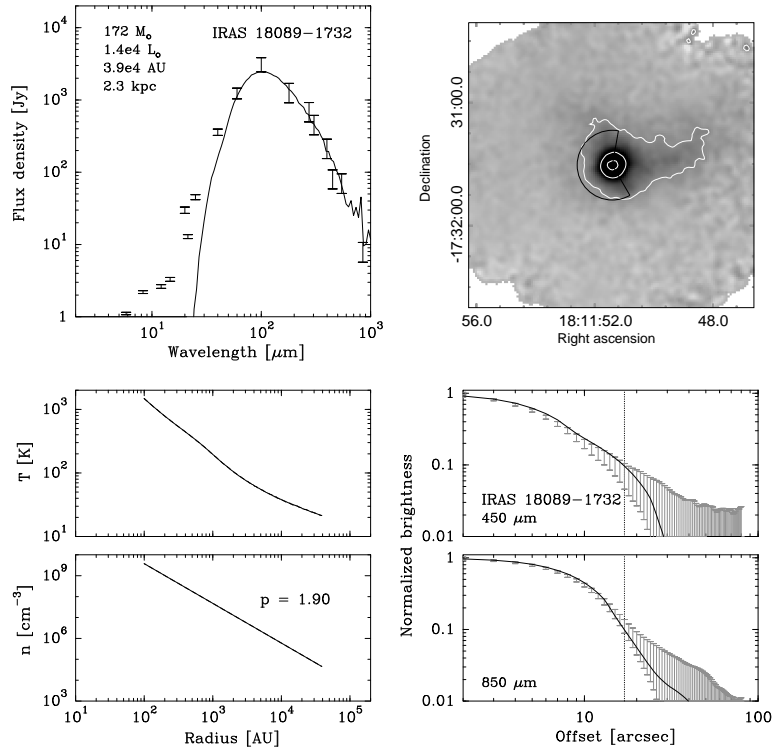}
\caption{As Figure~\ref{f:05358-model}, for IRAS 18089.}
\end{figure*} 

\begin{figure*}[t]
\centering
\includegraphics[width=10cm,angle=0]{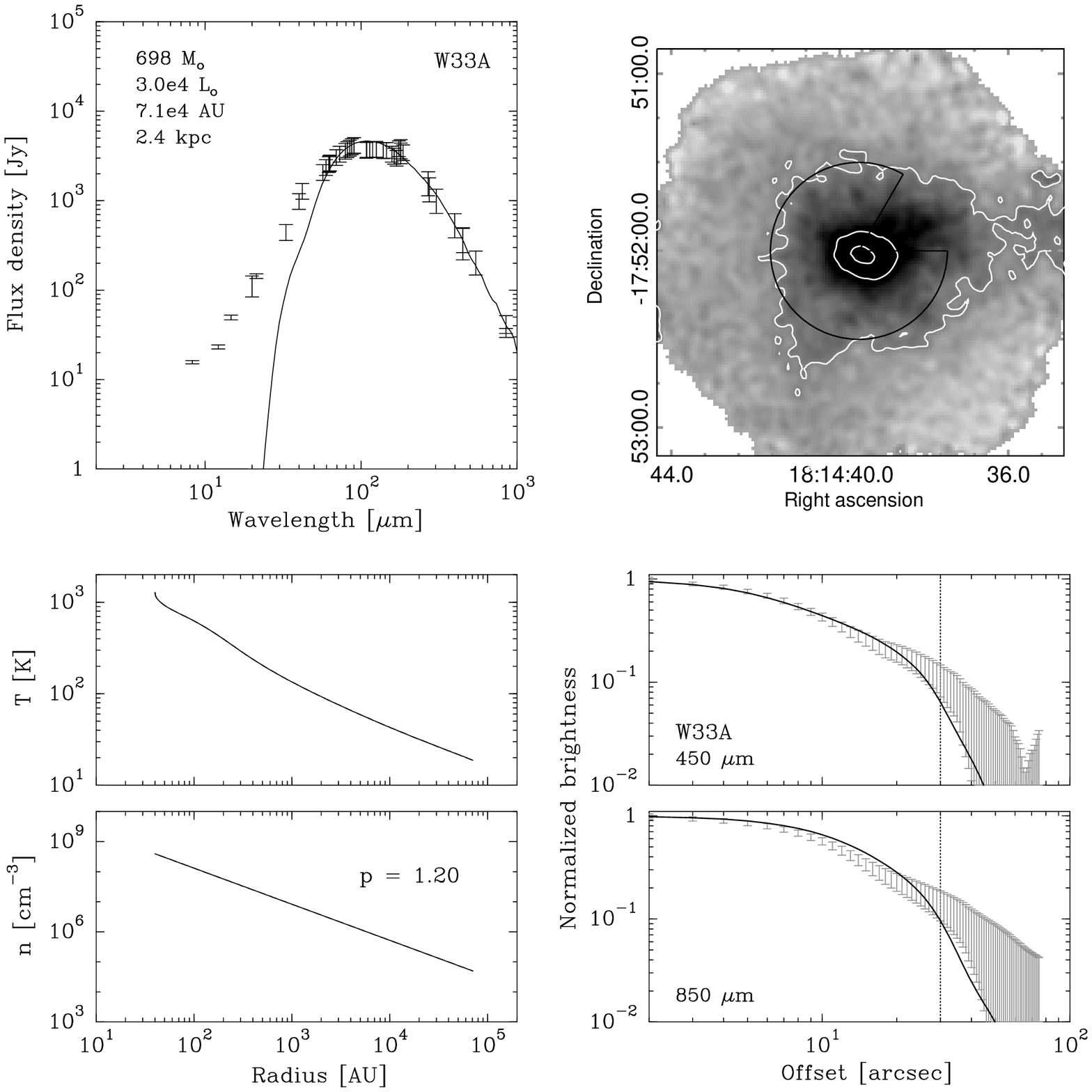}
\caption{As Figure~\ref{f:05358-model}, for W33A.}
\end{figure*} 

\clearpage
\newpage

\begin{figure*}[t]
\centering
\includegraphics[width=10cm,angle=0]{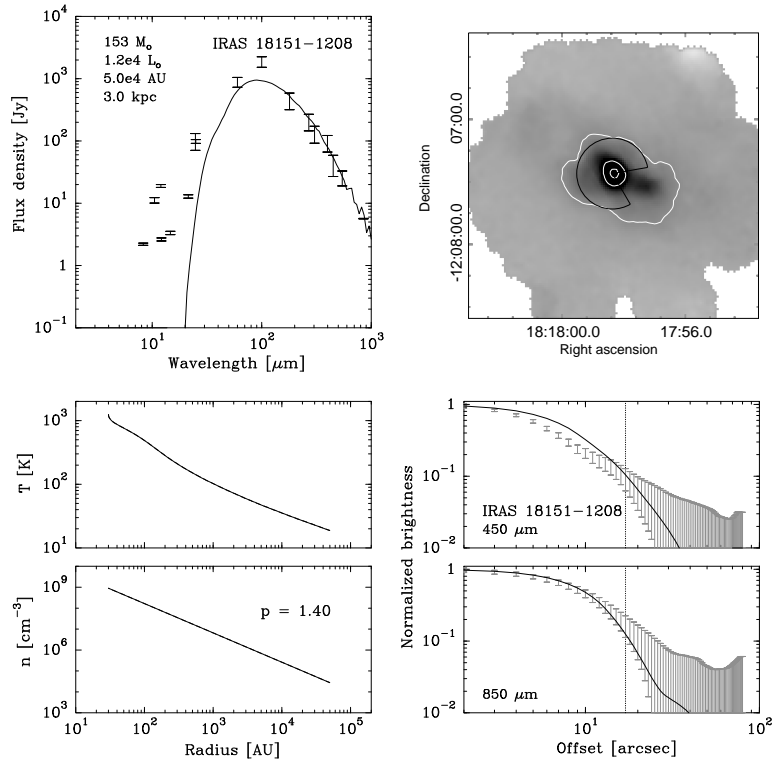}
\caption{As Figure~\ref{f:05358-model}, for IRAS 18151.}
\end{figure*} 

\begin{figure*}[t]
\centering
\includegraphics[width=10cm,angle=0]{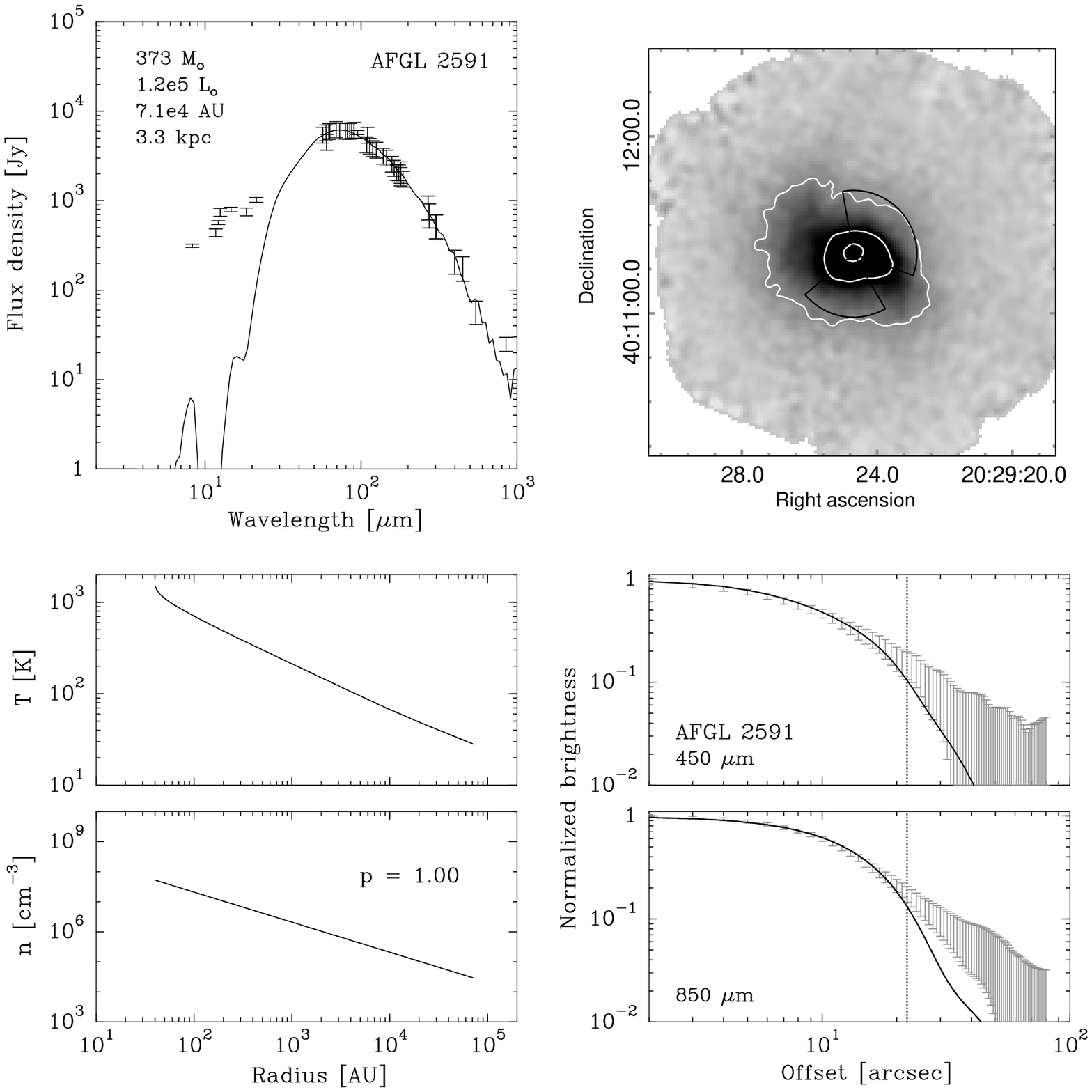}
\caption{As Figure~\ref{f:05358-model}, for AFGL 2591.}
\end{figure*} 

\clearpage
\newpage

\begin{figure*}[t]
\centering
\includegraphics[width=10cm,angle=0]{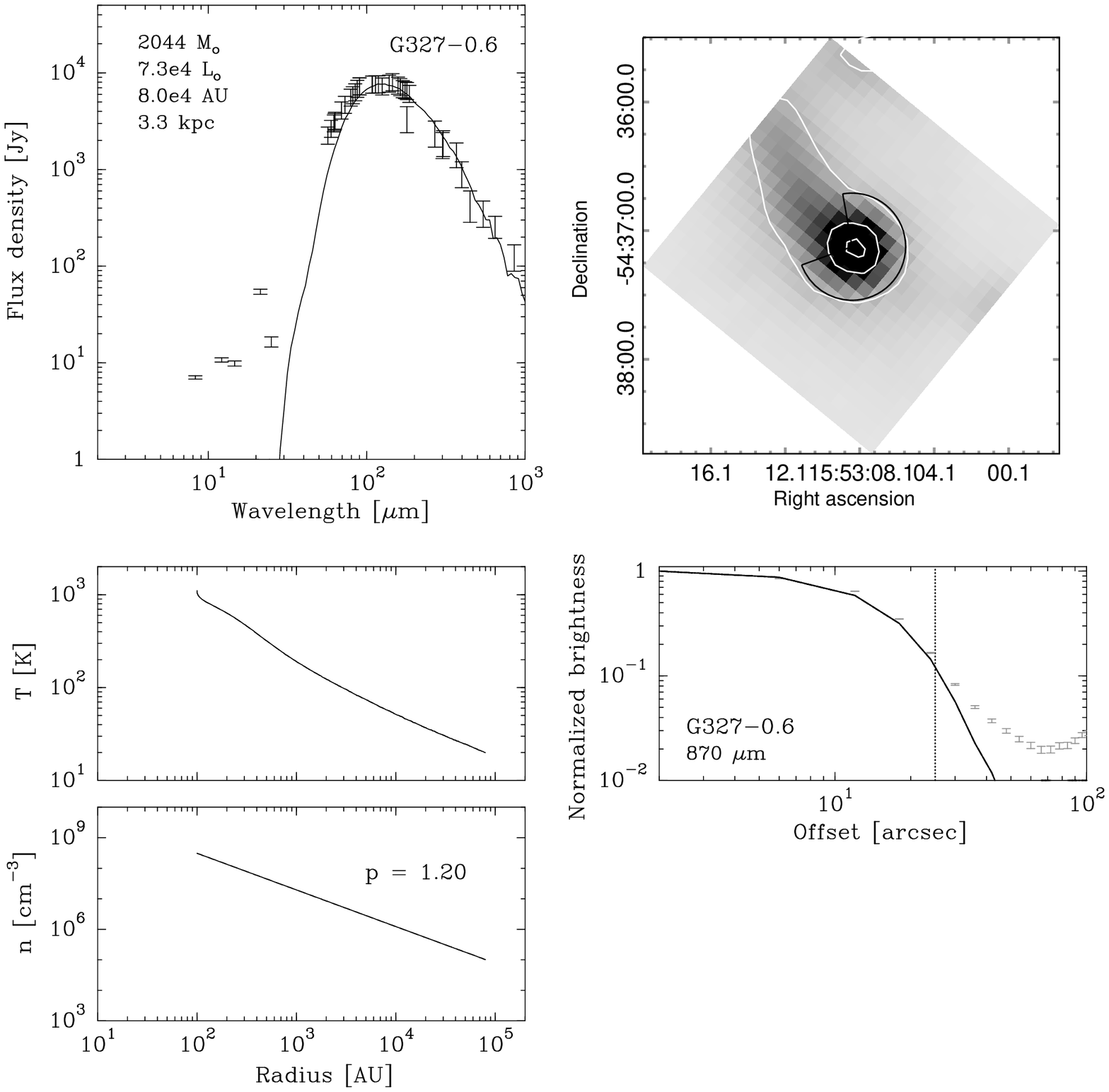}
\caption{As Figure~\ref{f:05358-model}, for G327--0.6.}
\end{figure*} 

\begin{figure*}[t]
\centering
\includegraphics[width=10cm,angle=0]{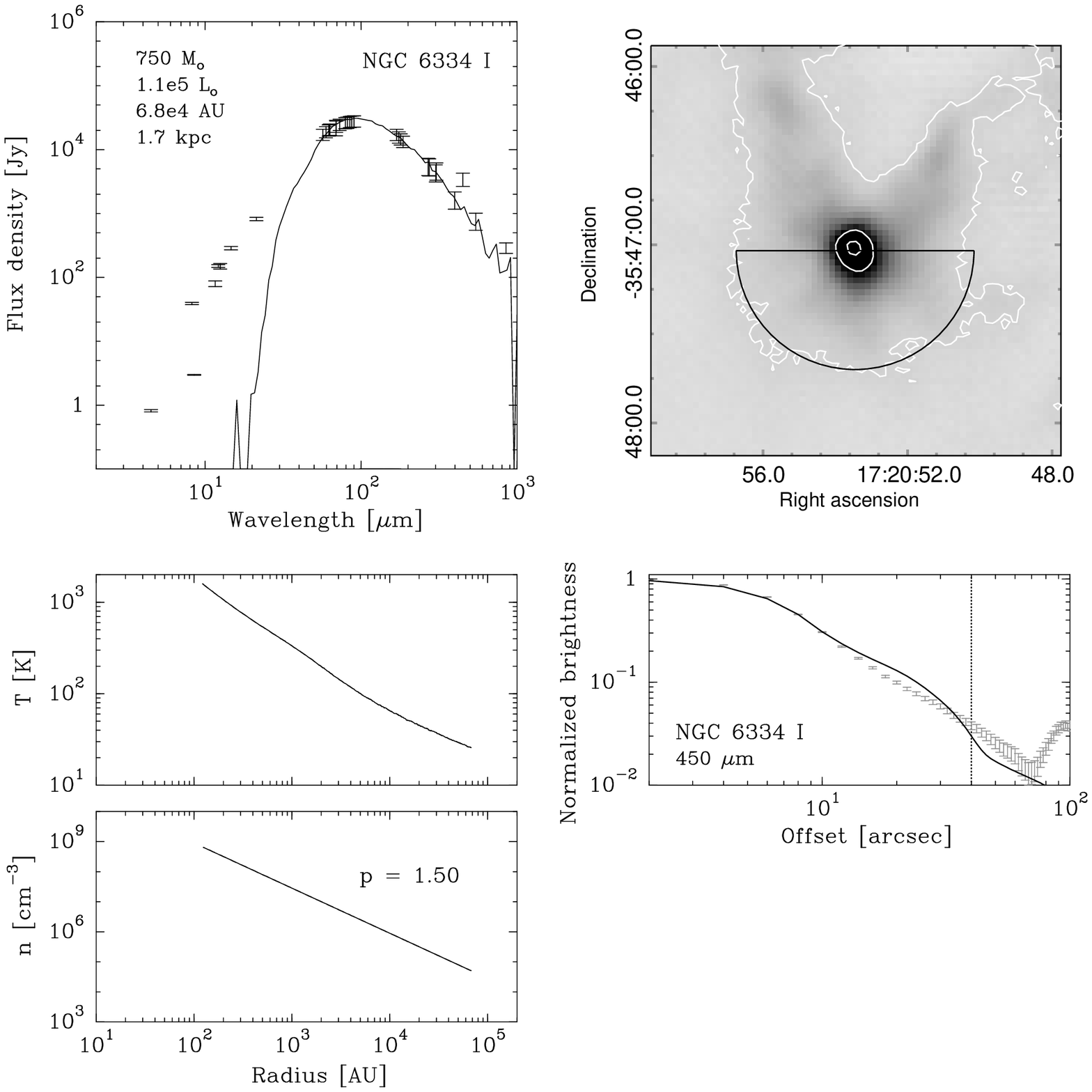}
\caption{As Figure~\ref{f:05358-model}, for NGC 6334I.}
\end{figure*} 

\clearpage
\newpage

\begin{figure*}[t]
\centering
\includegraphics[width=10cm,angle=0]{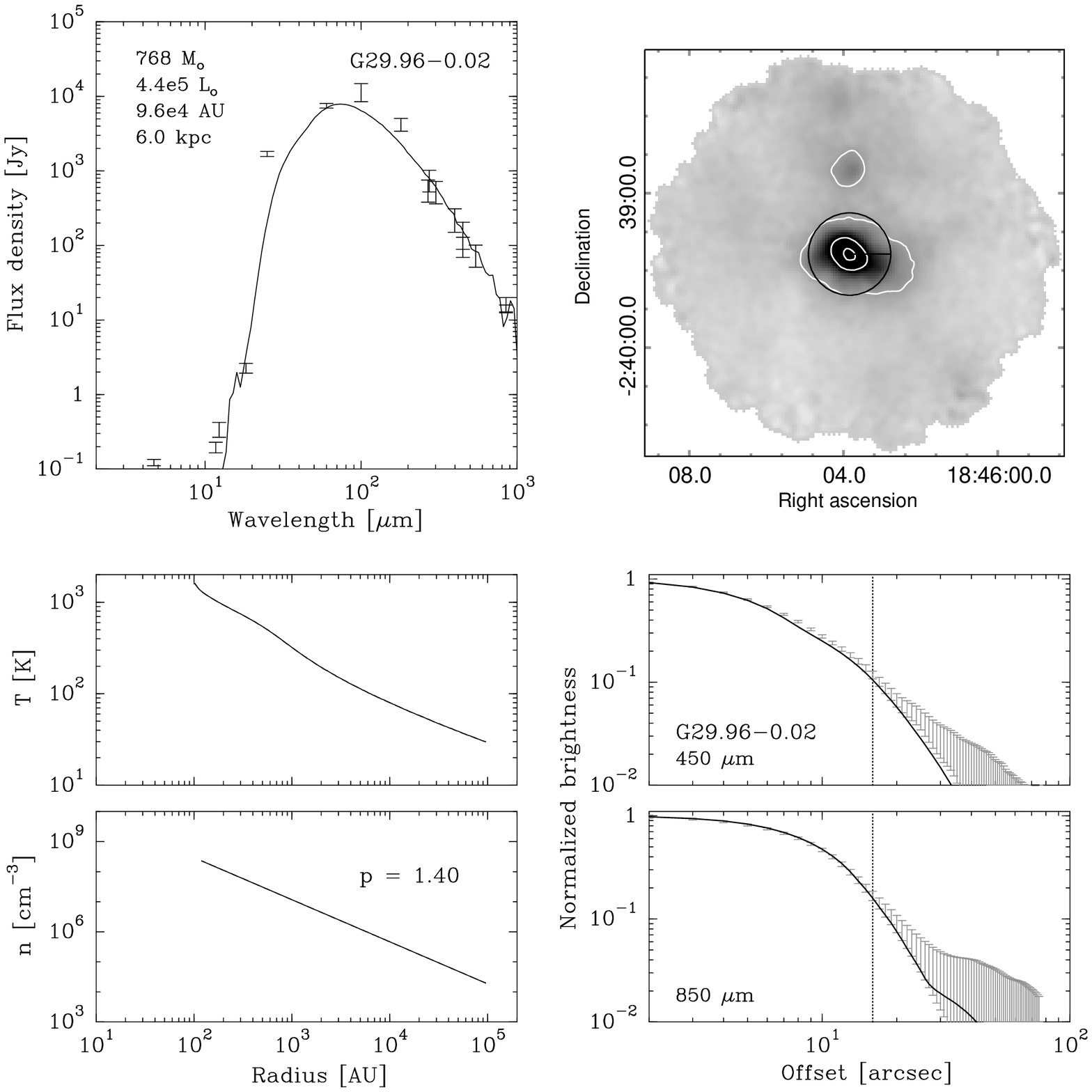}
\caption{As Figure~\ref{f:05358-model}, for G29.96.}
\end{figure*} 

\begin{figure*}[t]
\centering
\includegraphics[width=10cm,angle=0]{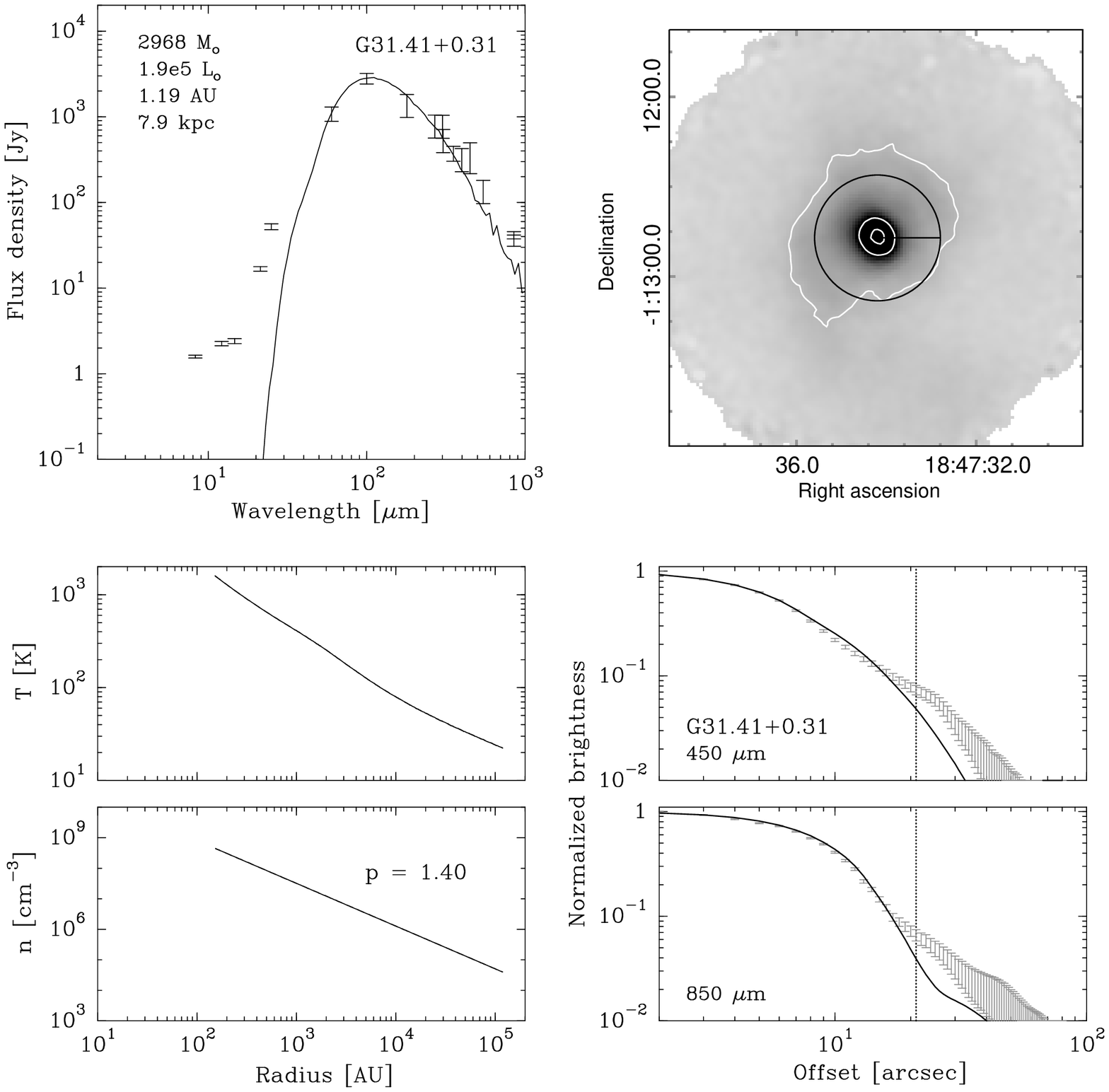}
\caption{As Figure~\ref{f:05358-model}, for G31.41.}
\end{figure*} 

\clearpage
\newpage

\begin{figure*}[t]
\centering
\includegraphics[width=10cm,angle=0]{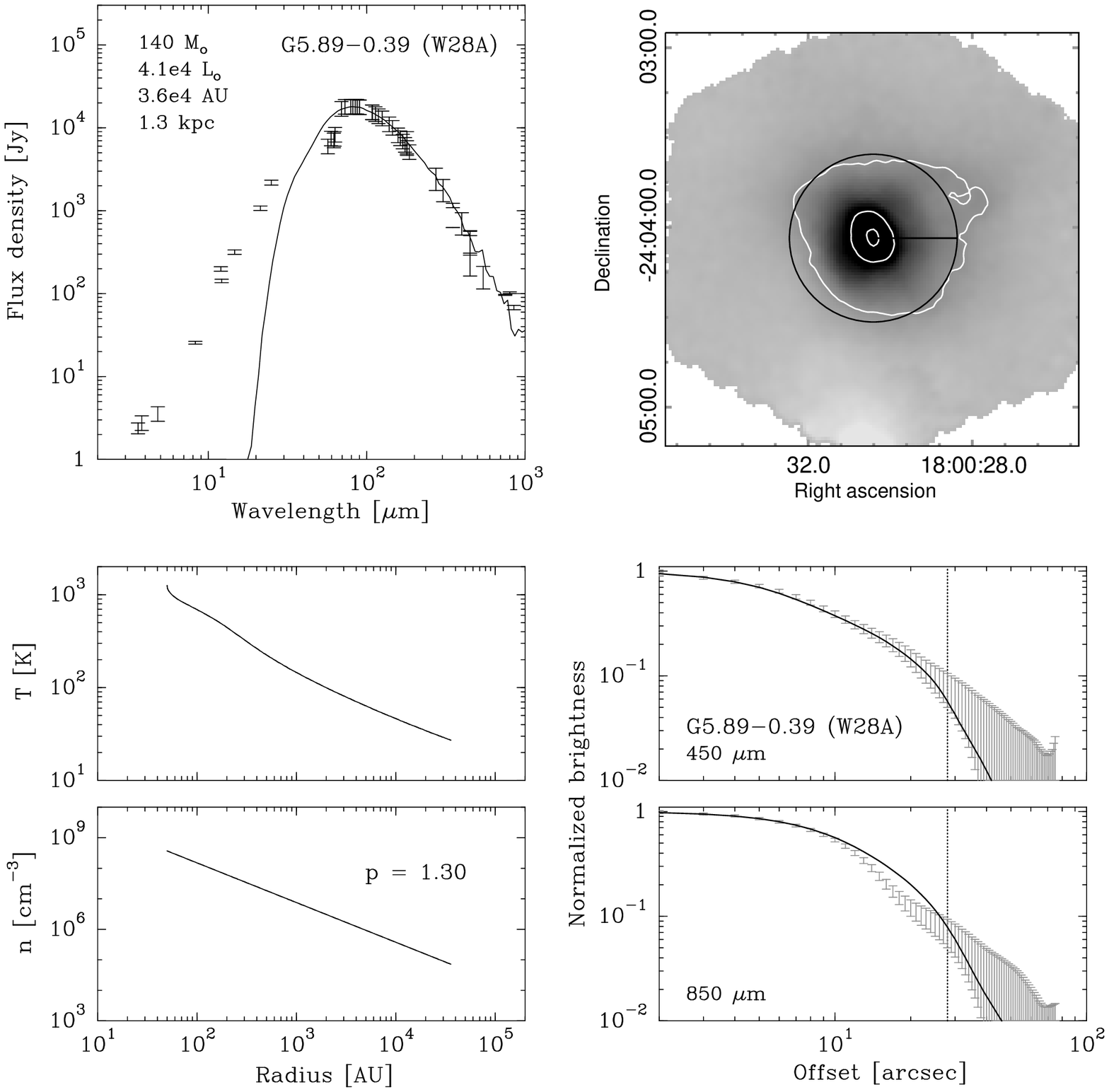}
\caption{As Figure~\ref{f:05358-model}, for G5.89.}
\end{figure*} 

\begin{figure*}[t]
\centering
\includegraphics[width=10cm,angle=0]{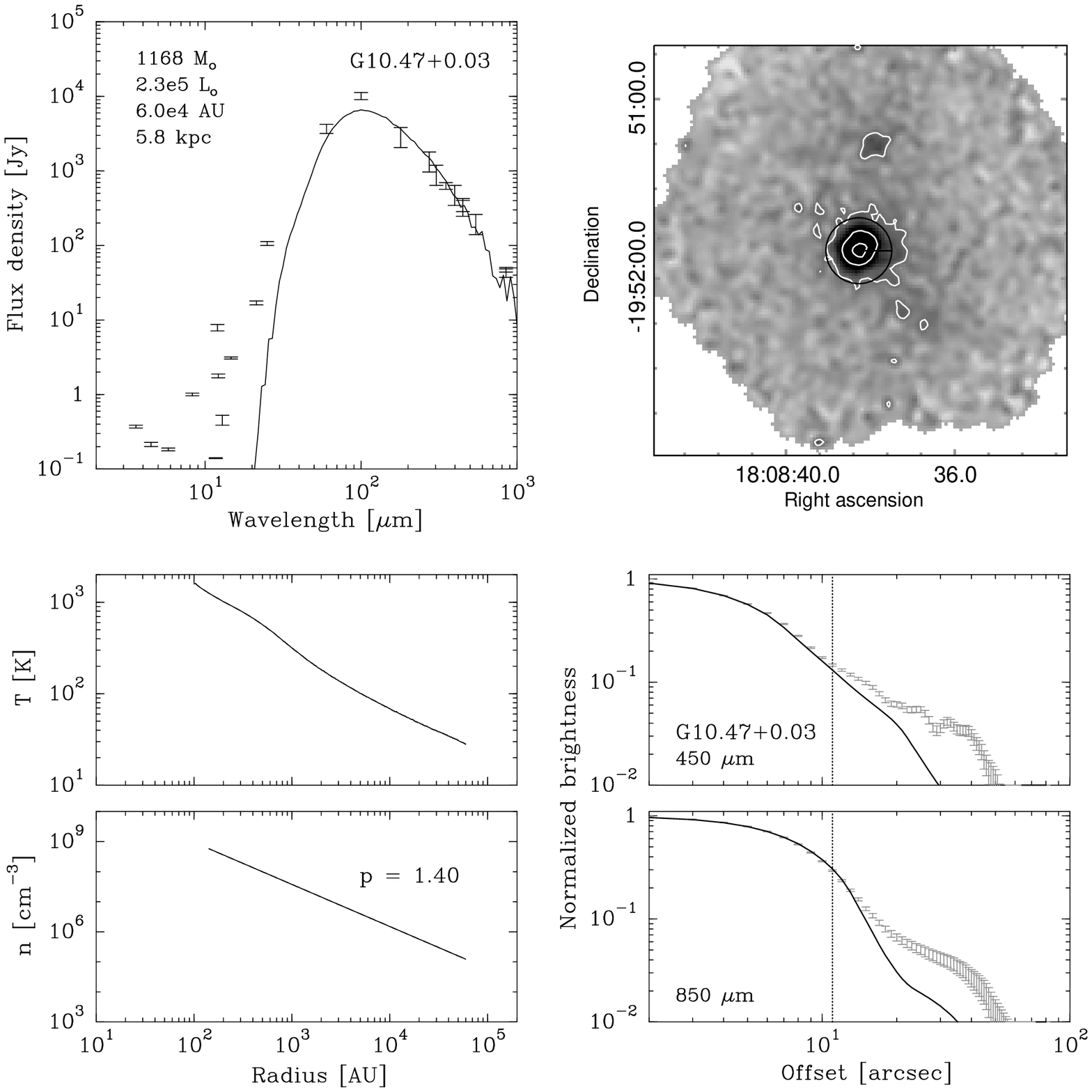}
\caption{As Figure~\ref{f:05358-model}, for G10.47.}
\end{figure*} 

\clearpage
\newpage

\begin{figure*}[t]
\centering
\includegraphics[width=10cm,angle=0]{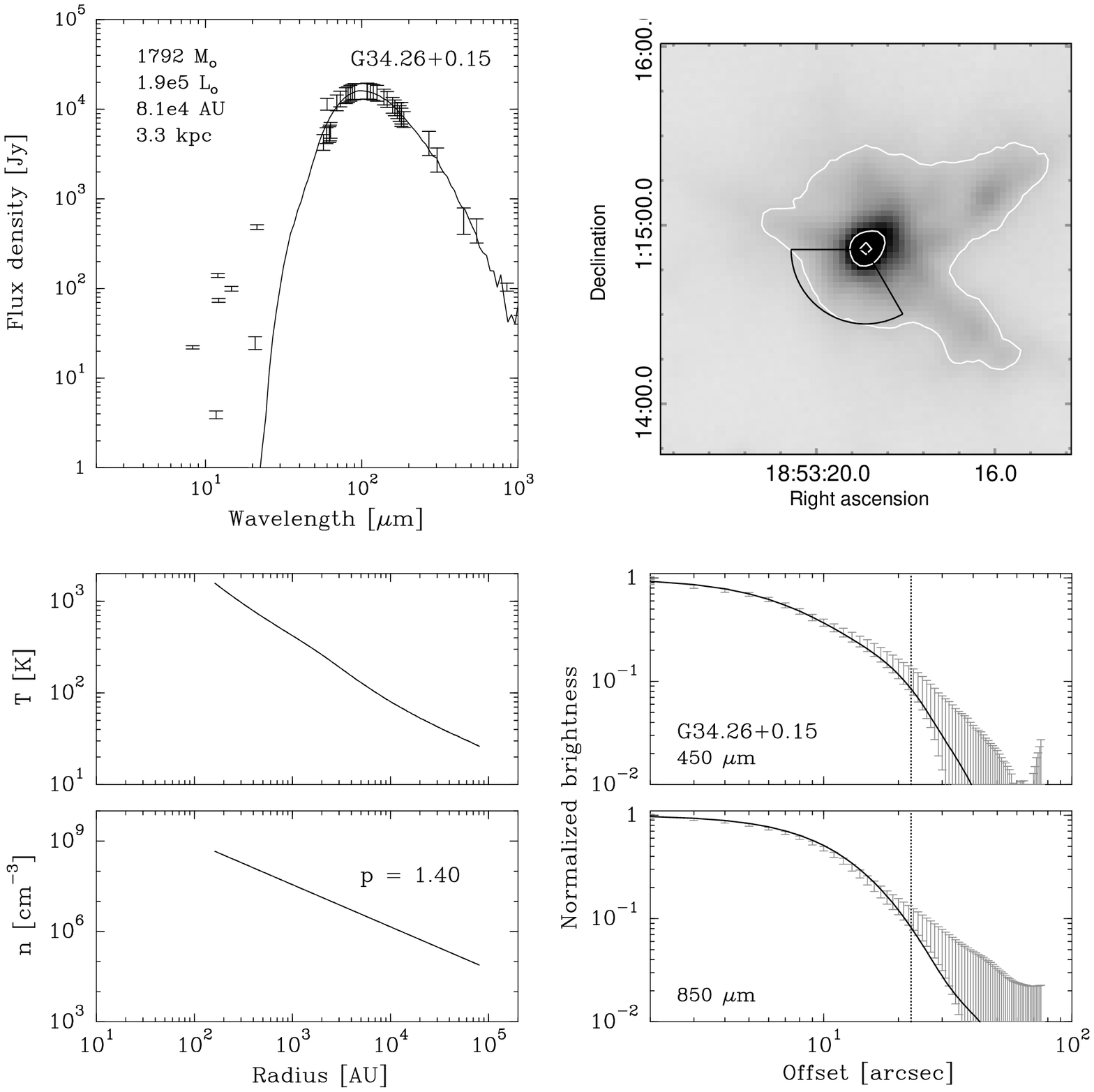}
\caption{As Figure~\ref{f:05358-model}, for G34.26.}
\end{figure*} 

\begin{figure*}[t]
\centering
\includegraphics[width=10cm,angle=0]{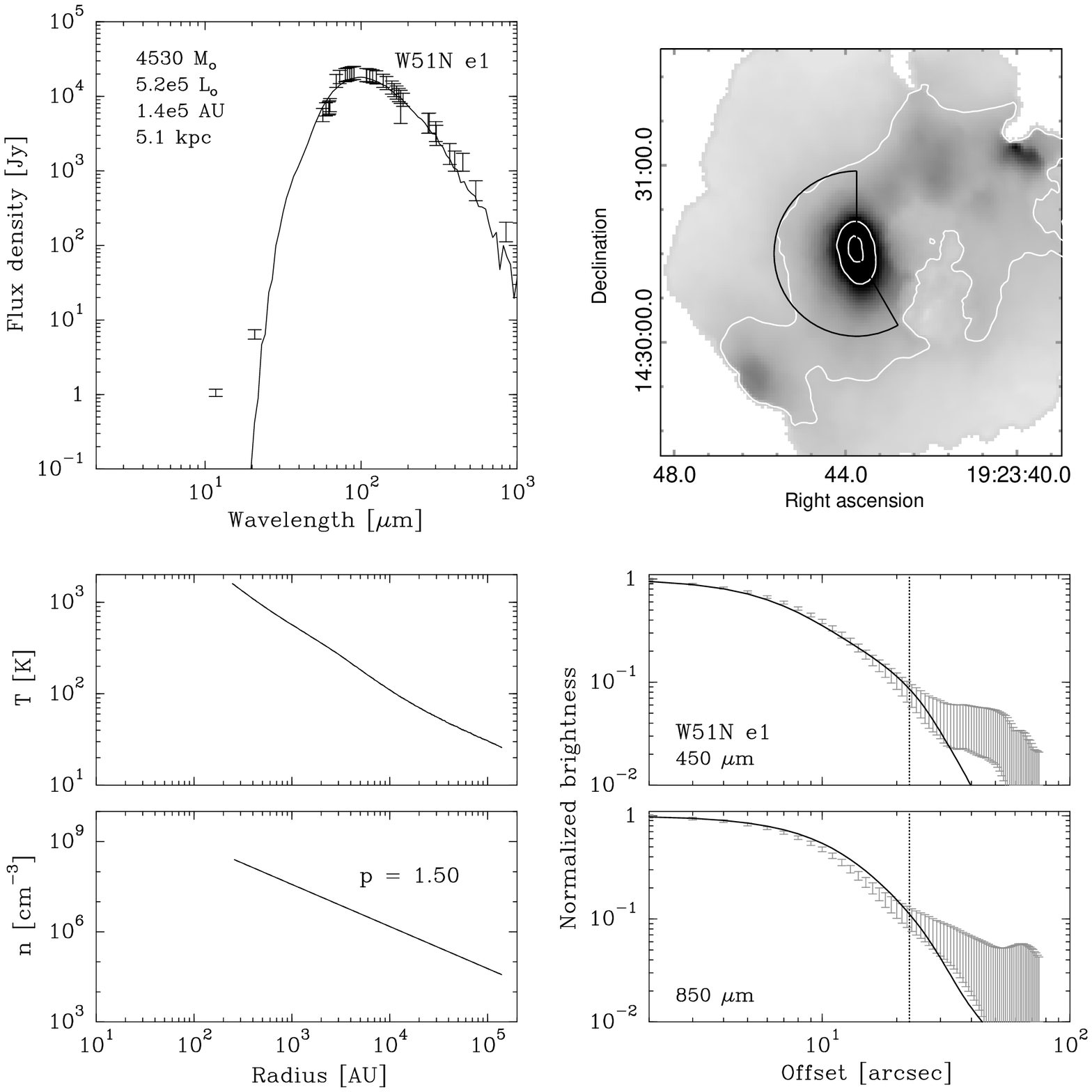}
\caption{As Figure~\ref{f:05358-model}, for W51e.}
\end{figure*} 

\clearpage
\newpage

\begin{figure*}[t]
\centering
\includegraphics[width=10cm,angle=0]{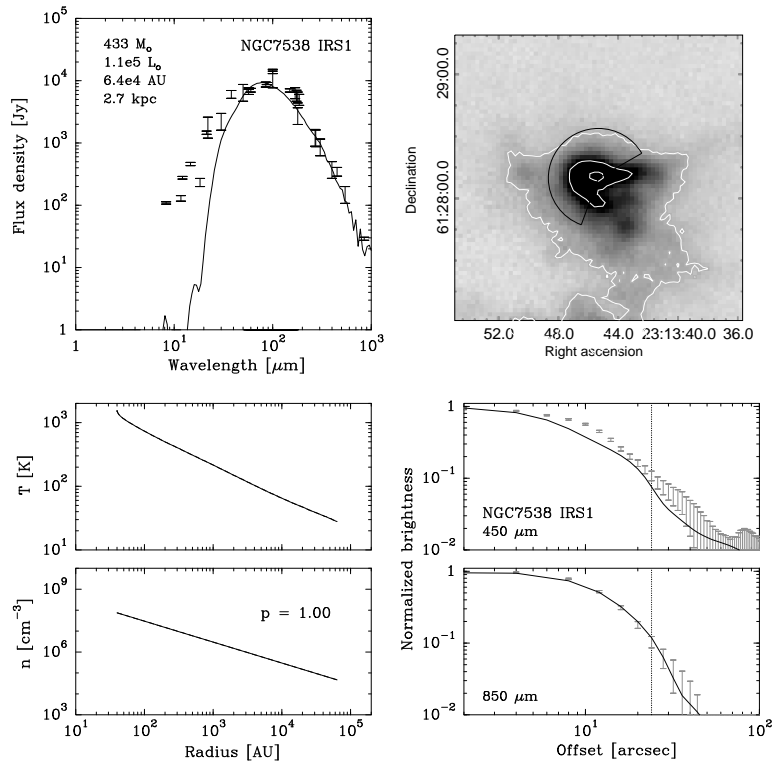}
\caption{As Figure~\ref{f:05358-model}, for NGC 7538 IRS1.}
\end{figure*} 

\clearpage
\newpage

\section{Correlation analysis}
\label{app:corr}

\begin{figure}[p]
\centering
\includegraphics[width=7cm,angle=0]{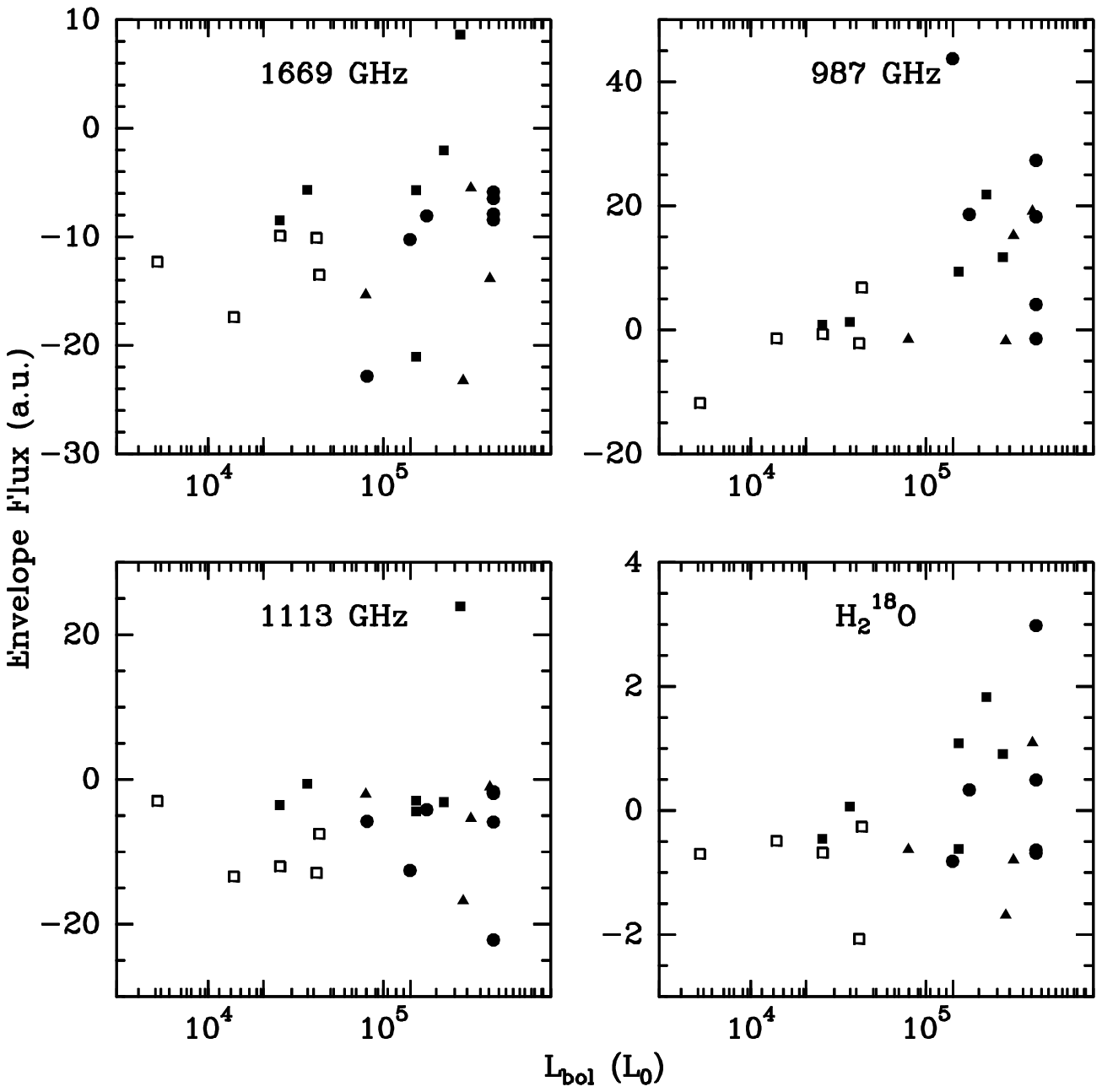}
\bigskip
\includegraphics[width=7cm,angle=0]{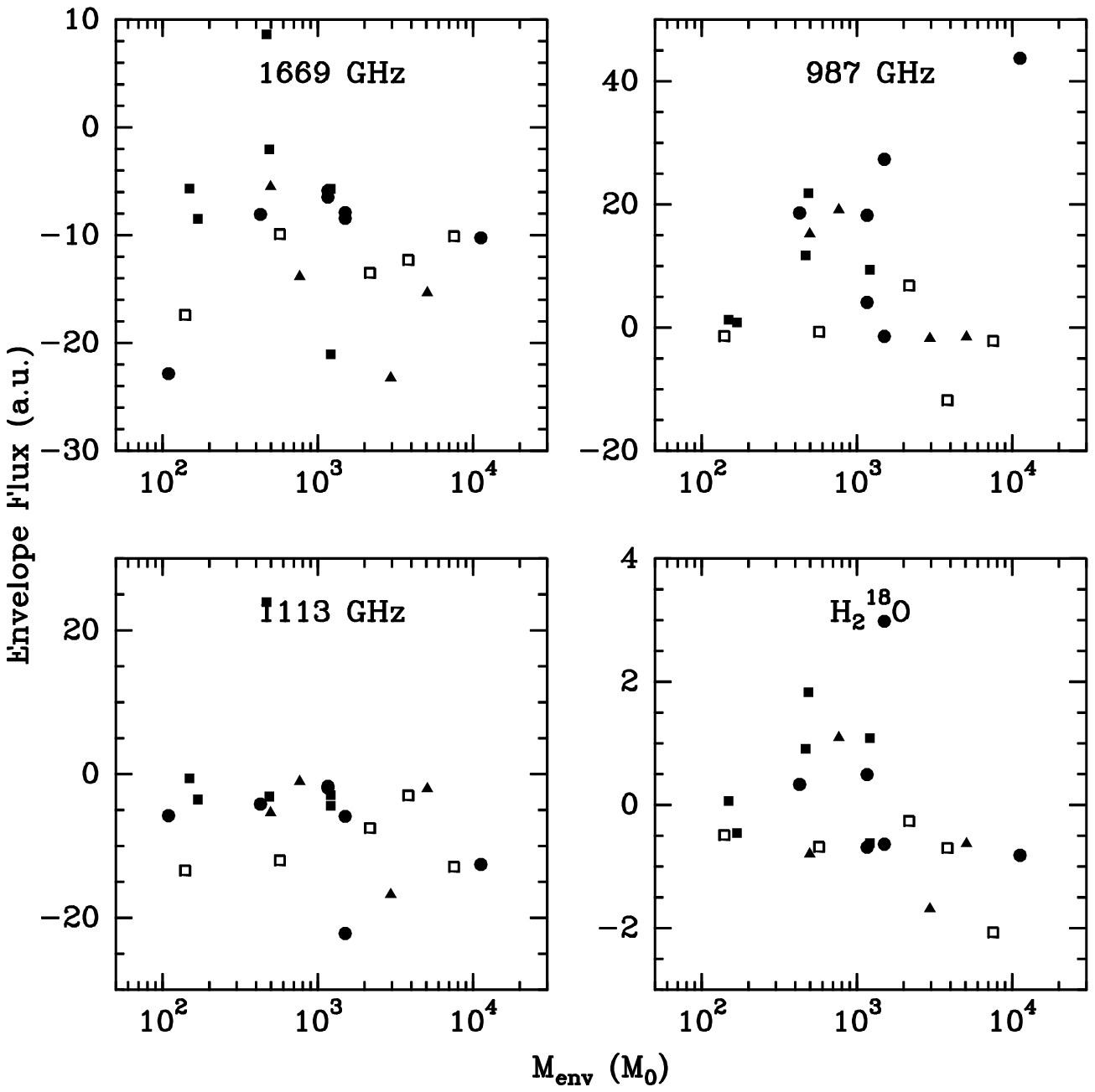}
\bigskip
\includegraphics[width=7cm,angle=0]{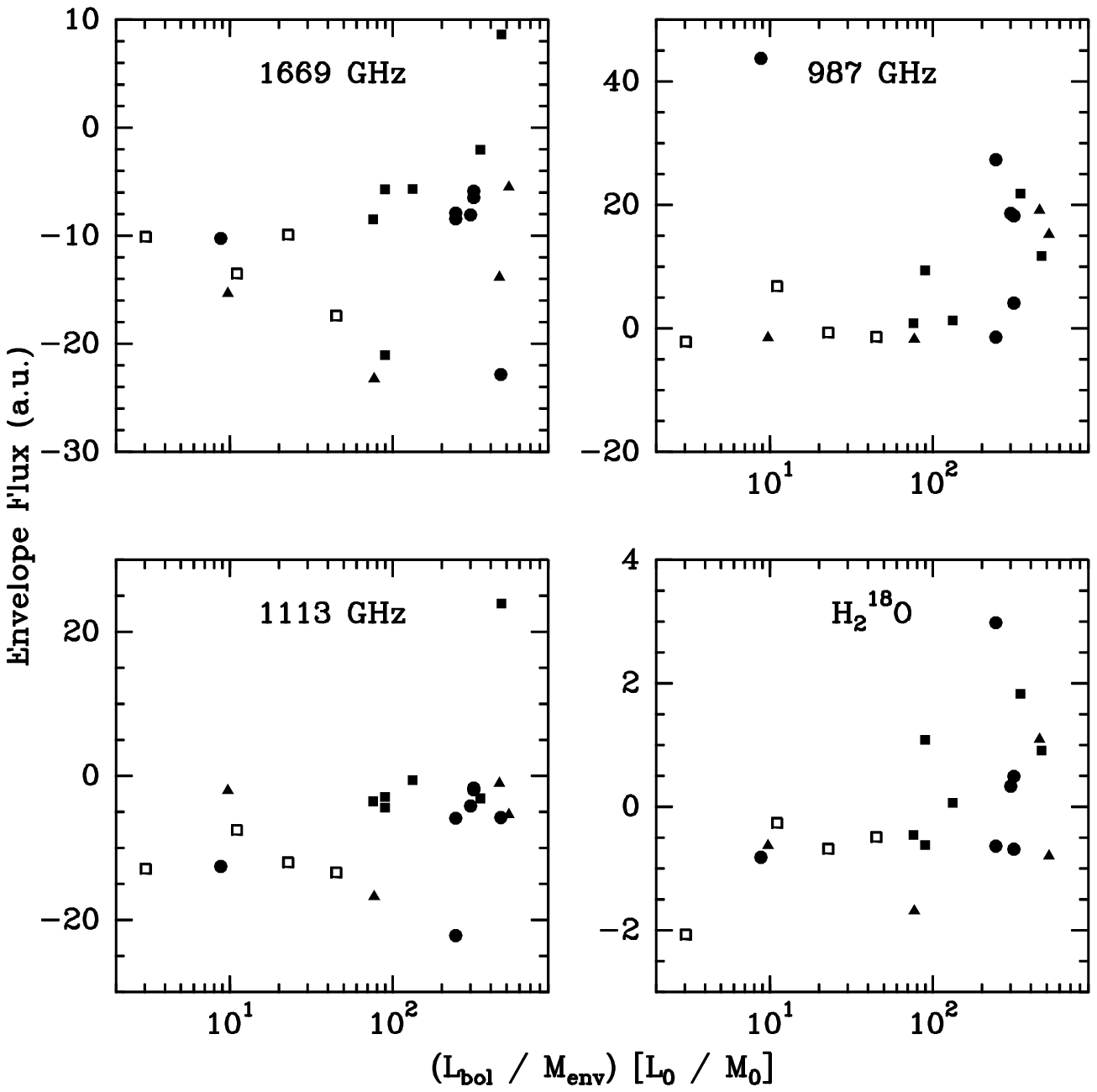}
\caption{Observed line fluxes (positive: emission) or absorbances (negative: absorption) versus bolometric luminosity $L$ of the source (top), versus envelope mass $M$ (middle), and versus the ratio $L/M$ (bottom). Open squares: mid-IR-quiet HMPOs; filled squares: mid-IR-bright HMPOs; triangles: hot molecular cores; circles: ultracompact H{\sc II} regions. Uncertainties in line fluxes are $\approx$10\%, while the typical uncertainty of masses, luminosities and $L/M$ ratios is a factor of 2.}
\label{f:env-flux-corr}
\end{figure} 

\begin{figure}[p]
\centering
\includegraphics[width=7cm,angle=0]{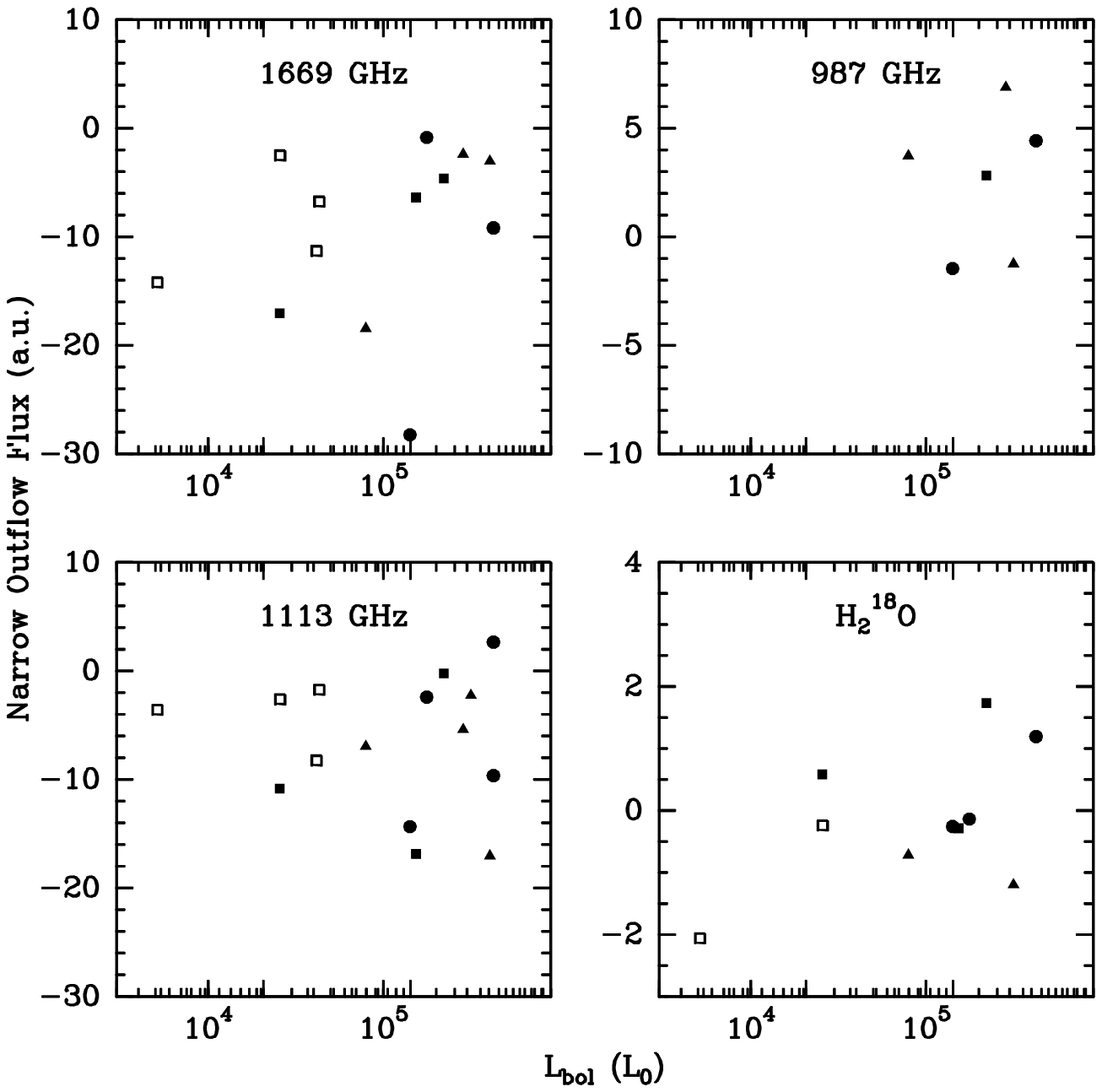}
\bigskip
\includegraphics[width=7cm,angle=0]{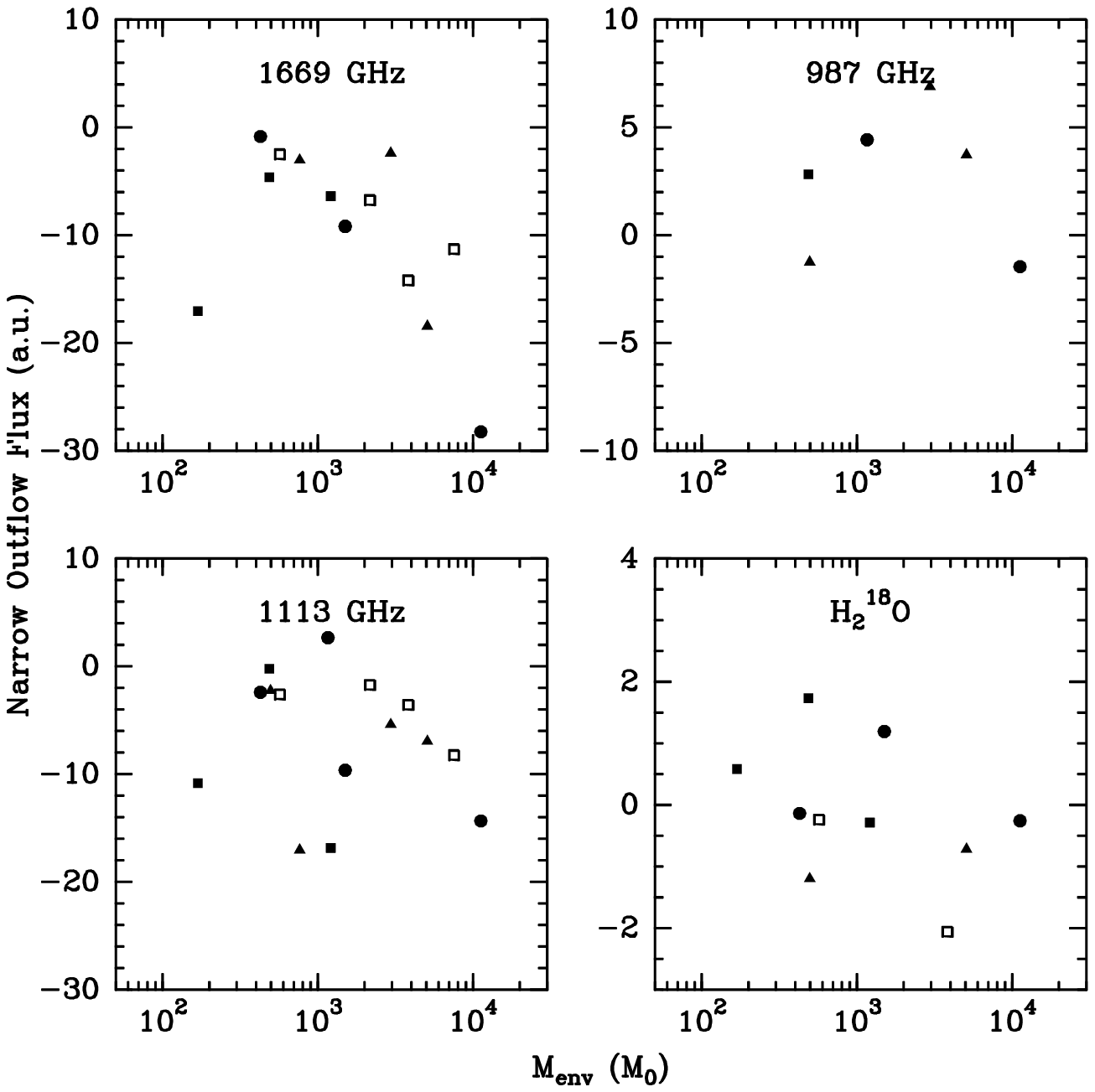}
\bigskip
\includegraphics[width=7cm,angle=0]{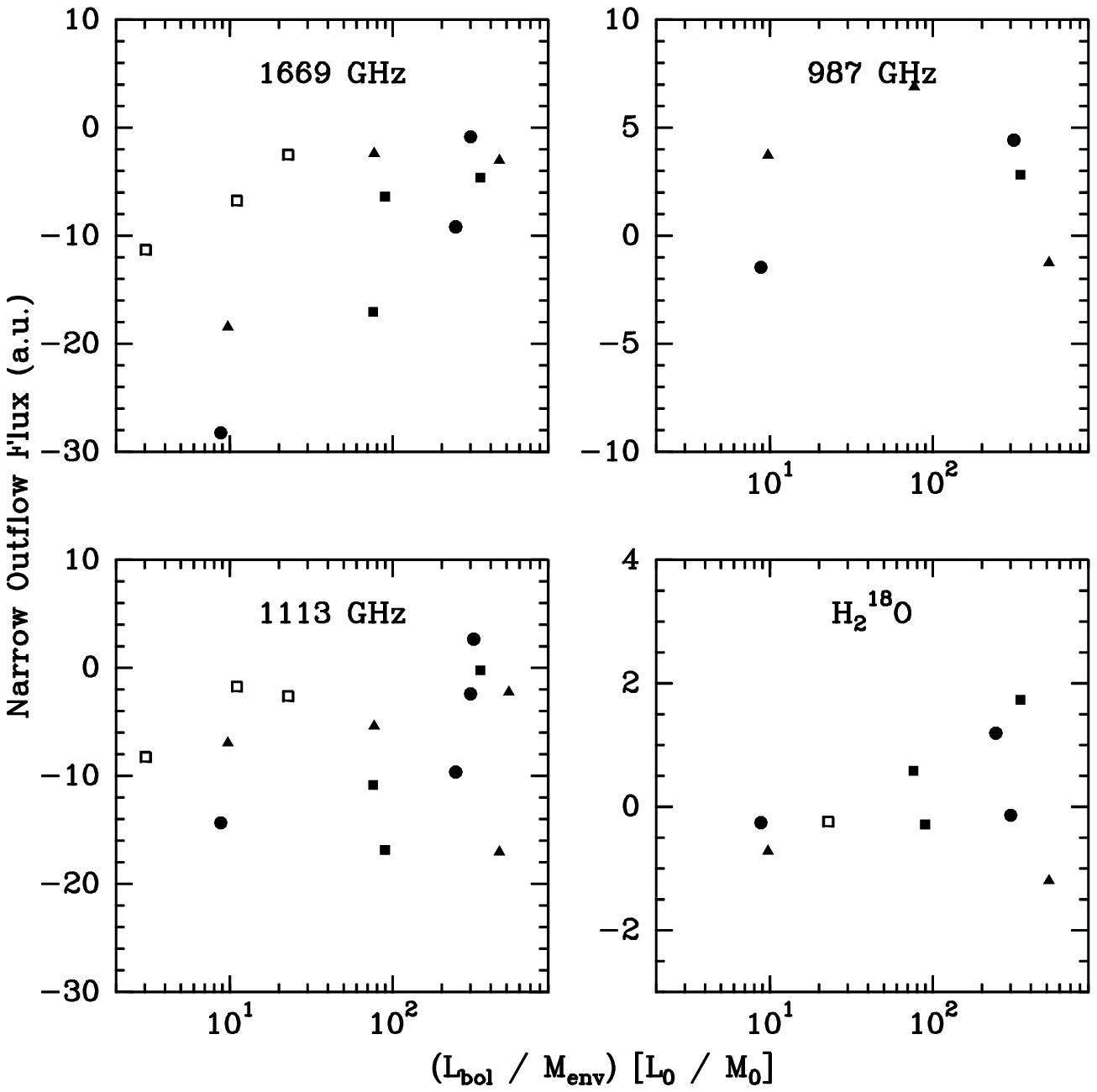}
\caption{As Figure~\ref{f:env-flux-corr}, for the narrow outflow component.}
\label{f:medium-flux-corr}
\end{figure} 

\begin{figure}[p]
\centering
\includegraphics[width=7cm,angle=0]{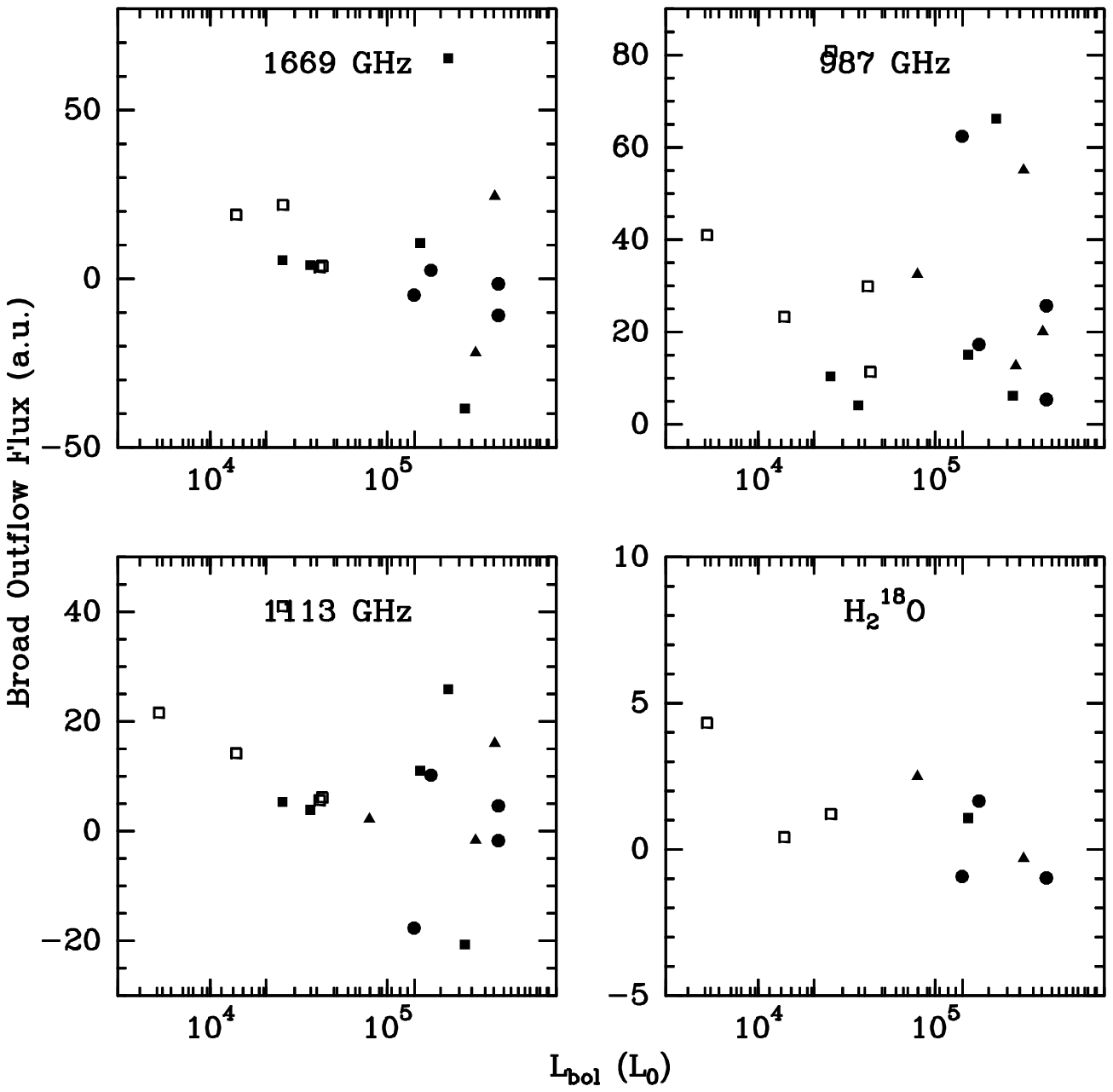}
\bigskip
\includegraphics[width=7cm,angle=0]{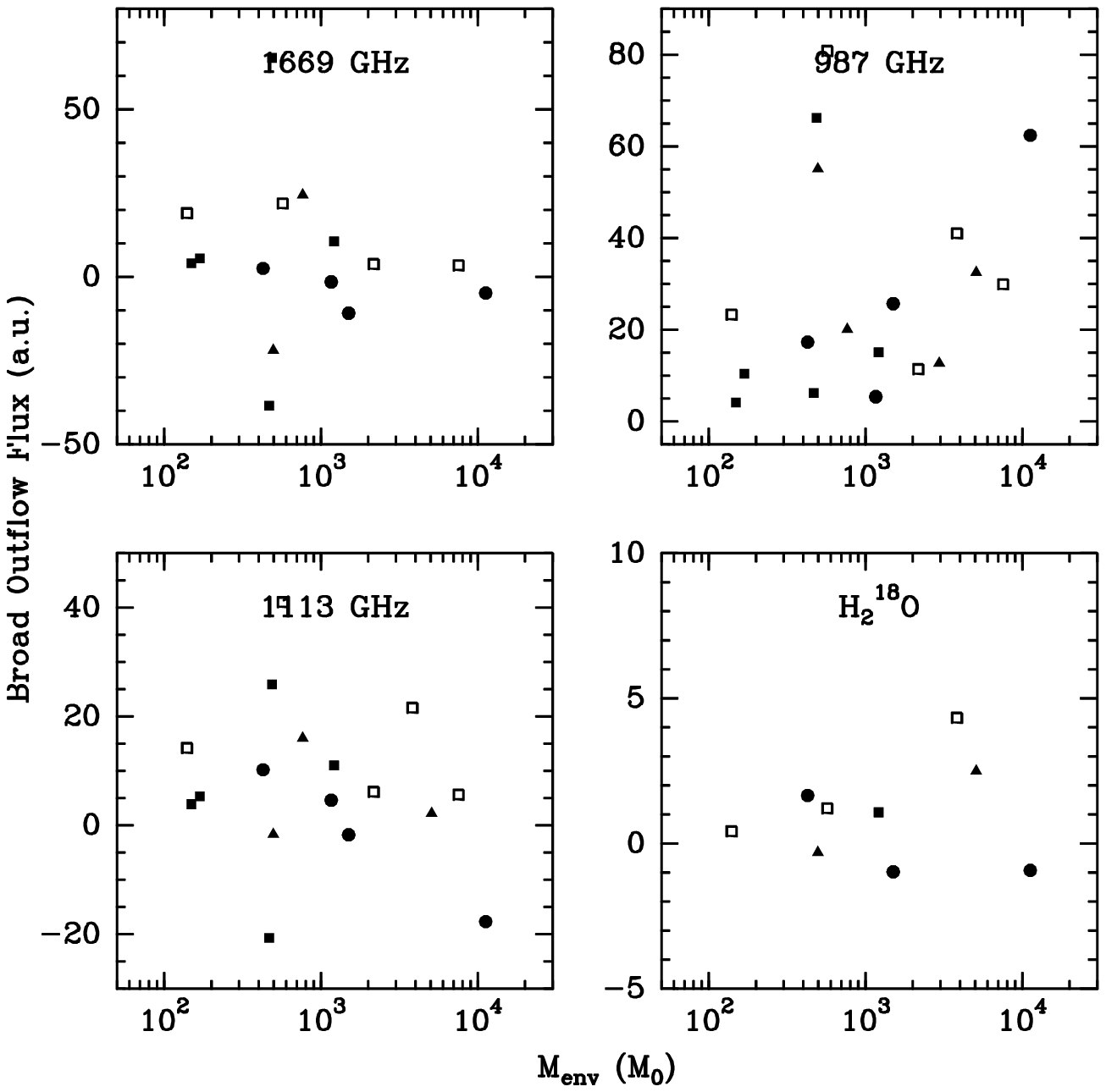}
\bigskip
\includegraphics[width=7cm,angle=0]{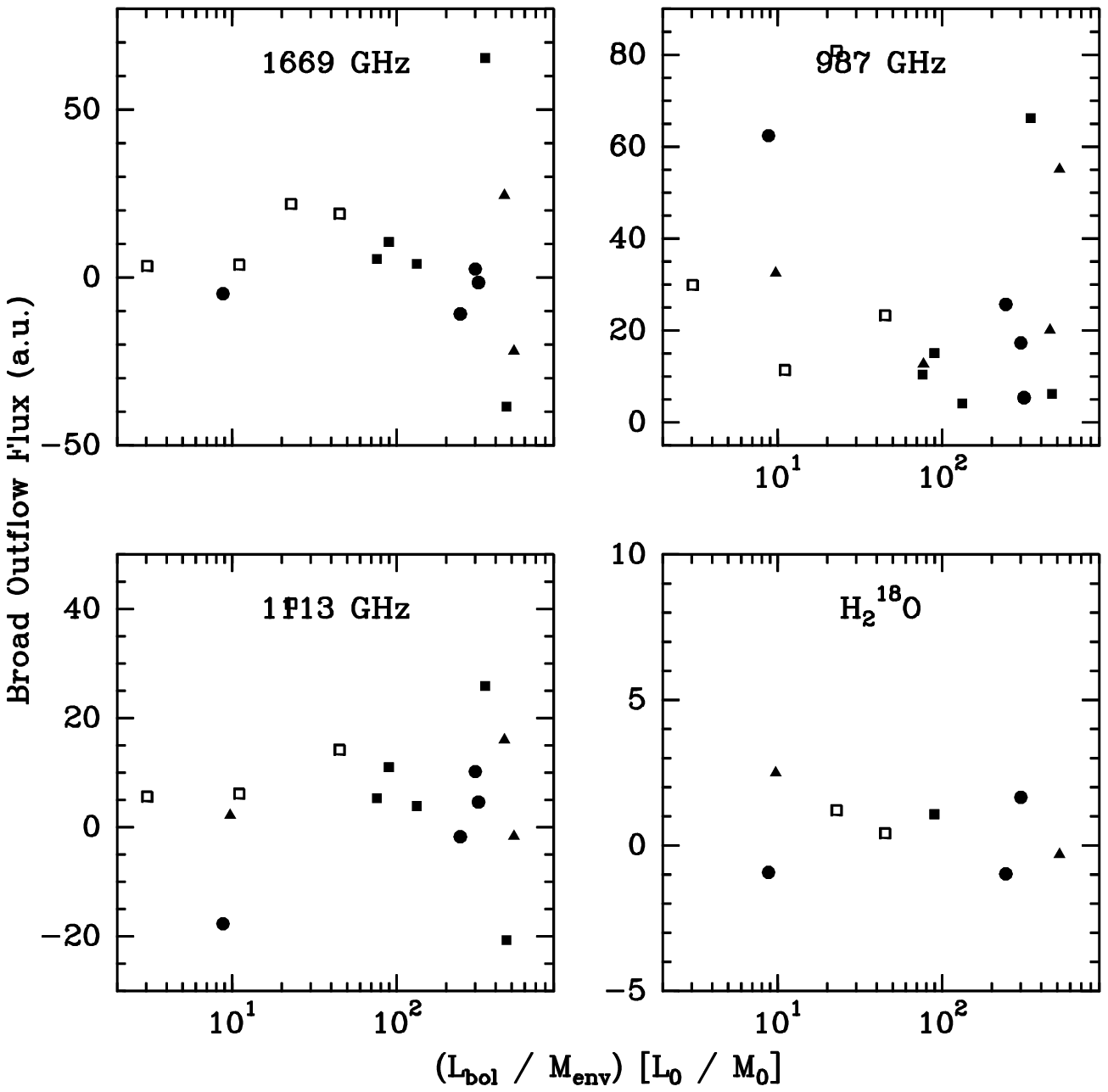}
\caption{As Figure~\ref{f:env-flux-corr}, for the broad outflow component.}
\label{f:broad-flux-corr}
\end{figure} 

\begin{figure}[p]
\centering
\includegraphics[width=7cm,angle=0]{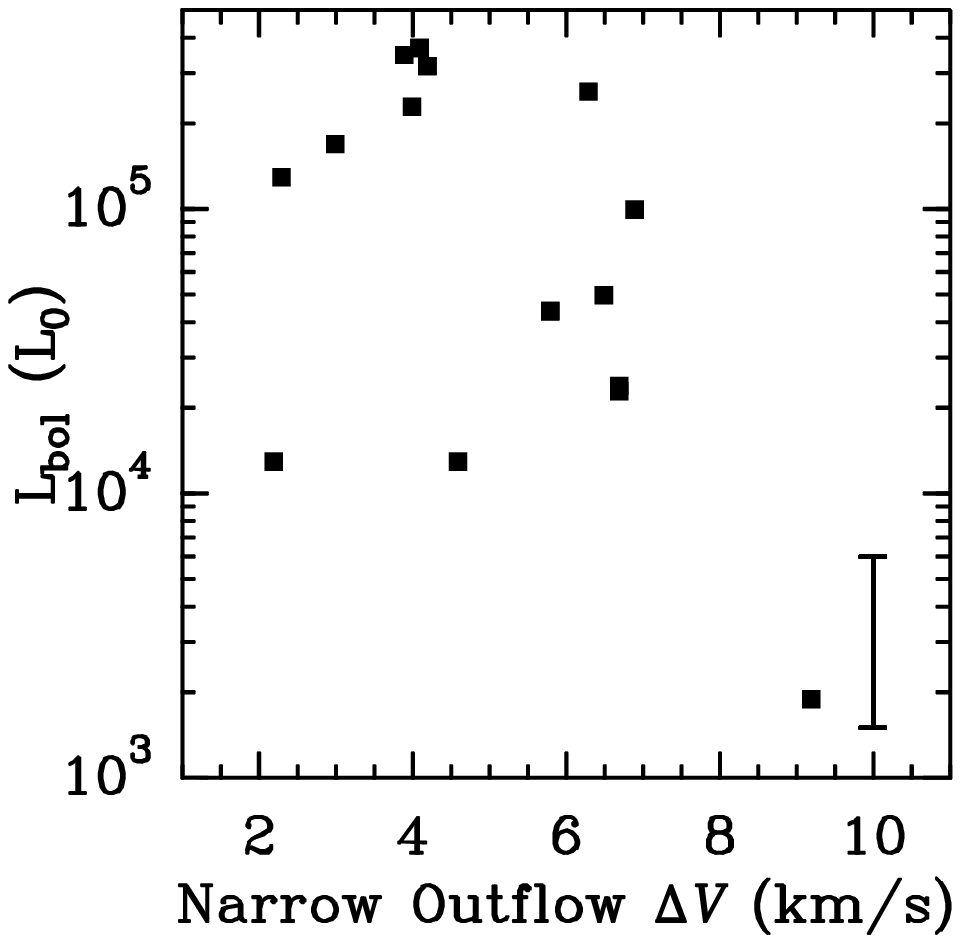}
\bigskip
\includegraphics[width=7cm,angle=0]{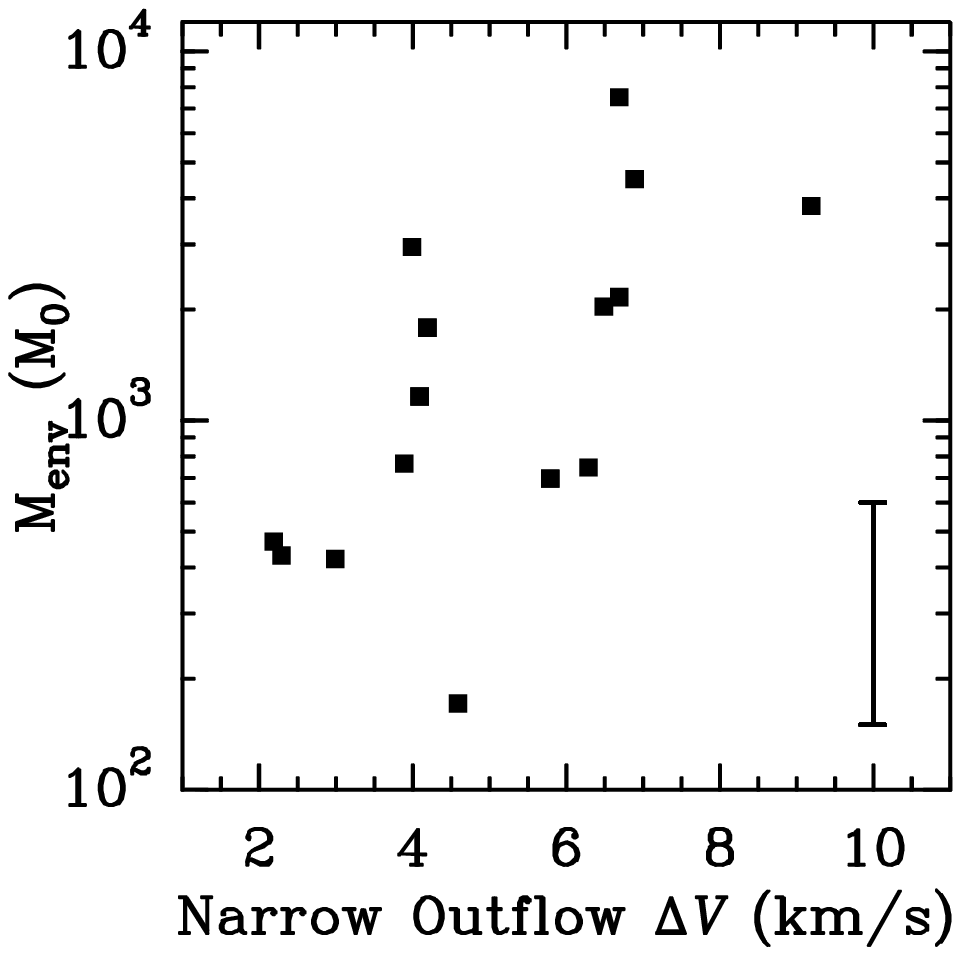}
\bigskip
\includegraphics[width=7cm,angle=0]{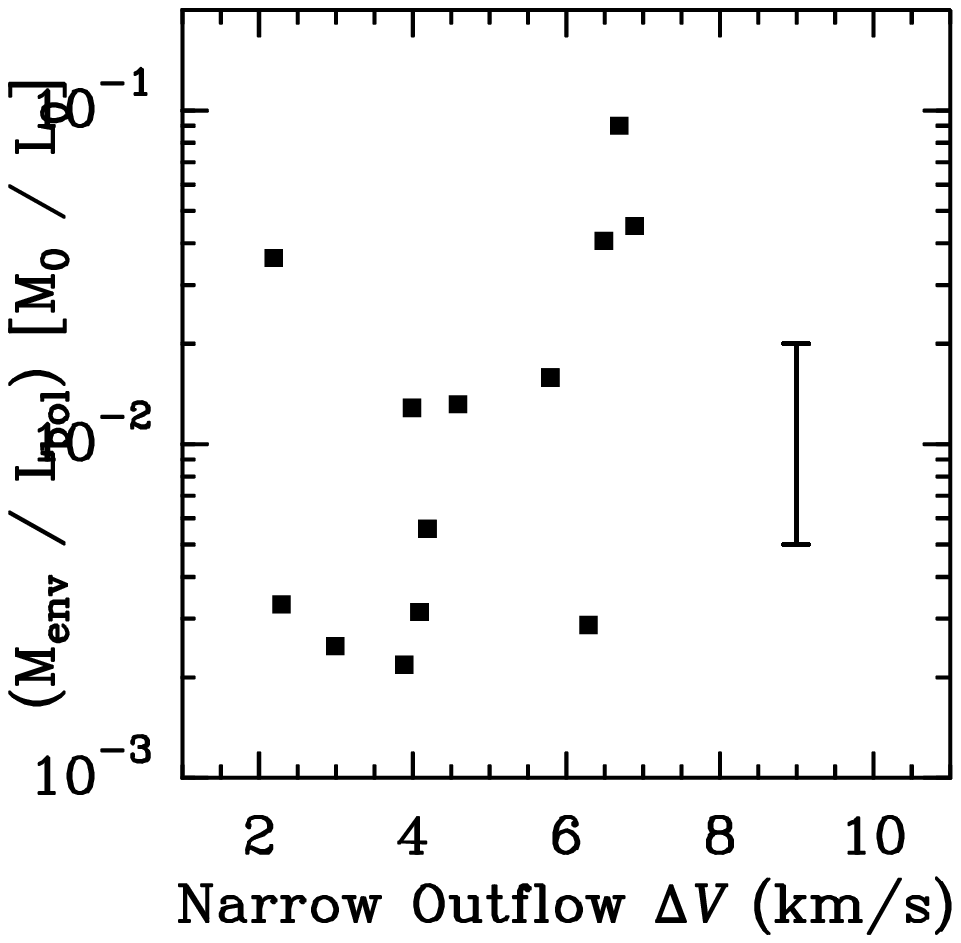}
\caption{Line width of the narrow outflow component versus bolometric luminosity $L$ of the source (top), versus envelope mass $M$ (middle), and versus the ratio $L/M$ (bottom). The error bar denotes the typical uncertainty in mass, luminosity, and $L/M$ ratio; the uncertainty in line width is smaller than the symbol size.}
\label{f:medium-width-corr}
\end{figure} 

\begin{figure}[p]
\centering
\includegraphics[width=7cm,angle=0]{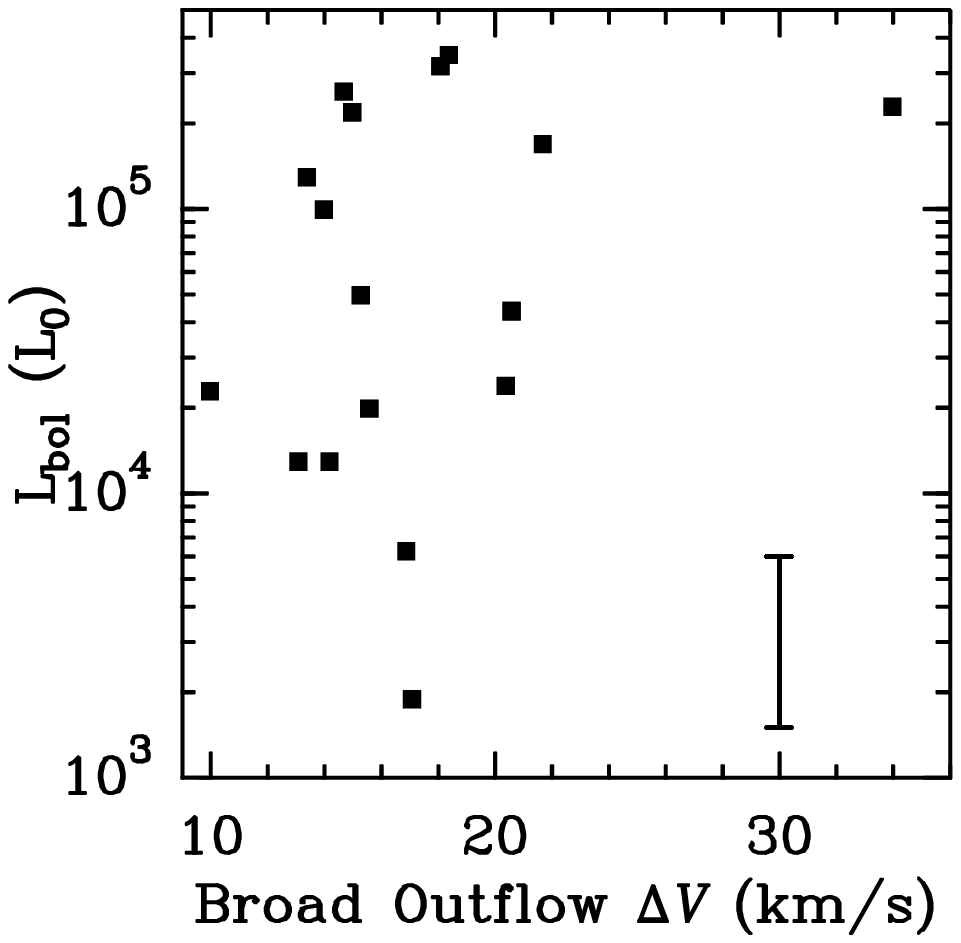}
\bigskip
\includegraphics[width=7cm,angle=0]{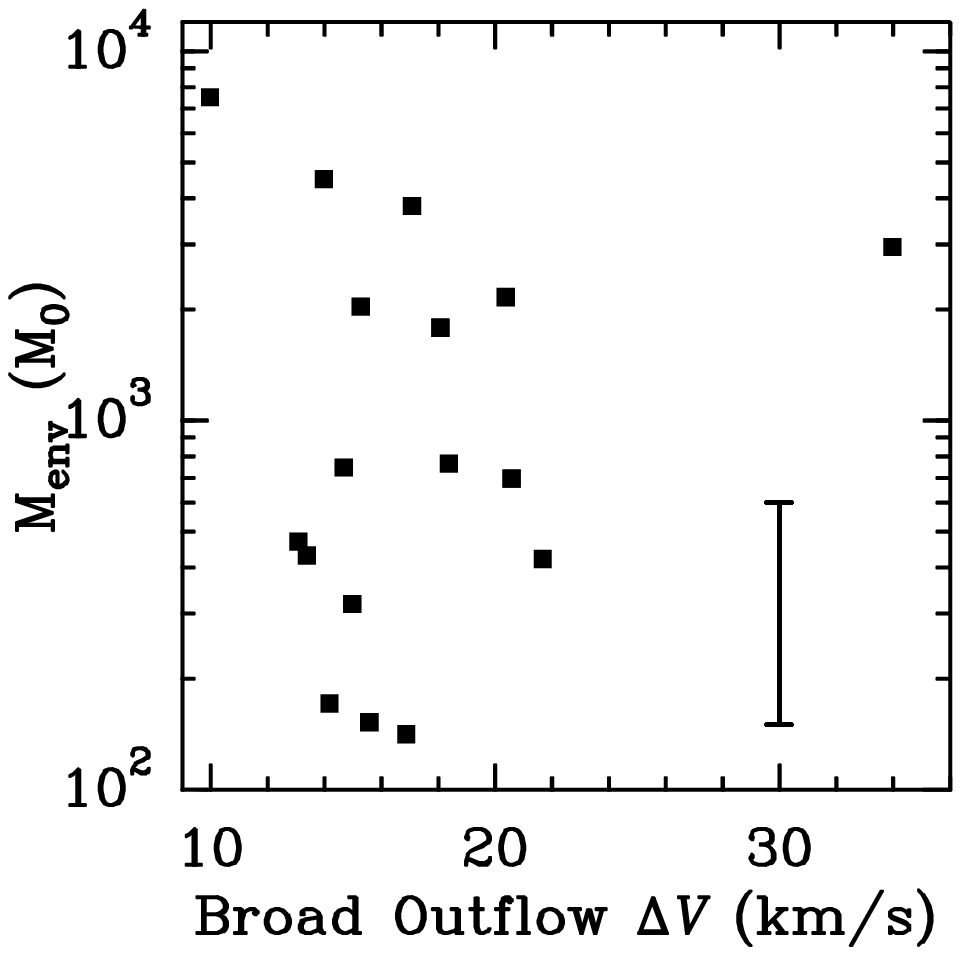}
\bigskip
\includegraphics[width=7cm,angle=0]{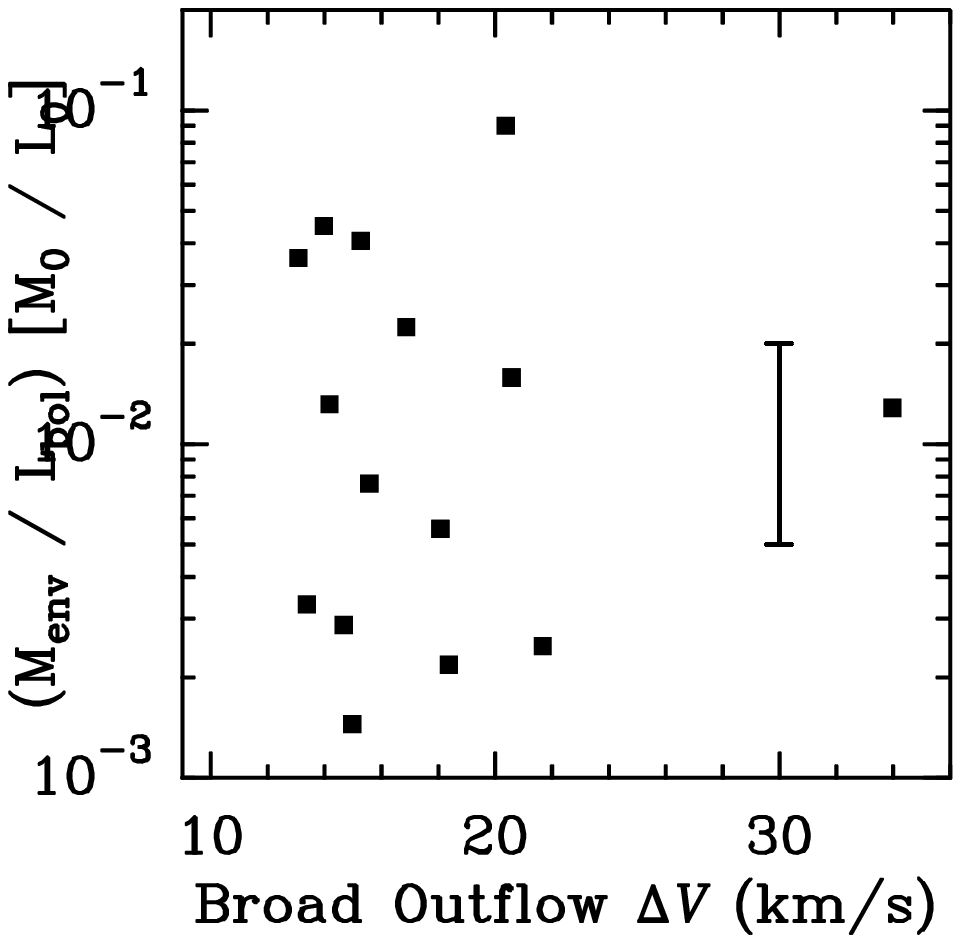}
\caption{As Figure~\ref{f:medium-width-corr}, for the broad outflow component.}
\label{f:broad-width-corr}
\end{figure}

\end{document}